\documentclass[longauth]{aa} 

\usepackage[utf8]{inputenc}
\usepackage{txfonts} 
\usepackage{graphicx} 
\usepackage{natbib} 
\usepackage{url} 
\usepackage{float}
\usepackage{graphicx}   
\usepackage{amsmath}    
\usepackage{amssymb}    
\usepackage[dvipsnames]{xcolor}
\usepackage{comment}

\usepackage[colorlinks = true,
            linkcolor = blue,
            urlcolor  = blue,
            citecolor = blue,
            anchorcolor = blue]{hyperref}

\usepackage{ulem}

\usepackage{longtable}

\usepackage{threeparttable}

\usepackage{colortbl}

\usepackage{subcaption}
\usepackage{mwe}

\usepackage{enumitem}

\usepackage{pifont}
\usepackage{dirtytalk}

\newcommand{\be}{\begin{equation}}
\newcommand{\ee}{\end{equation}}
\newcommand{\mstar}{M_{\star}}
\newcommand{\rstar}{R_{\star}}
\newcommand{\rp}{R_{\rm p}}
\newcommand{\rt}{R_{\rm t}}

\newcommand{\mbh}{M_{\rm BH}}

\newcommand{\rbb}{R_{\rm BB}}

\newcommand{\lbol}{L_{\rm bol}}

\def\msun{\, \mathrm{M}_{\hbox{$\odot$}}}
\def\rsun{\, \mathrm{R}_{\hbox{$\odot$}}}

\newcommand{\orcid}[1]{\href{https://orcid.org/#1}
{\includegraphics[width=10pt]{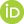}}}

\begin{document}

   \title{The fast transient AT~2023clx in the nearby LINER galaxy NGC~3799 as a tidal disruption of a very low-mass star}

    \titlerunning{The fast rising, nearby TDE AT~2023clx}

   \author{P. Charalampopoulos\inst{1}\fnmsep\thanks{Contact e-mail: \href{mailto:pachar@utu.fi}{pachar@utu.fi}}\orcid{0000-0002-0326-6715}\
          \and
          R. Kotak\inst{1}\
    \orcid{0000-0001-5455-3653}
    \and
    T. Wevers\inst{2,3}\
    \orcid{0000-0002-4043-9400}
          \
    \and
    G. Leloudas\inst{4}\
    \orcid{0000-0002-8597-0756}
    \and
    T. Kravtsov\inst{1,5}
    \orcid{0000-0003-0955-9102}
    \and
    M. Pursiainen\inst{6}
    \orcid{0000-0003-4663-4300}
    \and
    P. Ramsden\inst{7,8,9}
    \orcid{0009-0009-2627-2884}
    \and
    T. M. Reynolds\inst{10,1}
    \orcid{0000-0002-1022-6463}
    \and
    A. Aamer\inst{9}
    \orcid{0000-0002-9085-8187}
    \and
    J.~P. Anderson\inst{5,11}
    \orcid{0000-0003-0227-3451}
    \and
    I. Arcavi\inst{12}
    \orcid{0000-0001-7090-4898}
    \and
    Y.-Z. Cai\inst{13,14,15}
    \orcid{0000-0002-7714-493X}
    \and
    T.-W. Chen\inst{16}
    \orcid{0000-0002-1066-6098}
    \and
    M. Dennefeld\inst{17}
    \orcid{0000 0002 8197 5410 }
    \and
    L. Galbany\inst{18,19}
    \orcid{0000-0002-1296-6887}
    \and
    M. Gromadzki\inst{20}
    \orcid{0000-0002-1650-1518}
    \and
    C.P. Guti\'errez\inst{19,18}
    \orcid{0000-0003-2375-2064}
    \and
    N. Ihanec\inst{20,21}
    \and
    T. Kangas\inst{22,1}
    \orcid{0000-0002-5477-0217}
    \and
    E. Kankare\inst{1}
    \orcid{0000-0001-8257-3512}
    \and
    E. Kool\inst{22,1,23}
    \orcid{0000-0002-7252-3877}
    \and
    A. Lawrence\inst{24}
    \orcid{0000–0002–3134–6093}
    \and
    P. Lundqvist\inst{23}
    \orcid{0000-0002-3664-8082}
    \and
    L. Makrygianni\inst{25}
    \orcid{0000-0002-7466-4868}
    \and
    S. Mattila\inst{1,26}
    \orcid{0000-0001-7497-2994}
    \and
    T. E. M\"uller-Bravo\inst{18,19}
    \orcid{0000-0003-3939-7167}
    \and
    M. Nicholl\inst{9}
    \orcid{0000-0002-2555-3192}
    \and
    F. Onori\inst{27}
    \orcid{0000-0001-6286-1744}
    \and
    A. Sahu\inst{6}
    \orcid{0009-0007-6825-3230}
    \and
    S.J. Smartt\inst{28,9}
    \orcid{0000-0002-8229-1731}
    \and
    J. Sollerman\inst{23}
    \orcid{0000-0003-1546-6615}
    \and
    Y. Wang\inst{29,30}
    \orcid{0000-0003-3207-5237}
    \and
    D. R. Young\inst{9}
    \orcid{0000–0002–1229–2499}
          \
          }

   \institute{Department of Physics and Astronomy, University of Turku, FI-20014 Turku, Finland
   \and
   Space Telescope Science Institute, 3700 San Martin Drive, Baltimore, MD 21218, USA
   \and
   European Southern Observatory, Alonso de Córdova 3107, Vitacura, Santiago, Chile
   \and
   DTU Space, National Space Institute, Technical University of Denmark, Elektrovej 327, DK-2800 Kgs. Lyngby, Denmark
   \and
   European Southern Observatory, Alonso de C\'ordova 3107, Casilla 19, Santiago, Chile
   \and
   Department of Physics, University of Warwick, Gibbet Hill Road, Coventry, CV4 7AL, UK
   \and
   Institute for Gravitational Wave Astronomy, University of Birmingham, Birmingham B15 2TT, UK
   \and
   School of Physics and Astronomy, University of Birmingham, Birmingham B15 2TT, UK
    \and
   Astrophysics Research Centre, School of Mathematics and Physics,
    Queen’s University Belfast, Belfast BT7 1NN, UK
    \and
    Cosmic Dawn Center (DAWN), Niels Bohr Institute, University of Copenhagen, 2200, Denmark
    \and
    Millennium Institute of Astrophysics MAS, Nuncio Monsenor Sotero Sanz 100, Off.
    104, Providencia, Santiago, Chile
    \and
    School of Physics and Astronomy, Tel Aviv University, Tel Aviv 69978, Israel
    \and
    Yunnan Observatories, Chinese Academy of Sciences, Kunming 650216, P.R. China
    \and
    Key Laboratory for the Structure and Evolution of Celestial Objects, Chinese Academy of Sciences, Kunming 650216, P.R. China
    \and
    International Centre of Supernovae, Yunnan Key Laboratory, Kunming 650216, P.R. China  
    \and
    Graduate Institute of Astronomy, National Central University, 300 Jhongda Road, 32001 Jhongli, Taiwan
    \and
    IAP/Paris, CNRS and Sorbonne University
    \and
    Institute of Space Sciences (ICE, CSIC), Campus UAB, Carrer de  
    Can Magrans, s/n, E-08193 Barcelona, Spain
    \and
    Institut d’Estudis Espacials de Catalunya (IEEC), E-08034  
    Barcelona, Spain
    \and
    Astronomical Observatory, University of Warsaw, Al. Ujazdowskie 4, PL-00-478 Warszawa, Poland
    \and
    Isaac Newton Group of Telescopes, ING, 38700, La Palma (S.C. Tenerife), Spain
    \and
    Finnish Centre for Astronomy with ESO (FINCA), FI-20014 University of Turku, Finland
    \and
    The Oskar Klein Centre, Department of Astronomy, Stockholm University, AlbaNova, 10691, Stockholm, Sweden
    \and
    Institute for Astronomy, University of Edinburgh, Royal Observatory, Blackford Hill, Edinburgh EH9 3HJ, UK
    \and
    Department of Physics, Lancaster University, Lancaster LA1 4YB, UK
    \and
    School of Sciences, European University Cyprus, Diogenes Street, Engomi, 1516 Nicosia, Cyprus
    \and
    INAF-Osservatorio Astronomico d'Abruzzo, via M. Maggini snc, I-64100 Teramo, Italy
    \and
    Department of Physics, University of Oxford, Keble Road, Oxford, OX1 3RH, UK 
    \and
    National Astronomical Observatories, Chinese Academy of Sciences, 20A Datun Road, Beijing 100101, China
    \and
    Physics \& Astronomy, University of Southampton, Southampton, Hampshire SO17~1BJ, UK
             }

   \date{Received - ; accepted -}

  \abstract{
We present an extensive analysis of the optical and ultraviolet (UV) properties of AT~2023clx, the closest optical/UV tidal disruption event (TDE) to date ($z=0.01107$), which occurred in the nucleus of the interacting low-ionization nuclear emission-line region (LINER) galaxy, NGC~3799. After correcting for the host reddening ($\rm E(B-V)_{h}$ = 0.179 mag), we find its peak absolute $g$-band magnitude to be $-18.03 \pm 0.07$ mag, and its peak bolometric luminosity to be ${L_{\rm pk}=(1.57\pm0.19)\times10^{43} \rm\,erg\,s^{-1}}$. AT~2023clx displays several distinctive features: first, it rose to peak within  $10.4\pm2.5$ days, making it the fastest rising TDE to date. 
Our SMBH mass estimate of  $\overline{M}_{BH}\approx10^{6.0} \msun$ ---estimated using several standard methods---  rules out the possibility of an intermediate-mass BH as the reason for the fast rise. 
  Dense spectral follow-up reveals a blue continuum that cools slowly and broad Balmer and \ion{He}{II} lines as well as weak \ion{He}{I} $\lambda\lambda$5876,6678 emission features that are typically seen in TDEs. The early, broad (width $\sim 15\,000\rm\,km\,s^{-1}$) profile of H$\alpha$ matches theoretical expectations from an optically thick outflow. A flat Balmer decrement ($L_{H\alpha}$/$L_{H\beta} \sim 1.58$) suggests that the lines are collisionally excited rather than being produced via photoionisation, in contrast to typical active galactic nuclei. A second distinctive feature, seen for the first time in TDE spectra, 
  is a sharp, narrow emission peak at a rest wavelength of $\sim$ 6353 \AA. This feature is clearly visible up to 10\,d post-peak; we attribute it to clumpy material preceding the bulk outflow, which manifests as a high-velocity component of H$\alpha$ ($-9\,584\rm\,km\,s^{-1}$). Its third distinctive feature is the rapid cooling during the first $\sim$ 20 days after peak, reflected as a break in the temperature evolution. Combining these findings, we propose a scenario for AT~2023clx involving the disruption of a very low-mass star ($\lesssim0.1\msun$) with an outflow launched in our line of sight and with disruption properties that led to efficient circularisation and prompt accretion disc formation, observed through a low-density photosphere.}
   

   \maketitle
%

\section{Introduction} \label{sec:intro}
Black holes (BHs) are thought to reside in the nuclei of galaxies and span a wide range of masses, from supermassive (SMBHs) with $\mbh\gtrsim10^{5}\msun$ to intermediate (IMBHs) with $\mbh\approx10^{2}-10^{5}\msun$ \citep{Kormendy1995,Kormendy2013,Greene2020}. When a star falls in the vicinity of such BHs, it gets tidally ripped apart by the immense gravitational force. More specifically, when the pericenter distance $\rp$ of the star (with mass $\mstar$ and radius $\rstar$) is smaller than the tidal radius ---the disruption distance from the BH---, which is defined as $\rt \approx \rstar(\mbh/\mstar)^{1/3}$, the star is destroyed \citep{Hills1975} and around half of the stellar debris escapes in unbound orbits, while the rest becomes bound and stretched into a thin elongated stream that starts circularising around the BH into highly eccentric orbits \citep{Rees1988,Evans1989}. The above process is called a tidal disruption event (TDE) and leads to a strong ($\lbol \sim 10^{41-45}\,\rm erg\,s^{-1}$), transient, nuclear flare \citep{Lacy1982,Rees1988,Evans1989,Phinney1989} that sometimes surpasses the Eddington luminosity \citep{Strubbe2009,Lodato2011}.

Tidal disruption
events were first discovered in the X-ray regime (\citealt{Komossa1999}; see \citealt{Auchettl2017,Guolo2023} for more recent X-ray TDEs); however, during the last decade, discoveries of many TDEs were made at ultraviolet (UV) and optical wavelengths (\citealt{Gezari2012,Arcavi2014,VanVelzen2021,Yao2023}, and see reviews from \citealt{VanVelzen2020} and \citealt{Gezari}), taking advantage of the capabilities of surveys such as the Galaxy Evolution Explorer (GALEX; \citealt{Siegmund2004}) in the UV, and numerous wide-field sky surveys in the optical, such as the Sloan Digital Sky Survey (SDSS; \citealt{Gunn2006}), the Palomar Transient Factory (PTF; \citealt{Law2009}) and its successor the intermediate Palomar Transient Factory (iPTF), PanSTARRS1 (PS1; \citealt{Kaiser2002}), the All-Sky Automated Survey for Supernovae (ASAS-SN; \citealt{Shappee2014}), the Asteroid Terrestrial impact Last Alert System (ATLAS; \citealt{Tonry2018}), and the Zwicky Transient Facility (ZTF; \citealt{Bellm2019,Masci2019,Graham2019}). Furthermore, there are TDEs discovered in the mid-infrared \citep{Mattila2018,Kool2020,Jiang2021a,Reynolds2022} and others that launch relativistic jets and outflows leading to bright gamma and X-ray \citep{Zauderer2011,Andreoni2022,Pasham2023} as well as radio emission \citep{VanVelzen2016,Mattila2018,Alexander2020,Goodwin2022,Cendes2023}. 

The broad range of properties of TDEs (some being bright in the X-rays, some in the optical and UV, and others in both regimes) has led to the proposition of different theoretical scenarios regarding the emission mechanism at play. It was suggested that such strong optical and UV emission might arise from shocks during the self intersection of the debris stream \citep{Piran2015,Jiang2016} or from the reprocessing of X-rays arising from an accretion disc \citep{Guillochon2013,Metzger2016,Roth2016}. In the latter scenario, the {`reprocessing layer'} is suggested to be formed from optically thick outflows arising from super-Eddington accretion \citep{Miller2015,Dai2018,Thomsen2022} or a {`collision-induced outflow'} arising from the self-intersection point \citep{Lu2020,Bonnerot2021,Charalampopoulos2023a}. 

Studies of how the BH mass of the host galaxy affects the light curve properties of optical and UV TDEs are ongoing. There have been some events that are faint and fast ---rising to a peak on shorter timescales--- and have dimmer peak magnitudes relative to the bulk of the TDE population (see extensive discussion about their properties in \citealt{Charalampopoulos2023b}). \citet{Blagorodnova2017} suggested that the reason for faint and fast TDEs could be the low SMBH mass of the host galaxy; however, TDEs discovered later had estimated BH masses not significantly lower compared to the other more slowly evolving TDEs \citep{Nicholl2020}. At the opposite extreme, it has been suggested that TDEs might occur around BHs with masses of greater than $10^{8}\msun$, something not possible for non-spinning BHs, as the disruption of the star (hence the electromagnetic flare) would happen within the event horizon. However, for a highly rotating BH, a TDE outside the event horizon becomes possible \citep{Kesden2012}. Such a scenario was proposed to explain the superluminous transient ASASSN-15lh \citep{Leloudas2016}, which showed a spectacular rebrightening in the UV wavelengths approximately two months after the transient peak.
Spectroscopically, optical and UV TDEs typically show strong and broad Balmer emission lines and sometimes strong helium emission lines as well; they show a vast spectroscopic diversity as a class, with different line profiles seen in different TDEs, and dramatic profile evolution within the same event (e.g. see \citealt{Charalampopoulos2022}). Some of these peculiar line profiles (e.g. \citealt{Hung2019,Nicholl2020,Hung2020a}) have been attributed to the existence of outflows (\citealt{Roth2017}).

Demographically, optical and UV TDEs are usually discovered in quiescent galaxies. The rare subset of quiescent Balmer-strong E+A galaxies seem to be over-represented as hosts of TDEs by a factor of 30–35 \citep{Arcavi2014,French2016,Graur2018,French2020}, and lately a rate enhancement in green-valley galaxies has been discovered \citep{Yao2023}. However, TDE candidates have been discovered in the nuclei of active galaxies (e.g. \citealt{BLANCHARD2017,Kankare2017,Nicholl2020,Hung2020a,Cannizzaro2020,Cannizzaro2021a,Tadhunter2021,Cannizzaro2022,Wevers2022,Petrushevska2023,Homan2023}). It is harder to elucidate the nature of such transients as the pre-TDE activity and variability is entangled with the transient flux; it is therefore not trivial to ascribe a sudden change in nuclear brightness to a TDE or, for example, to a changing-look or a Seyfert active galactic nucleus (AGN), or 
a low-ionization nuclear emission-line region (LINER; \citealt{Frederick2019, Frederick2020, Huang2023}).

\section{Discovery and background} \label{sec:background}

On 2023 Feb 23 the ASAS-SN collaboration reported on the Transient Name Server (TNS\footnote{The TNS is the official IAU mechanism for reporting new astronomical transients \url{https://wis-tns.weizmann.ac.il/}}) the discovery of a transient (ASASSN-23bd, IAU name: AT~2023clx) in the nucleus of the nearby galaxy NGC~3799 and noted in the remarks that it could potentially be a TDE \citep{Stanek2023}. In 2023 Feb 26 the transient was classified as a TDE \citep{Taguchi2023} based on a SEIMEI telescope spectrum. It is important to note that the Sherlock package\footnote{\url{https://github.com/thespacedoctor/sherlock}} \citep{Young2023,Smith2020} (that predicts the classification of a reported new transient by cross-matching data against a library of historical and on-going astronomical survey data e.g. nearby galaxies, known CVs, AGNs, etc) used by the Lasair\footnote{\url{https://lasair-ztf.lsst.ac.uk/}} broker \citep{Smith2019} identified this transient as being within 0.5 arcseconds of the core of NGC~3799. At this separation, Sherlock would classify the source either as an NT (nuclear transient) or AGN depending on what information exists in galaxy catalogues. NGC~3799 is labelled as a potential AGN in the Milliquas catalogue \citep{Flesch2023} and hence an AGN label was applied both in the Lasair stream for the ZTF detection and in the ATLAS database. This means that every querying transient filter that disregards discovered transients because they occur in the nucleus of a known AGN or because of the Sherlock contextual classification, would have also disregarded AT~2023clx. To further complicate the matter, the ATLAS team sent the ATLAS initial source detection (2023 March 04) to TNS automatically, triggered after the ASAS-SN discovery. A detection of the source (either a spurious subtraction artefact or real variability) was made in 2018 (ATLAS18bcno), which caused an object under the name AT~2018meh to be created on the TNS. ZTF also had a variability detection in 2018 (ZTF18aabkvpi). To be clear, a comment was added on the TNS to note these are not likely real sources. However, there was a second classification (again as a TDE) on 2023 May 12 by \citet{Johansson2023} based on a Keck telescope spectrum, associated with AT~2018meh on the TNS. From hereon, we will refer to the transient as AT~2023clx. In an earlier work, \citet{Zhu2023} presented the early time properties of AT~2023clx, concluding it is the faintest and closest optical TDE discovered.

\begin{figure}
\centering
\includegraphics[width=0.4 \textwidth]{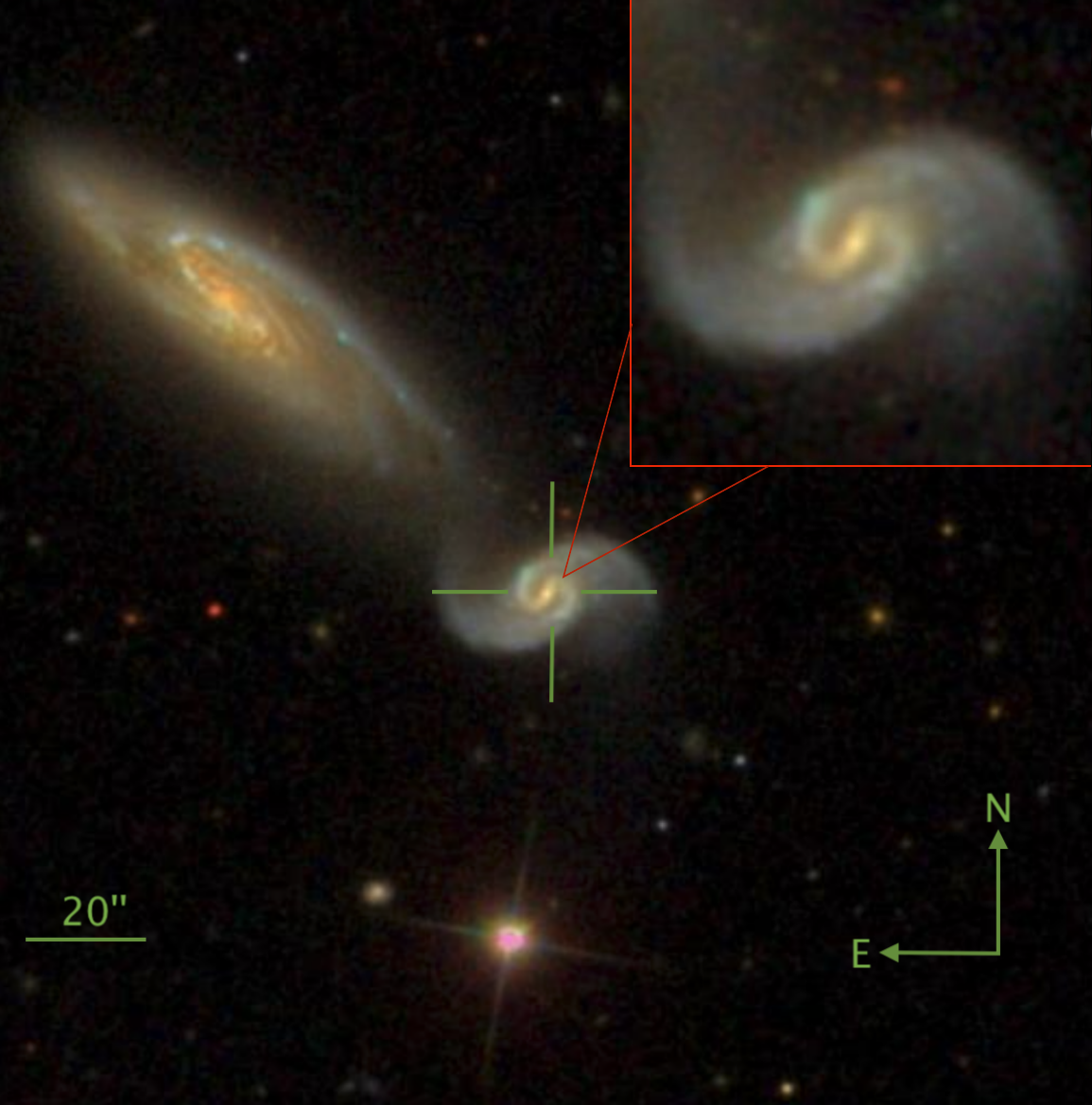}
\caption{SDSS image of host galaxy NGC~3799 before the occurrence of AT~2023clx. The nucleus of its host galaxy is marked with a green cross and is also shown in a zoomed-in inset in the top right corner. We note that a potential merger is underway with the nearby galaxy NGC~3800.}
\label{fig:host}
\end{figure}

In the NASA/IPAC Extragalactic Database (NED\footnote{\url{https://ned.ipac.caltech.edu/}}) and in \textit{SIMBAD} (\texttt{SIMBAD}\footnote{\url{https://simbad.cds.unistra.fr/}}), the host is classified as a LINER and  has a spectroscopic redshift of ${z = 0.01107 \pm 0.00001}$, reported in SDSS based on the archival host galaxy spectrum. This corresponds to a distance of 47.8\,Mpc assuming a $\Lambda$CDM cosmology with \mbox{H$_{0}$ = 67.4\,km\,s$^{-1}$ Mpc$^{-1}$}, ${\Omega_{\rm m}}$ = 0.315 and ${\Omega_{\rm \Lambda}}$ = 0.685 \citep{Aghanim2020}, which we use throughout this work. We note that distances to NGC~3799  that include corrections for Virgo and Great Attractor infall as well as the Shapley supercluster, are all consistent within the errors with the above value.
An image of the host is shown in Fig. \ref{fig:host}. We note that the galaxy appears to be merging with a companion, the nearby galaxy NGC~3800. \citet{RamosPadilla2020} studied 189 nearby interacting galaxies and, in a merger-state scale of 1 -- 5, they classified the NGC~3799 system as Stage 3, that is, galaxies that are moderately interacting, have apparent tidal features, and display moderate morphological distortions.

\indent In this paper, we present the follow-up and thorough photometric and spectroscopic analysis of AT~2023clx. The discovery and background of the transient was presented in this section (Sect. \ref{sec:background}). We describe our observations and data reduction in Sect. \ref{sec:observations}. We present a comprehensive analysis of the photometric and spectroscopic properties of AT~2023clx in Sect. \ref{sec:analysis} and we discuss their implications in Sect. \ref{sec:discussion}. Sect. \ref{sec:conclusion} contains our summary and conclusions.

\section{Observations and data reduction} \label{sec:observations}

We infer a foreground Galactic extinction of $A_{V}$ = 0.0829 mag \citep{Schlafly2010}, and apply this to our data assuming the \citet{Cardelli1989} extinction law with $R_{V}$ = 3.1.
We also deredden for the host galaxy extinction (we discuss this in Sect. \ref{subsubsec:host_ex}) assuming a host visual extinction of $A_{V_h}$ = 0.55 mag.

\subsection{Ground-based imaging} \label{subsec:gbi}

Well-sampled host subtracted light curves of AT~2023clx were obtained by the ZTF public survey in the $g$ and $r$ bands, from the ATLAS survey in the $o$ and $c$ bands and from the ASAS-SN survey in the $g$ band. We performed forced point spread function (PSF) photometry to extract precise flux measurements through the ZTF forced-photometry service \citep{Masci2019}. The ATLAS o and c band light curves were generated using the ATLAS Forced Photometry service \citep{Tonry2018,Smith2020,Shingles2021}\footnote{\url{https://fallingstar-data.com/forcedphot/}}. The ASAS-SN light curves were collected from the ASAS-SN Sky Patrol Photometry Database\footnote{\url{http://asas-sn.ifa.hawaii.edu/skypatrol/}}. Near-infrared (NIR) imaging of AT~2023clx was obtained with the Nordic Optical Telescope (NOT) as both part of the NUTS2 program (Nordic-optical-telescope Un-biased Transient Survey) and a systematic effort to observe nearby TDEs in the NIR, using the NOTCam instrument. The NOTCam data were reduced using a slightly modified version of the NOTCam \texttt{quicklook} v2.5 reduction package (e.g. to increase the field of view). The reduction process included flat-field correction, a distortion correction, bad pixel masking, sky subtraction and finally stacking of the dithered images. PSF photometry was performed with the \texttt{autophot} pipeline \citep{Brennan2022} after template subtraction and the resulting magnitudes were calibrated using the 2MASS catalogue \citep{Skrutskie2006}. Images were aligned for template subtraction using standard IRAF tasks and template subtraction was performed with the \texttt{hotpants} package\footnote{\url{https://github.com/acbecker/hotpants}}, an implementation of the image subtraction algorithm by \citet{Alard1998}. NIR imaging follow-up of AT~2023clx consisted of four epochs of $JHKs$ imaging. Template images were obtained on 2023 Dec 27 and 2023 Dec 28 under good conditions to obtain similar seeing and depth to the best quality follow-up images. The template epochs were observed at $\sim$305 rest frame days after the observed optical peak of the TDE, which is early enough that there could be residual infrared (IR) emission from a slowly evolving IR echo of the TDE \citep[e.g.][]{Mattila2018,Reynolds2022}. To check for such an IR echo, we performed image subtractions of our NOT template images against available template images of the field obtained in 2020 by the \texttt{UKIRT} large area survey \citep{Lawrence2007}. We did not detect a residual in any of these subtractions. Through injection and recovery of sources in the template subtracted images, we find 3$\rm\sigma$ upper limits in our NOT templates of 20.5, 20 and 19 mag for $J$, $H$ and $K$ respectively. Additionally, if a long lasting IR echo was present, we would expect the $K$ band flux to be brighter in the template than in our earliest NIR data, leading to negative residuals in the subtractions, which we do not observe. We conclude that there is no evidence of any residual emission in our template data.

\subsection{Swift UVOT photometry} \label{subsec:uvot_phot}

Target-of-opportunity observations spanning 30 epochs (PIs: Leloudas, Wevers, Huang, Gomez, Sfaradi) were obtained with the UV-Optical Telescope (UVOT) and
X-ray Telescope (XRT) on board the \textit{Neil Gehrels Swift} Observatory
(\textit{Swift}). The UVOT data were reduced using the standard pipeline available in the \texttt{HEAsoft} software package\footnote{\url{https://heasarc.gsfc.nasa.gov/docs/software/heasoft/}}. Observation of every epoch was conducted using one or several orbits. We  discarded the data from the orbit images in which our observations were in the low throughput areas of the detector (small-scale sensitivity check\footnote{\url{https://www.swift.ac.uk/analysis/uvot/sss.php}}). To improve
the signal-to-noise ratio (S/N) of the observation in a given band in a particular epoch, we co-added all orbit-data for that corresponding epoch using the \texttt{HEAsoft} routine \texttt{uvotimsum}. We used the routine \texttt{uvotsource}
to measure the apparent magnitude of the transient (using the most recent UVOT photometric zero-points of \citealt{Breeveld2011} that updated the ones of \citealt{Poole2008}) by performing aperture photometry using a 5$''$ radius aperture for the source and a 25$''$ aperture for the background. In order to convert from magnitudes to fluxes we used the UVOT central wavelengths recommended by the \textit{Swift} team\footnote{\url{https://www.swift.ac.uk/analysis/uvot/filters.php}}.
Since there were archival images, in all bands except $B$, we performed aperture photometry using the same methods in order to estimate the host galaxy flux in each of the UVOT bands and subtract it from the TDE+host photometry. For the $B$-band, we scaled the archival SDSS spectrum with the archival UVOT photometry and then computed the synthetic photometry. The $V$ band photometry quickly reaches the host level so we do not take it into account for the rest of our analysis. 

The complete, host subtracted and dereddened UV and optical light curves from Swift, ZTF, ATLAS and ASAS-SN are shown in Fig. \ref{fig:LC_pl} and all magnitudes are tabulated in Table \ref{tab:phot} of the Appendix. In Fig. \ref{fig:colours}, we show the early rapid decline in the bluer bands and the colour evolution with respect to the $g-$band.

\begin{figure}[h!t]
\centering
\includegraphics[width=0.5 \textwidth]{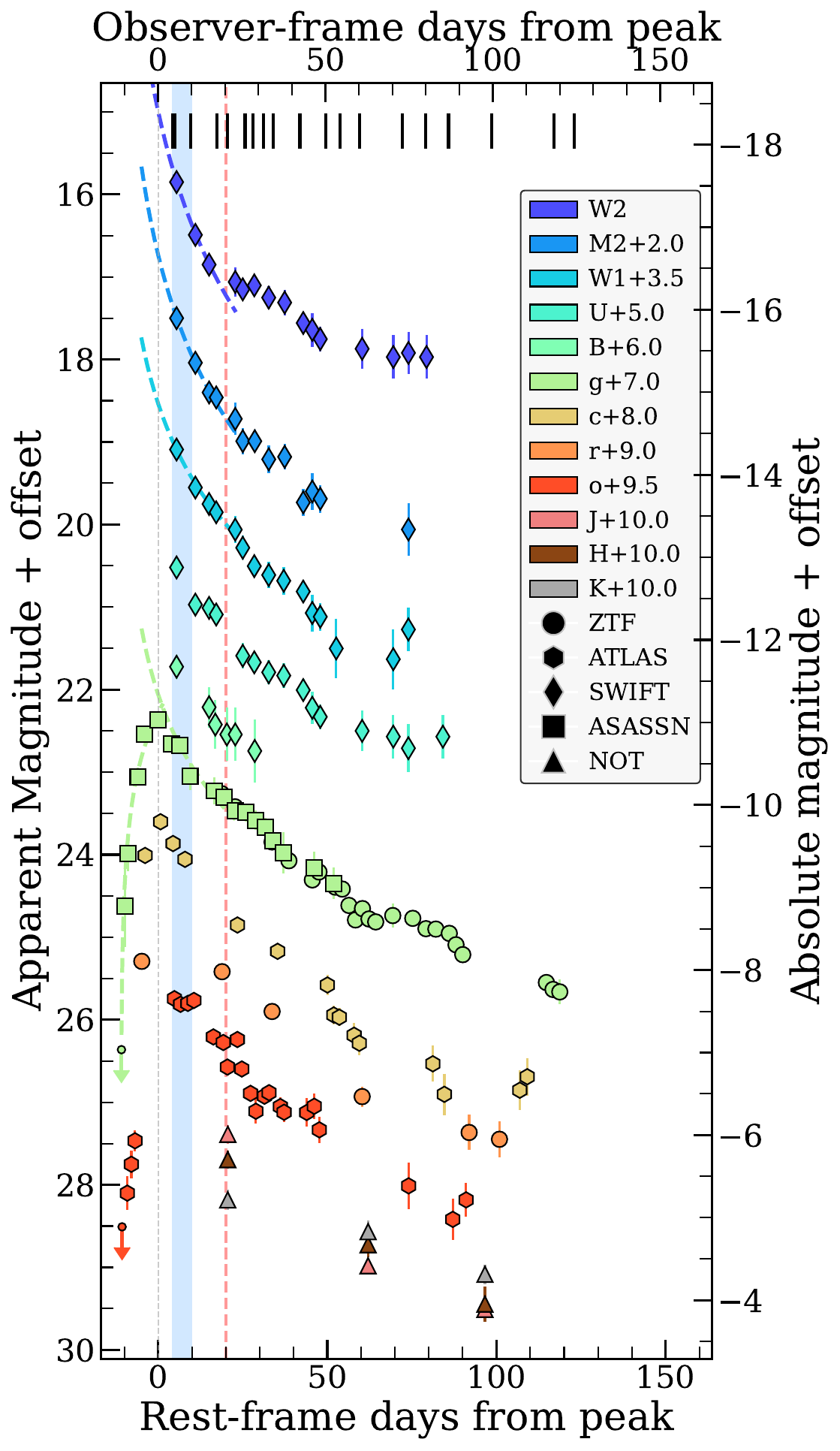}
\caption{Host-subtracted and dereddened light curves of AT~2023clx. The peak epoch is constrained to MJD 59997.2 from fits to the $g$-band. The short black vertical bars denote the epochs when spectra were taken (Table \ref{tab:spec_log}). Non-detections are shown as small downward-facing arrows. The blue shaded region denotes the epochs during which the sharp, narrow emission feature at $\sim$ 6353\,\AA\, is present in the spectra.  Power-law fits to the data over the first 20 days post-peak are shown for the $UVW2$, $UVM2$, $UVW1,$ and $g$ bands. For the latter, we also show a power-law fit to the rising part of the light curve. The red dashed vertical line denotes the end of the rapid cooling phase (see Fig. \ref{fig:colours} and \S\ref{subsec:UV_dip})}.
\label{fig:LC_pl}
\end{figure}

\begin{figure}[h!t]
\centering
\includegraphics[width=0.5 \textwidth]{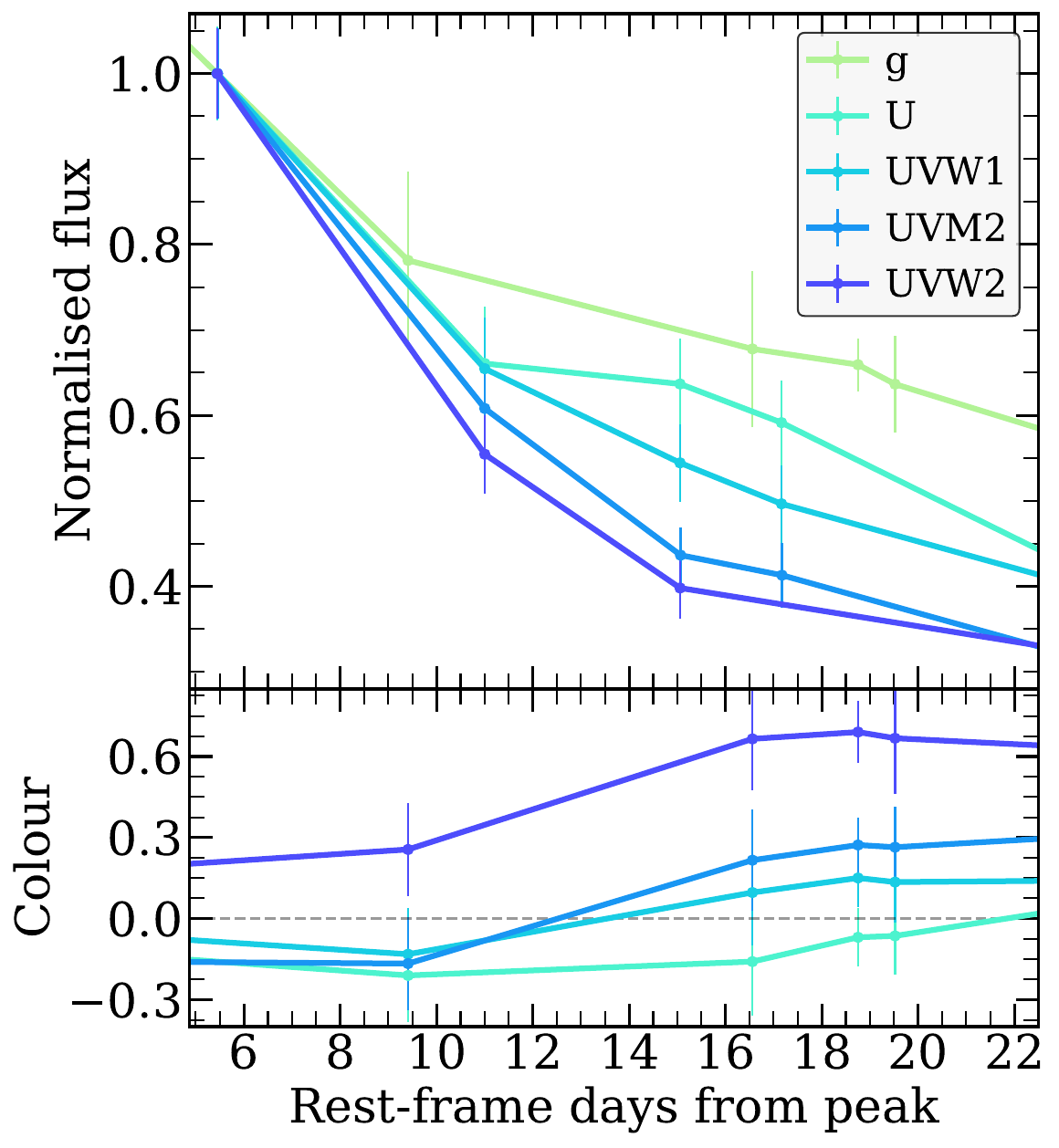}
\caption{Focus on the early evolution of the light curves and colours of AT~2023clx. Top panel: Early evolution of the light curves in the bluer bands for comparison with Fig. \ref{fig:LC_pl} in flux density space, normalised so that they are all equal to one at the first UVOT epoch. There is a clear wavelength dependence in the decline rate during the first $\sim$20 days post-peak: the bluer the band, the faster the decline. Bottom panel: Colour evolution of the light curves shown in the top panel with respect to the $g-$band. There appears to be a `rapid cooling' (unusual for TDEs) that occurs during the first $\sim$ 20 days post-peak.}
\label{fig:colours}
\end{figure}

\subsection{X-ray upper-limits} \label{subsec:uplims}

 We looked for X-ray emission from the position of the source in the XRT data and we report that AT~2023clx was not bright in X-rays. After stacking all available observations (30 epochs; 129 days) we find a marginal detection at $3.87\times10^{-4}$ counts per second that we consider a 3$\rm\sigma$ upper-limit. Using the \texttt{webPIMMS} tool\footnote{\url{ https://heasarc.gsfc.nasa.gov/cgi-bin/Tools/w3pimms/w3pimms.pl}} and assuming a Galactic column density $N_{\rm H}$ of $2.51\times10^{20}$~cm$^{-2}$, we convert the count rate to an unabsorbed flux upper-limit of $1.29\times 10^{-14}\,\rm erg\,cm^{-2}\,s^{-1}$ (0.3–10~keV, assuming a blackbody model with $kT=0.1$ keV) which translates to an upper-limit for the X-ray luminosity of $\sim 3.53\times 10^{39}\,\rm erg\, s^{-1}$ (a power-law model with $n=1.75$ results in an upper-limit of $\sim 4.41\times10^{39}\,\rm erg\, s^{-1}$). 
 We note here that \citet{Hoogendam2024} reported a late time ( $+97.7$d post-peak) XMM-NEWTON detection of $\sim 4.2\times 10^{39}\,\rm erg\, s^{-1}$.

\subsection{Imaging polarimetry} \label{subsec:im_pol}

We obtained two epochs of imaging polarimetry with the Multi-colour OPTimised Optical Polarimeter (\texttt{MOPTOP}) mounted on the Liverpool Telescope (LT). \texttt{MOPTOP} is a dual-beam, dual camera polarimeter with the beams sent to a pair of low-noise fast-readout imaging cameras. It utilises a continuously rotating half-wave plate, with 16 rotation positions per cycle. Images were obtained in `slow mode' with four-second exposures and each set of 56 images was stacked per rotation angle. Exposures undergo bias subtraction, dark subtraction and flat fielding using an automated facility pipeline. LT observed AT~2023clx on 2023 Mar 13 (MJD 60016.95; $+19.7$\,d) and 2023 Apr 18 (MJD 60052.90; $+55.7$\,d) using \texttt{MOPTOP} in the L band. Unfortunately, the observing conditions were not optimal as the seeing was very high in the first epoch ($\sim$4$''$) and varied a lot in the second (1$''$-- 3.4$''$). Images from 2023 Mar 13 were affected by cloud cover resulting in two exposures being removed from the final stacked images. Observations also commenced on 2023 Apr 17 but were aborted due to high cloud cover and are not subsequently analysed. Fluxes were extracted with a 10 pixel aperture, corresponding to approximately twice the full-width at half-maximum (FWHM), using the \texttt{ASTROPY PHOTUTILS} package \citep{Bradley2019}. Background fluxes were calculated using the median value of a patch of the exposure close to source due to interference from the host and neighbouring galaxy close to the target. The S/N of the target was $\sim$120 on 2023 Mar 13 and $\sim$130 on 2023 Apr 18. Polarisation measurements were calculated using the two-camera technique described by \citet{Shrestha2020}. This method uses four rotation positions from each camera to produce an independent set of q, u pairs. All four sets of q, u parameters were combined and averaged to produce a final q, u pair per night of observation. 

For the first epoch (2023 Mar 13), we measure $Q=-5.33\pm0.58\%$ and $U=4.28\pm0.60\%$ (leading to $P=7.43\pm0.64\%$). Due to the very large seeing and cloud cover we consider this measurement unreliable. For the second epoch (2023 Apr 18), we measure $Q=0.23\pm0.54\%$ and $U=-0.24\pm0.55\%$ (leading to $P=0.36\pm0.59\%$). This measurement is more trustworthy and is consistent with an intrinsic polarisation degree of zero. 
Due to the bad observing conditions and the low S/N, we decided not use the imaging polarimetry measurements in our discussion and conclusions about the object.

\subsection{Optical spectroscopy} \label{subsec:opt_spec}

We collected low-resolution spectra of AT~2023clx with the Alhambra Faint Object Spectrograph and Camera (ALFOSC) mounted on the NOT through our programme (PI Charalampopoulos; P67-021) and through the NUTS2 collaboration.
Spectra were also obtained with EFOSC2 on the New Technology Telescope
(NTT) in La Silla Observatory, Chile, as part of the extended Public ESO Spectroscopic Survey for Transient Objects (ePESSTO+) survey \citep{Smartt2014}. The NOT spectra were reduced using the \texttt{PyNOT} reduction pipeline\footnote{\url{https://github.com/jkrogager/PyNOT/}}, and the NTT spectra were reduced in a standard manner with the PESSTO pipeline \citep{Smartt2014}. In addition to the above low-resolution spectra, we also acquired four intermediate-resolution spectra with the X-shooter mounted on European Southern Observatory's (ESO) Very Large Telescope (VLT) in Cerro Paranal, Chile through Director's Discretionary Time (DDT; PI Wevers; 110.25AX). The observations were performed using the standard nod-on-slit mode, but each single arm spectrum was reduced using the `stare' mode reduction, using \texttt{ESOreflex} and the standard X-shooter pipeline \citep{Goldoni2006,Modigliani2010}. Finally, we retrieved
archival Very Large Telescope/Multi Unit Spectroscopic Explorer (VLT/MUSE) integral-field spectrograph data of NGC 3799, taken before the transient occurred ($-$ 719\,d). In order to reduce the data, we used \texttt{IFUANAL}\footnote{\url{https://github.com/Lyalpha/ifuanal}} \citep{Lyman2018}. This package incorporates spectral pixel (spaxel) binning algorithms and fits stellar continuua (using \texttt{STARLIGHT}; \citealt{CidFernandes2011}) and emission-lines in these spaxel bins to discern spatially resolved properties of galaxies in IFU data.

We performed host subtraction (following the methods described in \citealt{Charalampopoulos2022}) using the SDSS archival spectrum. This is not a trivial process; first we degraded the SDSS spectrum to the resolution of the NOT and NTT spectra by binning the flux values of the former to the wavelengths of the latter (or alternatively, degraded the X-shooter spectra to match the resolution of the SDSS spectrum). We then scaled the spectra to match the photometry (including the one of the host) before carrying out the subtraction. All spectra were taken at the parallactic angle. Typically TDEs are found in elliptical galaxies so the orientation of the slit does not change which parts of the galaxy are included in the slit. However NGC~3799 is a spiral galaxy. A consequence of using the parallactic angle is that the background contamination changes with each epoch (see Fig. \ref{fig:host}). To minimise the effect of this, we exercised extreme care in our choice of aperture when extracting the spectra so that only the nuclear region was included ($\sim$ $2-4$$''$ aperture width was used to extract the 1-D spectra depending on the observing conditions). As expected, the subtraction was not perfect, and residuals corresponding to narrow lines from the host galaxy were apparent in the host subtracted spectra, especially around the H$\alpha$ and the [\ion{N}{II}] $\lambda\lambda\,6548,6583$ lines. For these cases, we applied sigma-clipping of the residuals on top of the broad H$\alpha$ profile. The NIR arm X-shooter spectra were corrected for telluric features using telluric standard star observations taken the same nights with the spectra. Since there is no host galaxy spectrum in the NIR, we could not perform host subtraction.

\begin{figure*}
\centering
\includegraphics[width=0.9 \textwidth]{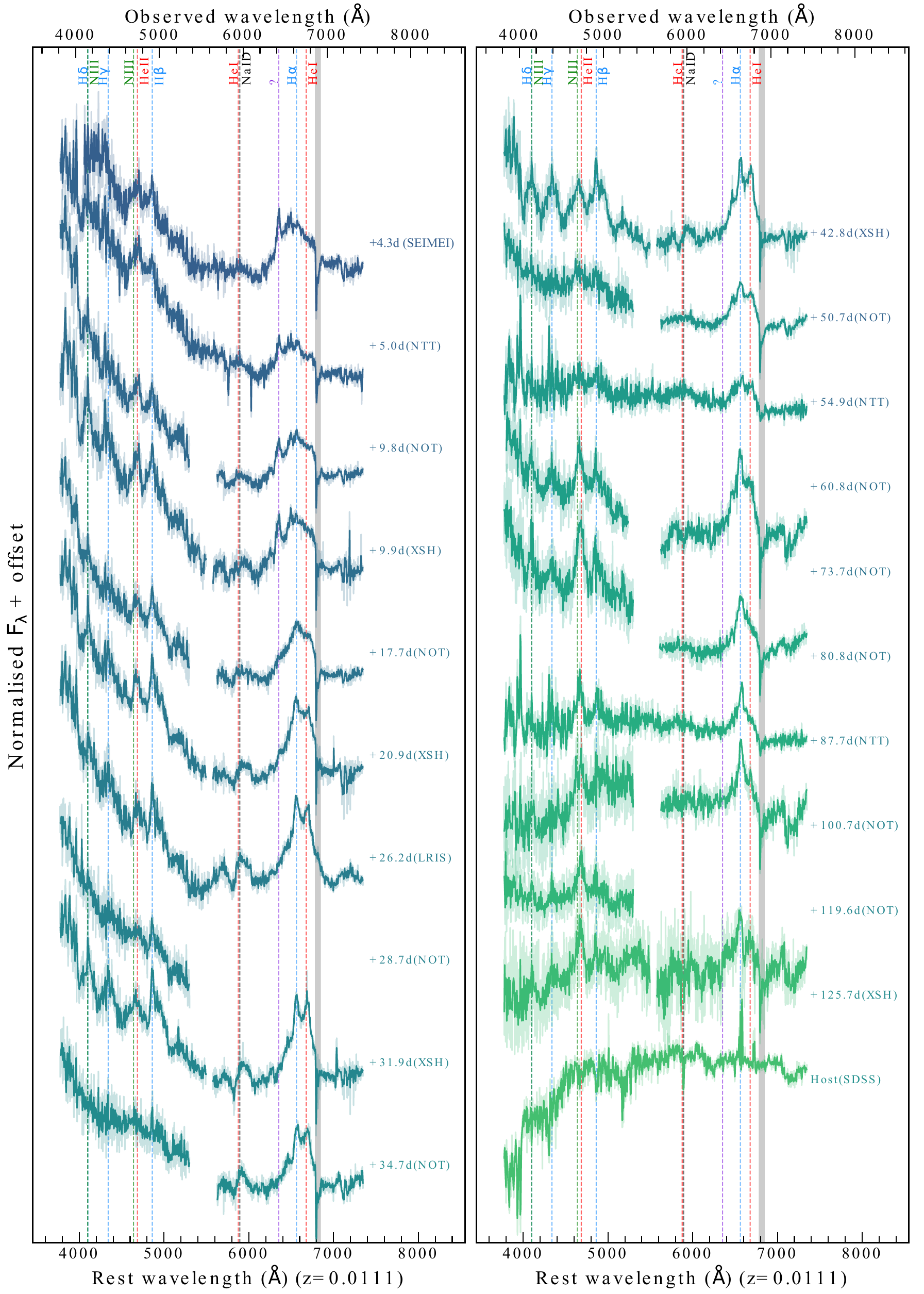}
\caption{Host-subtracted spectra of AT~2023clx (corrected for MW and host extinction). Emission lines are marked with vertical dashed lines. The sharp, narrow emission feature at $\sim$ 6353 \AA\, present in at least the
first four spectra, is labelled with a question mark. The dates are with respect to the $g$-band peak.}
\label{fig:spectra}
\end{figure*}

A spectroscopic log is provided in Table \ref{tab:spec_log}. The host subtracted spectral series are presented in Fig. \ref{fig:spectra}; the effect of host subtraction can be appreciated by comparing with Fig. \ref{fig:spectra_wo_host_sub} that displays the non-host subtracted spectral series. The NIR spectra are shown in Fig. \ref{fig:nir_spectra}.

\begin{table*} 
\renewcommand{\arraystretch}{1.2}
\setlength\tabcolsep{0.1cm}
\fontsize{10}{11}\selectfont
\begin{center}
\caption{Spectroscopic observations of AT~2023clx}\label{tab:spec_log}
\begin{tabular}{ccrcccccc}
\hline
      UT date & 
    MJD & 
    Phase$^{a}$ & 
    Telescope+Instrument &
    Grism/Grating &
        Slit Width &
        Airmass &
    Exposure Time  & Position Angle\\
    (yyyy-mm-dd)   &     &   (days)  &     &   &   (arcsec)  &     &  (s) & (deg) \\

2023-02-26      &   60001.58 & 4.3 & SEIMEI+KOOLS-IFU$^{b}$     & VPH-blue                         &  -    & - & 3600 & -     \\
2023-02-27      &   60002.29 & 5.0 & NTT+EFOSC2         & GR\#11                        &  1.0    & 1.45 & 1500 & $-$105.4     \\
2023-03-04      &   60007.02 & 9.8 & NOT+ALFOSC         & GR\#18, GR\#8                          &  1.0  & 1.07 & 1800, 1800 & $-$56.6     \\
2023-03-04      &   60007.13 &  9.9  & VLT+XSHOOTER     & UVB,VIS,NIR                           &  1.0, 1.0, 0.9  & 1.63 &  960, 1088, 600 & 136.6 \\
2023-03-11      &   60014.98 & 17.7 & NOT+ALFOSC        & GR\#18, GR\#8                          &  1.0  & 1.12 & 2400,2400 & $-$60.2, $-$51.6      \\
2023-03-15      &   60018.16 &  20.9  & VLT+XSHOOTER    & UVB,VIS,NIR                           &  1.0, 1.0, 0.9  & 1.35 &  1200, 1326, 500 & 154.4 \\
2023-03-20      &   60023.50 & 26.2 & Keck+LRIS$^{b}$   & -                     &  -      & - &  300 & -     \\
2023-03-22      &   60025.98 & 28.7 & NOT+ALFOSC        & GR\#18                        &  1.0    & 1.08 & 2400 & $-$50.3     \\
2023-03-26      &   60029.11 &  31.9  & VLT+XSHOOTER    & UVB,VIS,NIR                           &  1.0, 1.0, 0.9  & 1.42 &  1668, 1800, 600 & 148.4 \\
2023-03-28      &   60031.99 & 34.7 & NOT+ALFOSC        & GR\#18, GR\#8                          &  1.0  & 1.07 & 2400, 2400 & $-$35.4, 51.2     \\
2023-04-06      &   60040.09 &  42.8  & VLT+XSHOOTER    & UVB,VIS,NIR                           &  1.0, 1.0, 0.9  & 1.38 &  1974, 2100, 960 & 148.4 \\
2023-04-13      &   60047.98 & 50.7 & NOT+ALFOSC        & GR\#18, GR\#8                          &  1.0  & 1.04 & 2400, 2400 & 41.2, 7.7     \\
2023-04-18      &   60052.20 & 54.9 & NTT+EFOSC2        & GR\#13                        &  1.0    & 1.72 & 3600 & $-$120.5     \\
2023-04-24      &   60058.06 & 60.8 & NOT+ALFOSC        & GR\#18, GR\#8                          &  1.0  & 1.37 & 3000, 3000 & 61.7     \\
2023-05-06      &   60070.98 & 73.7 & NOT+ALFOSC        & GR\#18                        &  1.0    & 1.10 & 3600 & 46.8     \\
2023-05-14      &   60078.01 & 80.8 & NOT+ALFOSC        & GR\#8                         &  1.0    & 1.29 & 3600 & 61.2     \\
2023-05-20      &   60085.00 & 87.7 & NTT+EFOSC2        & GR\#13                        &  1.0    & 1.42 & 5400 & $-$77.9     \\
2023-06-02      &   60097.94 & 100.7 & NOT+ALFOSC       & GR\#18, GR\#8                          &  1.0  & 1.31 & 3600, 3600 & 59.5     \\
2023-06-21      &   60116.90 & 119.6 & NOT+ALFOSC       & GR\#18                        &  1.0    & 1.3 & 3600 & 61.3     \\
2023-06-28      &   60122.99 &  125.7  & VLT+XSHOOTER   & UVB,VIS,NIR                           &  1.0, 1.0, 0.9  & 1.48 &  440, 400, 300 & $-$149.6 \\
\hline
\end{tabular}
\\[-10pt]
\end{center}
$^{a}$With respect to the date of $g$-band maximum ($\mathrm{MJD} = 59997.2$) and given in the rest frame of AT~2023clx ($z=0.01107$).\\
$^{b}$These spectra were retrieved from TNS hence we do not have information on the slit width, airmass and position angle.
\end{table*}

\section{Analysis} \label{sec:analysis}

\subsection{Host-galaxy properties} \label{subsec:host}

\subsubsection{SED fit and host extinction} \label{subsubsec:host_ex}

We retrieved archival photometry of the host galaxy from the PanSTARRS catalogue\citep{Huber2015} in the $g,\,r,\,i,\,z,\,y$ filters we used the stacked Kron magnitudes, which are recommended for extended sources) and from the Sloan Digital Sky Survey (SDSS; \citealt{York2000}) in the $g,\,r,\,i,\,z$ filters (we used the default `modelMag' apertures\footnote{\url{https://live-sdss4org-dr12.pantheonsite.io/algorithms/magnitudes/\#mag_model}}). Furthermore, we retrieved archival photometry from the 2 Micron All Sky Survey (2MASS; \citealt{Skrutskie2006}) in the $J,\,H,\,Ks$ filters (we used the `total extrapolated photometry' aperture which combines the isophotal mag with an integration of the surface brightness profile capturing 10--20\% more flux) and from the Wide-field Infrared Survey Explorer (WISE; \citealt{Wright2010}) un-blurred images (un-WISE; \citealt{Lang2014}) in the $W1,\,W2,\,W3,\,W4$ filters. We  also performed aperture photometry in the UVOT archival images (see Sect. \ref{subsec:uvot_phot}) in filters $UVW2$, $UVM2$, $UVW1$ and $U$, with a 20$''$ aperture radius in order to include all the galaxy flux. The archival photometry is presented in Table \ref{tab:host_phot}.

\begin{figure}
  \centering
  \includegraphics[width=0.47 \textwidth]{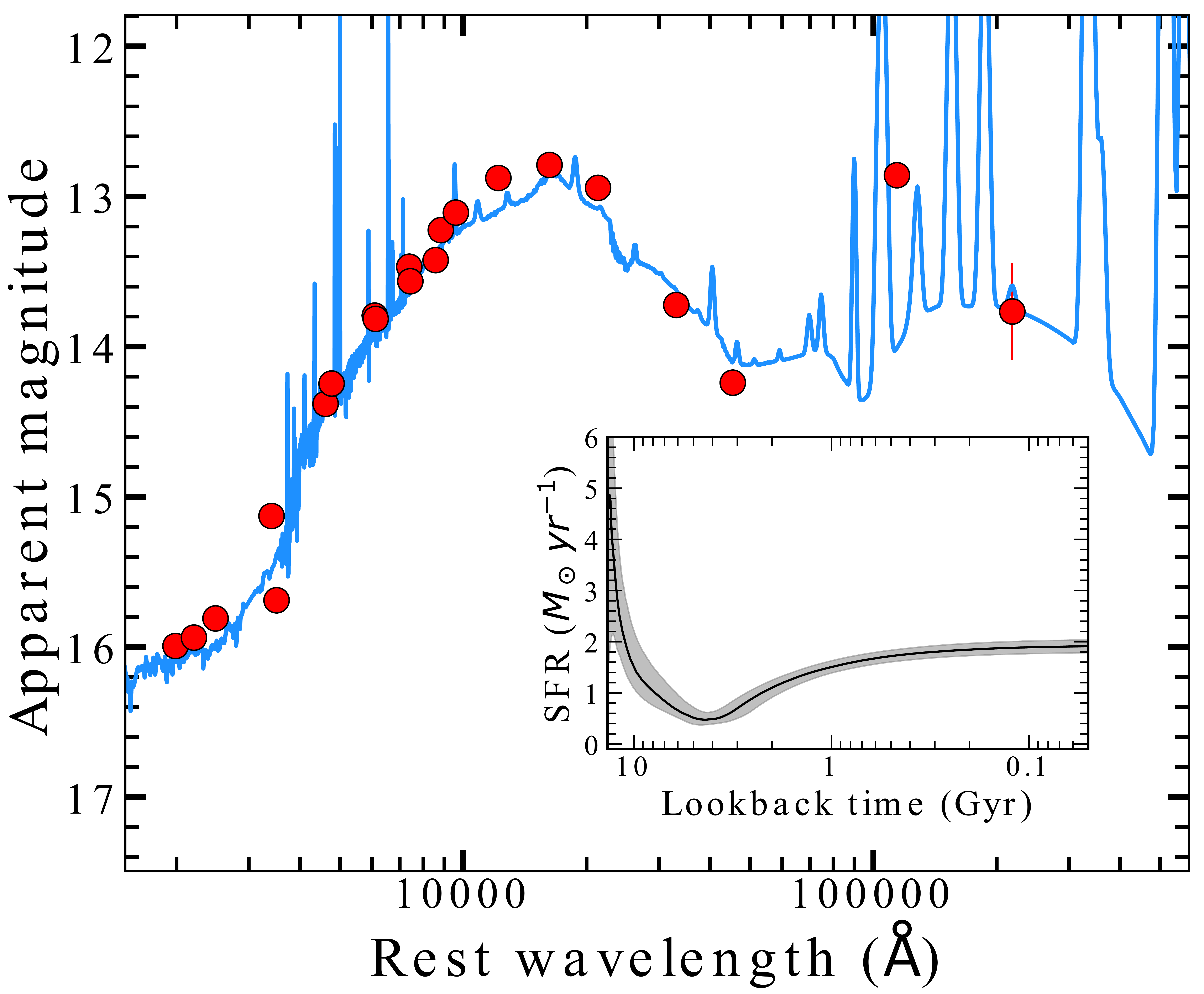}
  \caption{Host-galaxy SED fit with \texttt{PROSPECTOR}. The archival photometry is plotted in red circles and the best-fit template spectrum in blue. The inset shows the derived star-formation history, which shows a rise in the last 4 Gyr that reaches a plateau at $\sim$ 1.9 $\msun\,\rm yr^{-1}$.}
  \label{fig:sed_fit}
\end{figure}

We used \texttt{PROSPECTOR} \citep{Leja2017} to fit the archival photometry and produce the best-fitting spectral energy distribution (SED) model (see \citealt{Ramsden2022} for a description of the methods), which is shown in Fig. \ref{fig:sed_fit}. We find a stellar mass of ${\log (M_*/M_\odot) = 10.21^{+0.04}_{-0.04}}$
and a high specific star formation rate (SFR) of ${\log {\rm sSFR/\msun yr^{-1}}=-9.92^{+0.06}_{-0.05}}$. The reported values and uncertainties are the median and 16th and 84th percentiles of the marginalised posterior distributions. In the figure, we also show the star-formation history versus lookback time since the Big Bang. We find that the SFR of the galaxy around 4 Gyrs ago was $\sim$ 0.5  $\msun\,\rm yr^{-1}$, when a rise started reaching a plateau of $\sim$ 1.9 $\msun\,\rm yr^{-1}$. This confirms that this is indeed an unusual TDE host since post-starburst galaxies (typical TDE hosts) show a recent decline in star-formation. 

Furthermore, we find a high dust extinction of $\rm E(B-V)_{h}$ = 0.179 mag, which translates to $A_{V_h}$ = 0.55 mag\footnote{In order to retrieve an $\rm E(B-V)_{h}$ from the \texttt{PROSPECTOR} values we followed the discussion in Sect. 3.1.3 of \citet{Leja2017} and references therein. This assumes an $R^{\prime}_{V}=3.1$ for the host.}. Inspecting the SDSS spectrum of the host, we also see a very prominent \ion{Na}{I} D line which is also strong in the TDE spectra (see Fig. \ref{fig:spectra_wo_host_sub} in the Appendix), indicating a potential high host extinction, based on the Poznanski relation \citep{Poznanski2012}. We measure the equivalent width (EW) of the very prominent \ion{Na}{I} D line in the SDSS host spectrum to be $\sim$1.755 \AA. In our first X-shooter spectrum (where the resolution is higher and we can resolve the doublet), we measure 0.808 \AA\, for the \ion{Na}{I} D1 line and 0.863 \AA\, for the \ion{Na}{I} D2 line. The combined \ion{Na}{I} D line relation saturates for combined doublet EW values greater than $\sim$ 0.8\AA\, (as is the case here since we measure $\sim$ 1.7\AA), which means that we can derive a lower-limit for the extinction ($\rm E(B-V)_{h}$ > 0.3 mag) using the Poznanski relation, corroborating a large host extinction. However, correcting for host extinction using this limit makes it impossible to fit the TDE SED, because apart from requiring an unrealistically high temperature, the $UVM2$ data points become considerably brighter than the $UVW2$ ones, with a best-fit blackbody peak residing in much bluer wavelengths.
For this reason, we decide to use the \texttt{PROSPECTOR} estimate for $\rm E(B-V)_{h}$. The fact that two independent methods result in high host extinction values influenced our decision to deredden our data for the host extinction as well. Another reason that supports our choice is that without the correction, AT~2023clx seems to have a very low blackbody temperature compared to standard optical and UV TDEs (see Sect. \ref{subsubsec:Bol}). This is one of the first TDE studies to correct for the host extinction since TDEs are typically observed in passive galaxies and negligible host extinctions are assumed. Furthermore, the \ion{Na}{I} D lines of the hosts are typically weak suggesting a low host extinction, unlike AT~2023clx. After applying this correction, we find that AT~2023clx is not anymore the faintest TDE (see Sect. \ref{subsec:phot_analysis}). Undoubtedly, assuming an attenuation value and performing a host extinction correction based on that, introduces a large systematic uncertainty to the photometry (and especially in the UV) that propagates to the rest of the values derived by blackbody fitting (like bolometric luminosities and temperatures). However, we believe that assuming a non-zero value will lead to a more realistic estimate of those properties. Assuming negligible attenuation would lead to lower-limits for those properties.

\begin{table}
\centering
 \caption{Archival host-galaxy magnitudes}
 \label{tab:host_phot}
 \begin{tabular}{ccc}
  \hline
  Filter & Magnitude & Uncertainty\\
  \hline
  UVOT W2 & 15.994 & 0.02 \\
  UVOT M2 & 15.938 & 0.03 \\
  UVOT W1 & 15.810 & 0.02 \\
  UVOT U &  15.128 & 0.01 \\
  SDSS u & 15.686 & 0.006 \\
  SDSS g & 14.379 & 0.002 \\
  PS1 g & 14.241 & 0.001 \\
  SDSS r & 13.792 & 0.002 \\
  PS1 r & 13.813 & 0.001 \\
  SDSS i & 13.466 & 0.002 \\
  PS1 i & 13.562 & 0.001 \\
  SDSS z & 13.223 & 0.003 \\
  PS1 z & 13.425 & 0.001 \\
  PS1 y & 13.111 & 0.001 \\
  2MASS J & 12.877 & 0.026 \\
  2MASS H & 12.790 & 0.041 \\
  2MASS Ks & 12.942 & 0.041 \\
  un-WISE W1 & 13.722 & 0.001 \\
  un-WISE W2 & 14.239 & 0.003 \\
  un-WISE W3 & 12.860 & 0.011 \\
  un-WISE W4 & 13.767 & 0.324 \\
  \hline
 \end{tabular}\\
\begin{flushleft} Archival magnitudes of the host galaxy of AT~2023clx \mbox{NGC~3799} used for the \texttt{PROSPECTOR} fit. All magnitudes are presented in the AB system. \end{flushleft}
\end{table}

\subsubsection{Velocity dispersion and black hole mass} \label{subsubsec:vel_disp}

We use the penalized pixel fitting (\texttt{pPXF}) method \citep{Cappellari2004,Cappellari2017} to measure the line-of-sight velocity dispersion function and we convert it to BH masses using the M-$\rm\sigma$ relation from \citet{Ferrarese2005}, following the methods of \citet{Wevers2017}. We performed the above routine on both the SDSS archival spectrum and our last X-shooter spectrum ($+125.7$\,d), and we measure a BH mass of $\log_{10}(\mbh) = 6.49 \pm 0.39$ dex, and $\log_{10}(\mbh) = 5.71 \pm 0.40$ dex respectively. The uncertainty estimates include the systematic uncertainty of the relation which is 0.34 dex. The large difference between the measurements (still within 1-$\rm\sigma$) might arise from weak contribution of the transient, even though the emission lines were masked, to the continuum of the X-shooter spectrum.

Finally, we estimate the BH mass based on the empirical relation between the total galaxy stellar mass (which we derived from \texttt{PROSPECTOR}) and the BH mass \citep{Reines2015} and we find a mass of $\log_{10}(\mbh) = 6.62 \pm 0.69$ dex. Again, the uncertainty estimate includes the systematic uncertainty of the relation which is in this case is 0.55 dex. All the different black hole mass estimates are tabulated in Table \ref{tab:mbh_ledd} along with the corresponding Eddington luminosity estimate and the Eddington ratio based on the peak bolometric luminosity estimate (see Sect. \ref{subsubsec:Bol}).

\subsubsection{Galaxy emission lines and evidence for a LINER} \label{subsubsec:liner}
The host spectrum shows narrow emission lines (some of them evident in the TDE spectra as well) including H$\alpha$, [\ion{O}{III}] $\lambda5007$, [\ion{N}{II}] $\lambda\lambda\,6548,6583$ and [\ion{S}{II}] $\lambda\lambda\,6717,6732$. The presence of stellar absorption in the continuum often affects the flux of weaker lines. In order to recover the pure nebular fluxes, we fitted the SDSS host spectrum with the \texttt{STARLIGHT} software \citep{CidFernandes2011}. \texttt{STARLIGHT}
fits the stellar continuum, identifying the underlying stellar populations in terms of age and metallicity. By subtracting the output synthetic spectrum from the observed data, we isolate the pure nebular continuum, and we manage to identify also H$\beta$ in emission. We measure the fluxes of the emission lines in the nebular spectrum by fitting Gaussian profiles and we construct a  Baldwin, Phillips, and Terlevich (BPT) diagram \citep{Baldwin1981}, which is widely adopted to identify the level of nuclear activity in a galaxy. We use two line ratio diagnostics, [\ion{N}{II}]$\lambda6583$/H$\alpha$ and [\ion{S}{II}]$\lambda\lambda\,6717,6732$/H$\alpha$ against the [\ion{O}{III}]$\lambda5007$/H$\beta$ ratio. We also extracted a `nuclear region' spectrum with a $\sim 0.9''$ diameter aperture from the host galaxy MUSE cube and measured the same ratios in the nebular spectrum. The spectra are presented in Fig. \ref{fig:starlight} in the Appendix. The BPT diagrams are shown in Fig. \ref{fig:BPT}. In both diagrams, NGC~3799 falls in the LINER region (for both spectra used); however it is very close to the locus of AGN, LINER and star-forming regions. Nonetheless, we do confirm the LINER nature of the host\footnote{\url{https://simbad.cds.unistra.fr/simbad/sim-id?Ident=\%401845231\&Name=NGC\%20\%203799\&submit=submit}}.

Using the nebular MUSE spectrum of the host, we measured the Balmer decrement ($L_{H\alpha}$/$L_{H\beta}$) in the nuclear region and from that, make an estimate of the host extinction (see e.g. \citealt{Moustakas2006,Dominguez2013}). We measure a decrement of $3.47 \pm 0.14$ that translates to $E(B-V)_{h} \sim 0.16$ mags (i.e. $A_{V_{h}} \sim 0.51$ mags using $R_{V}$ = 3.1). This is in good agreement with our \texttt{PROSPECTOR} host extinction estimate.

\begin{figure}
  \centering
  \includegraphics[width=0.47 \textwidth]{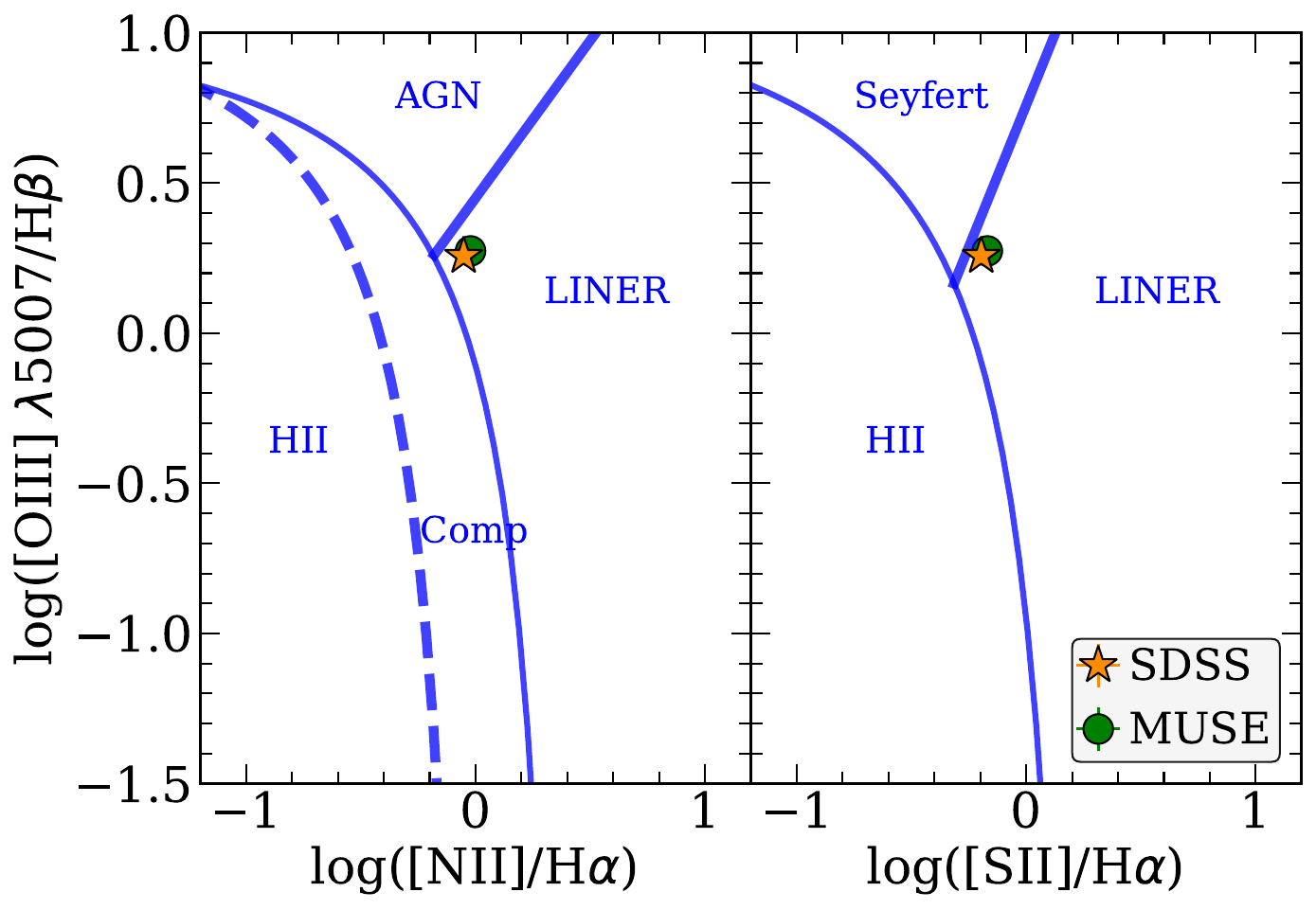}
  \caption{BPT diagram of NGC~3799 confirming the LINER nature of the host. The line ratios were measured using both the SDSS nebular spectrum (orange star), and
  using the MUSE data cubes to extract a spectrum of the nucleus (green circle). `Comp' (left panel) stands for composite region. }
  \label{fig:BPT}
\end{figure}

\subsection{Photometric analysis} \label{subsec:phot_analysis}

\subsubsection{Broadband light-curve evolution} \label{subsubsec:bb_lc}

From a fourth degree polynomial fit to the ASAS-SN $g$-band light curve, we constrain the peak date to MJD $59997.2 \pm 2.0$. The resulting peak magnitude is $15.37 \pm 0.07$ mag ($M=-18.03$ mag). Given that the last non-detection was on MJD 59\,986.3 and the first detection at MJD 59\,987.3, the rise time to peak is $10.4 \pm 2.5$ days, making AT~2023clx the fastest rising TDE to date\footnote{We note that \citet{Malyali2023} measured a rise time of $\sim 10$ d for TDE AT~2020uwq however with large uncertainties ($^{+6}_{-3}$ d).}. The rate of brightening from first detection to peak is estimated to be $\sim 0.27$ mag/day. Our first detection and peak epochs agree within 1-$\sigma$ with those of \citet{Hoogendam2024}.

Concerning the decline, we see that during the first $\sim$ 20 days post-peak, there is a wavelength dependence in the decline rate of the light curves; the bluer the band, the faster the decline. Hereon, we refer to this feature as {rapid cooling} and show it in the top panel of Fig. \ref{fig:colours}. The early colours of the near-UV (NUV) light curves with respect to the $g-$band are in the bottom panel of the same figure. In order to quantify this effect, we fit power laws ($L \propto t^{b}$) to the early decline, that is, until +20\,d post-peak, and we show the fits as dashed lines in Fig. \ref{fig:LC_pl}. The best-fit power-laws return the following indices: $b_{W2}$ = $-2.09\pm0.02$ in $UVW2$, $b_{M2}$ = $-1.86\pm0.05$ in $UVM2$, $b_{W1}$ = $-1.37\pm0.05$ in $UVW1$ and $b_{g}$ = $-0.87\pm0.06$ in $g$. The early decline becomes sharper as we move from the optical to the NUV. Further discussion about the nature of this `rapid cooling' is in Sect. \ref{subsec:UV_dip}.

\subsubsection{Bolometric light curve} \label{subsubsec:Bol}

\begin{figure}
  \centering
  \includegraphics[width=0.5 \textwidth]{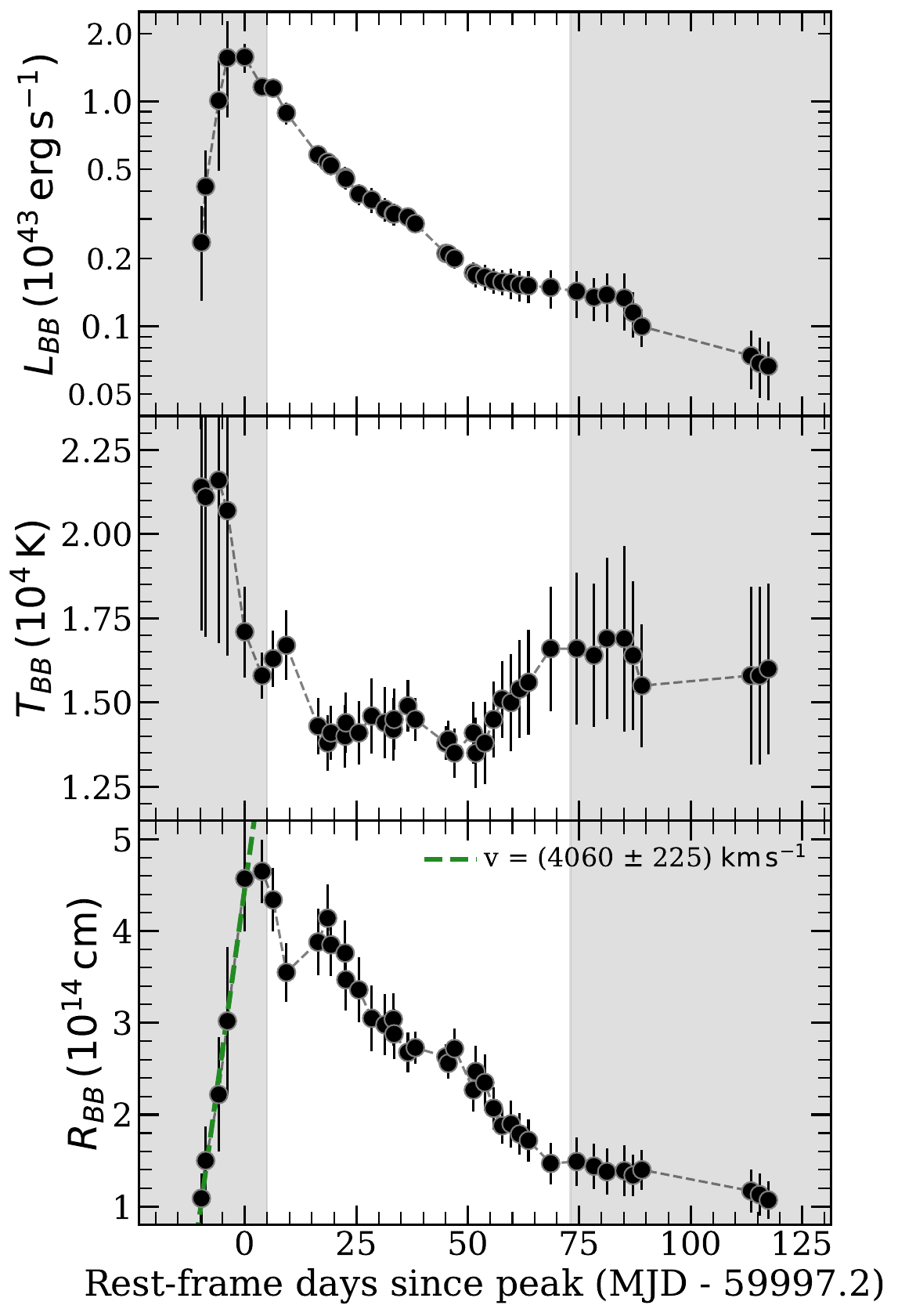}
  \caption{Bolometric light curve, blackbody temperature, and radius evolution. The luminosity (top panel) is plotted in a log-linear scale. The shaded regions indicate the time regions where constant colour was assumed for extrapolation purposes for some of the bands. There is a break in the temperature evolution (middle panel) around 10--20 days post-peak. This is when \ion{He}{II} recombines and we see the emergence of \ion{He}{I} in the spectra. The early expansion of the photosphere (bottom panel) is best described (green dashed line) with a constant velocity $v\approx4\,000\,\rm km\,s^{-1}$ (i.e. $\sim$ 0.014c), one of the fastest expansion velocities recorded in TDEs.}
  \label{fig:bol}
\end{figure}

We construct the bolometric light curve of AT~2023clx by interpolating or extrapolating the host subtracted and dereddened photometry of each band in time, to any epoch with data in the $g$ band (defined as the reference filter), using \texttt{SUPERBOL} \citep{Nicholl2018a}. We interpolate the light curves using polynomials of third to fifth order and we integrate under the SED of each epoch (from 1 to 25000 \AA) to get the luminosity. We fit a blackbody function to the SED in order to estimate the temperature and the radius, and also to calculate the missing energy outside of the observed wavelength range. Since our first UVOT epochs were observed $\sim$ five days post-peak, we extrapolated before this epoch assuming constant colours. We base this choice on the fact that TDEs having a generally constant colour evolution (e.g. \citealt{Holoien2018,VanVelzen2021,Yao2023}) and because of the persistent blue colour of the spectra in all the early epochs (see Fig. \ref{fig:spectra}). The bolometric light curve, temperature and radius evolution are plotted in Fig. \ref{fig:bol}. The epochs for which we used the constant colour extrapolation method are shaded with grey colour and the effect is also evident in the large error-bars. The blackbody fits are presented in Fig. \ref{fig:BB_fits} in the Appendix.

The bolometric light curve analysis results in a blackbody temperature in the range $\sim14\,000\,\rm K-20\,000$~K. However, there is a break in the temperature evolution around 10--20 days post-peak, driven by the `{rapid cooling'} around these epochs (see Fig. \ref{fig:LC_pl}, \ref{fig:colours}, and Sect. \ref{subsubsec:bb_lc} as well as Sect. \ref{subsec:UV_dip} for further discussion). Interestingly, this is when we see the emergence of \ion{He}{I} in the spectra, most likely associated with the sudden temperature drop, as \ion{He}{II} recombines to \ion{He}{I}. During this phase, the temperature drops to a minimum of $\sim13\,500$~K around $+20$\,d post-peak, and after that remains almost constant or slowly rises. Concerning the radius of the blackbody photosphere, it rises to a value of $\rbb\sim4.7\times10^{14}$~cm $\sim$ peak, then it starts contracting before showing a secondary peak ($\rbb\sim4.4\times10^{14}$~cm) in the epochs where the temperature drops, and after that it keeps contracting. We fit the expansion of the radius before the peak (the first four data points) with a constant velocity and we find a best-fit value of $(4\,060 \pm 225)$~$\rm\,km\, s^{-1}$ (i.e. $\sim$ 0.014c). The velocity of the expanding photosphere is very fast compared to other TDEs where the expansion of their photosphere was fit with a constant velocity: $2\,200\rm\,km\, s^{-1}$ for AT~2019qiz \citep{Nicholl2020}, $2\,900\rm\,km\, s^{-1}$ for AT~2020zso \citep{Wevers2022}, $1\,300\rm\,km\, s^{-1}$for AT~2020wey \citep{Charalampopoulos2023b}. For the fit, we used the \texttt{lmfit}\footnote{\url{https://lmfit.github.io/lmfit-py/}} package \citep{Newville2016}, with a Levenberg--Marquardt algorithm (i.e. least-squares method).

\begin{figure}
  \centering
  \includegraphics[width=0.5 \textwidth]{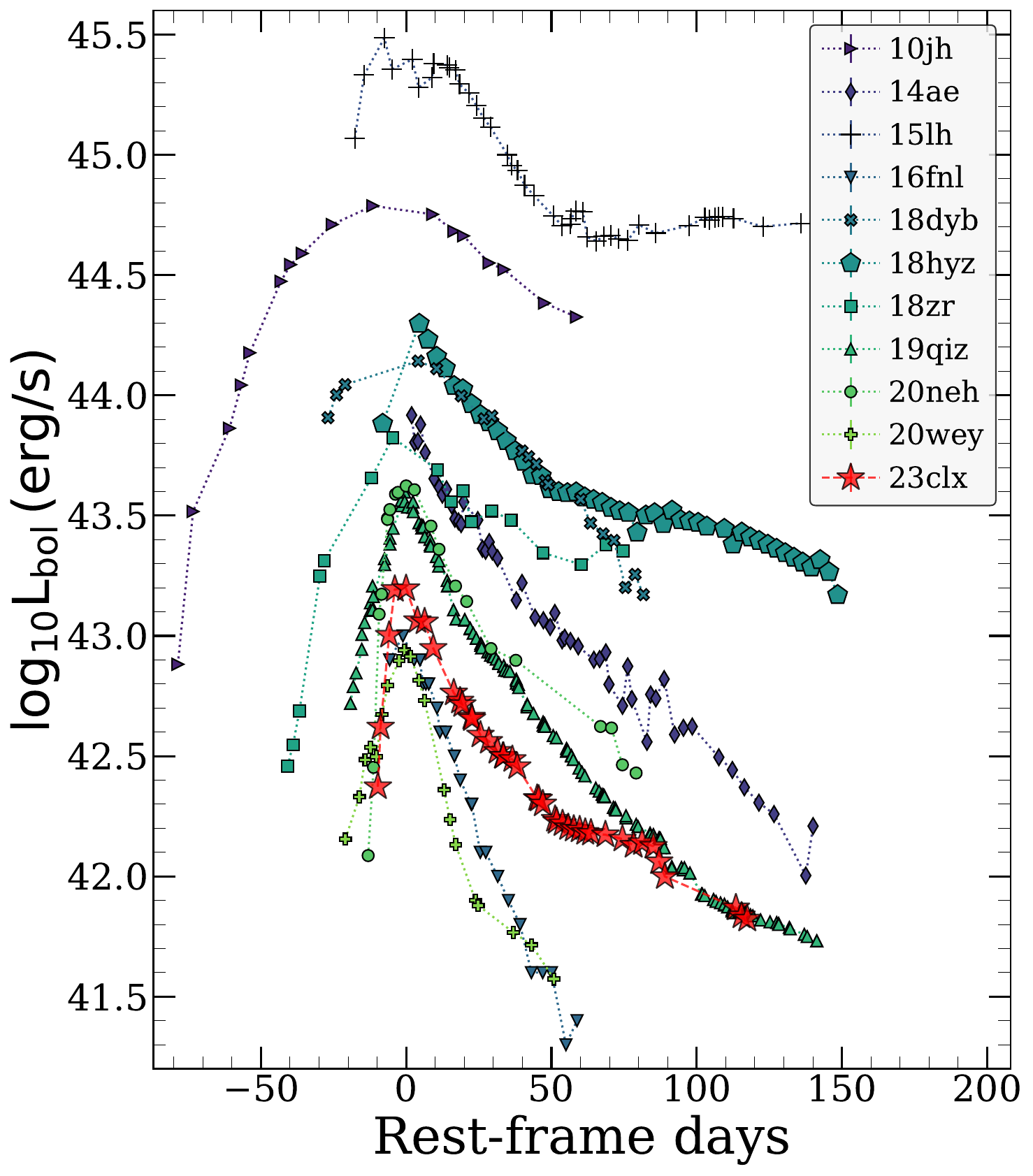}
  \caption{Comparison of the bolometric light curve of AT~2023clx with other TDEs and including ASASSN-15lh in this grouping. The shape of the light curve resembles that of AT~2018hyz.}
  \label{fig:bol_comp}
\end{figure}

In Fig. \ref{fig:bol_comp}, we plot the bolometric light curve compared to other well observed TDEs studied in \citet{Nicholl2020} and \citet{Charalampopoulos2023b}, as well as the super-luminous transient ASASSN-15lh (data from \citealt{Leloudas2016}). AT~2023clx is in the faint to intermediate luminosity region with a peak bolometric luminosity of ${L_{\rm pk}=(1.57\pm0.19)\times10^{43} \rm\,erg\,s^{-1}}$, having a peak luminosity between the faint ones like iPTF16fnl \citep{Blagorodnova2017} and AT~2020wey \citep{Charalampopoulos2023b} and the intermediate luminosity ones like AT~2019qiz \citep{Nicholl2020,Hung2020a,Short2023} and AT~2020neh \citep{Angus2022}. It also shows similar rise times with the latter, which was an exceptionally fast TDE candidate. Further discussion on the fast timescale can be found in Sect. \ref{subsec:fast_rise}. AT~2023clx also declines very fast during the first 30 days post-peak, when it shows a small `bump' and then declines slower. Similar behaviour is seen in AT~2018hyz and ASASSN-15lh. We note here that the rebrightening seen 70-90 days post peak is driven by a small bump in the $g$-band light curve, which is artificially enhanced by the constant colour extrapolation in the UV bands. Hence we do not trust it is real and we do not emphasise on it. Based on the different black hole mass estimates (see Sect. \ref{subsubsec:vel_disp} and \ref{subsubsec:mosfit}) and on the peak bolometric luminosity above, the Eddington ratio ranges from $\sim$ 0.03 to 0.25, hence sub-Eddington luminosities but still around typical values of optical and UV TDEs. We note here that if we had not applied the host extinction dereddening, those values would range from 0.01 to 0.08, somewhat low compared to typical TDE values. On a similar note, in Fig. \ref{fig:tbb_comp} of the Appendix, we plot the blackbody temperature evolution of AT~2023clx with and without the host galaxy extinction dereddening, compared to the TDEs presented in \citet{Hinkle2021a}. Without applying the host galaxy extinction, the temperature of the TDE would be very low, further supporting the need for correcting the data for the host galaxy extinction. On the other hand (as already noted in Sect. \ref{subsubsec:host_ex}), applying a correction for an extinction moderately higher than the one that we apply (i.e. $A_{V_{h}}=0.55$), leads to unrealistically high temperatures and oddly shaped SEDs, where the flux in the $UVM2$ filter is considerably higher than the one of $UVW2$.

Concerning the NIR imaging, we detect IR emission in the first three epochs of our NOTcam observations, as shown in Fig. \ref{fig:LC_pl}. In order to assess the nature of the emission, we compare the IR fluxes to a blackbody fit to the UV and optical fluxes, interpolated to the same epoch as the NIR observations through the power-law fits presented in Fig. \ref{fig:LC_pl}. For the first NIR epoch, the IR fluxes are consistent with the tail of the UV and optical blackbody Fig. \ref{fig:bol}, which at this phase of $+21$\,d has a temperature of $13\,000$~K $\pm 10\%$. However, for the second and third epochs ($+61$\,d and $+97$\,d respectively), the IR detections are in clear excess of the UV and optical blackbody. 
We simultaneously fit the UV, optical and IR SED with two blackbodies and derive the associated parameters and uncertainties using the \texttt{emcee} python implementation of the Markov Chain Monte Carlo (MCMC) method \citep{Foreman-Mackey2013}. We find that the additional blackbody has a temperature of $1\,100 \pm 200$~K and a radius of $(3 \pm 1) \times 10^{16}$~cm at $+61$\,d and a temperature of $962 \pm 164$~K and a radius of $(3 \pm 2) \times 10^{16}$~cm at $+97$\,d. In both cases, the UV and optical blackbody parameters are consistent with a single blackbody. The decreasing temperature and increasing radius is consistent with an IR echo for this emission (e.g. \citealt{Lu2016,VanVelzen2016a}). The luminosity of the IR blackbody from the Stefan–Boltzmann law, gives $L_{IR} = (9 \pm 8) \times 10^{41}$ and $(7 \pm 7) \times 10^{41}\,\rm \,erg\,s^{-1}$ at the second and third epochs respectively, with large uncertainties arising from the uncertainty in the radius. These luminosities are $\sim (7 \pm 7)$\% of the UV and optical blackbody luminosity at those epochs, indicating a high covering factor of dust compared to previous observations of TDEs \citep{Jiang2021}; however the large uncertainties do not allow for robust conclusions.

\begin{table}[h]
\renewcommand{\arraystretch}{1.4}
\setlength\tabcolsep{0.15cm}
\fontsize{9}{11}\selectfont
\centering
 \caption{Black hole mass estimates (second column) based on different methods (first column), and corresponding Eddington luminosites (third column) and Eddington ratios (fourth column).} \vspace{-0.2cm}
 \label{tab:mbh_ledd}
 \begin{tabular}{c|c c c}
  \hline
  Method & $\log_{10}(\mbh)$ ($\msun$) & $L_{Edd}$ ($10^{44}\,erg\,s^{-1}$) & $L_{pk}/L_{Edd}$\\
  \hline
M-$\rm\sigma$ (SDSS) & 6.49$\pm$0.39 & 3.88$^{+5.65}_{-2.30}$ & 0.04$^{+0.06}_{-0.02}$ \\
M-$\rm\sigma$ (X-shooter) & 5.71$\pm$0.40 & 0.64$^{+0.97}_{-0.39}$ & 0.24$^{+0.37}_{-0.15}$ \\
$\mbh$-$M_{stellar\_gal}$ & 6.62$\pm$0.69 & 5.24$^{+20.43}_{-4.17}$ & 0.03$^{+0.12}_{-0.02}$ \\
\texttt{MOSFiT} & 5.83$\pm$0.28 & 0.83$^{+0.75}_{-0.39}$ & 0.19$^{+0.17}_{-0.09}$ \\
  \hline
  \hline
 \end{tabular}\\
\end{table}

\subsubsection{TDE model fit} \label{subsubsec:mosfit}

We fit our host subtracted, multi-band light curves using the Modular Open Source Fitter for Transients \citep[\texttt{MOSFiT};][]{Guillochon2018} with the TDE model from \citet{Mockler2019} (see details for the parameters therein) in order to estimate physical parameters of the disruption. This model assumes a mass fallback rate derived from simulated disruptions of polytropic stars by a SMBH of $10^6\,\mathrm{M}_{\odot}$ \citep{Guillochon2014}, and uses scaling relations and interpolations for a range of black hole masses, star masses, and encounter parameters. We ran \texttt{MOSFiT} using dynamic nested sampling with \texttt{DYNESTY} \citep{Speagle2020} in order to evaluate the posterior distributions of the model. 

\begin{figure}
  \centering
  \includegraphics[width=0.5 \textwidth]{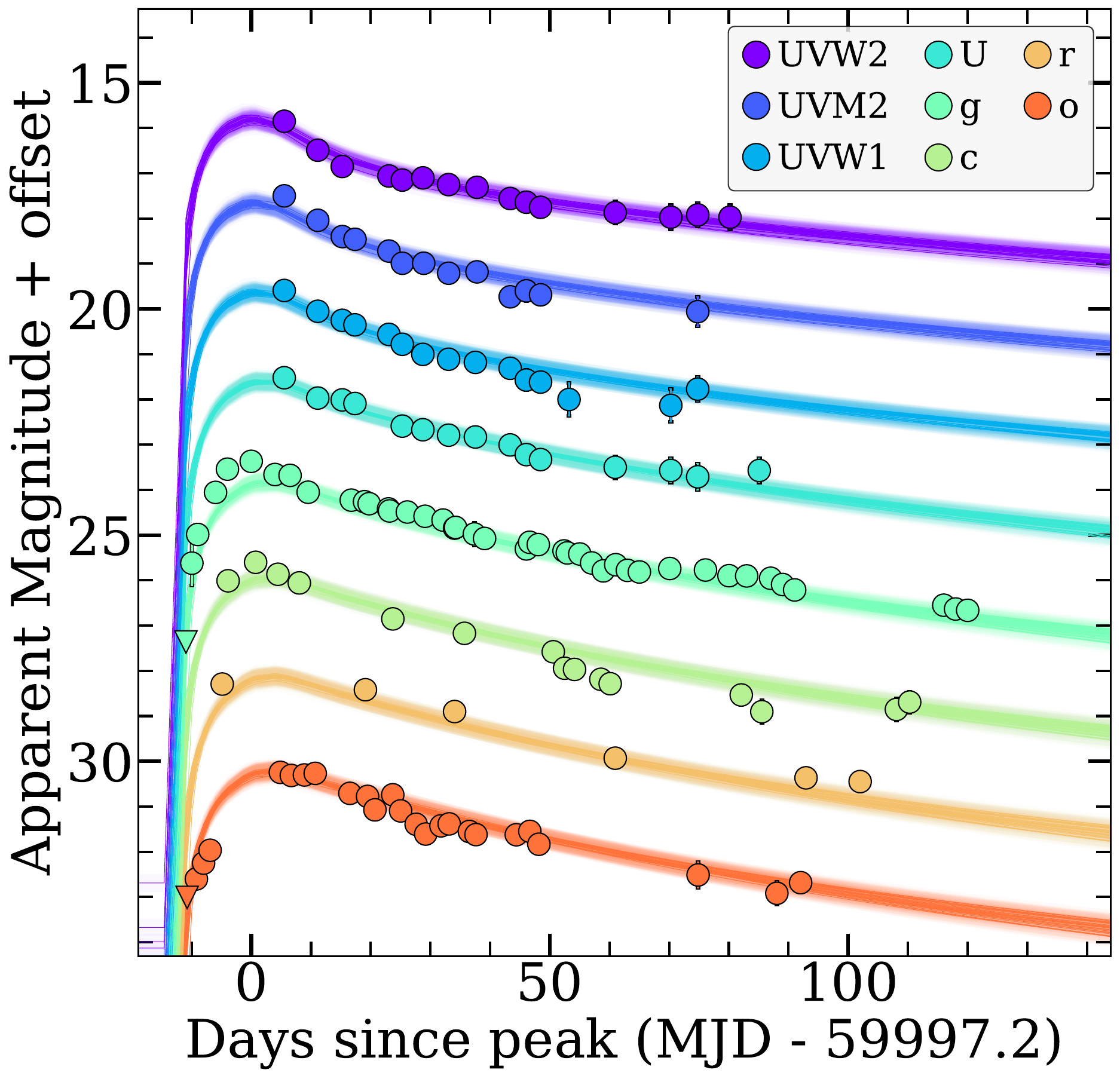}
  \caption{Fits to the multi-colour light curve using the TDE model in \texttt{MOSFiT} \citep{Guillochon2018,Mockler2019}. The relevant parameters are listed in Table \ref{tab:mosfit}.}
  \label{fig:mosfit}
\end{figure}

For our \texttt{MOSFiT} runs, we used several different configurations and combination of priors. For example, we considered fits with and without correction for host extinction, or fits where excluding sparsely sampled bands. In all cases, the returned stellar mass parameter ($\mstar$) invariably converged to $\sim$ 0.1 $\msun$. In order to investigate this behaviour further, we fixed the stellar mass to  
$\mstar\geq1$ and allowed it to vary only between $0.01\leq\mstar (\msun) \leq0.09$.
In the former case, it was clear that the match was poor in the sense that neither the rise time, nor the peak was reproduced. In the latter case, the fit converged at the higher bound (i.e. 0.09\,$\msun$) and the fits were virtually indistinguishable from those when the stellar parameter was left free to vary.
For the parameters listed in Table \ref{tab:mosfit}, the model light curves reproduce the salient features in each band (see Fig. \ref{fig:mosfit}).  We list the free parameters of the model (as defined by \citealt{Mockler2019}), their priors and their posterior probability distributions in Table \ref{tab:mosfit}, with two-dimensional posteriors shown in Fig. \ref{fig:corner}. \citet{Mockler2019} calculate the systematic uncertainties of \texttt{MOSFiT} TDE fitting and report a $\pm$0.2 dex uncertainty in the BH mass and a $\pm$0.66 dex in the stellar mass. The model returns a SMBH mass of $\log_{10}(\mbh) = 5.83 \pm 0.28$ dex, in between (and within 1-$\rm\sigma$) the M-$\rm\sigma$ relation estimates using the X-shooter spectrum ($\log_{10}(\mbh) = 5.71 \pm 0.40$ dex) and the SDSS spectrum ( $\log_{10}(\mbh) = 6.49 \pm 0.39$ dex). 

\begin{table}
  \centering
  \begin{tabular}{cccc}
  \hline
  Parameter & Prior & Posterior & Units\\
  \hline
$ \log{(M_\bullet )}$ & $[5, 8]$ & $ 5.83^{+0.06}_{-0.07} $ & M$_\odot$ \\
$ M_*$ & $[0.01, 100]$ & $ 0.10 ^{+0.01}_{-0.00} $  & M$_\odot$ \\
$ b$ & $[0, 2]$ & $ 1.03^{+0.06}_{-0.06}$ &   \\
$ \log(\epsilon) $ & $[-4, -0.4] $ & $ -3.22 ^{+0.05}_{-0.04}$ &   \\
$ \log{(R_{\rm ph,0} )} $ & $[-4, 4] $ & $ 1.90 ^{+0.10}_{-0.11} $  &   \\
$ l_{\rm ph}$ & $[0, 4]$ & $ 1.10 ^{+0.06}_{-0.06} $  &   \\
$ \log{(T_v )} $ & $[-3, 3] $ & $ -1.08^{+1.09}_{-1.24} $ & days  \\
$ t_0 $ & $[-500, 0]$ & $  -3.25 ^{+0.47}_{-0.61}$ & days  \\
$ \log{\sigma} $ & $[-4, 2] $ & $ -0.70  ^{+0.03}_{-0.03} $  &   \\
  \hline
\end{tabular}
  \caption{Priors and marginalised posteriors for the \texttt{MOSFiT} TDE model. The posterior results are the median of each distribution, and the uncertainties are the 16th and 84th percentiles. The errors are purely statistical and further estimates of the systematic uncertainty are provided in \citet{Mockler2019}.}
  \label{tab:mosfit}
\end{table}

\subsection{Spectroscopic analysis} \label{subsec:spec_analysis}

\subsubsection{Line identification} \label{subsubsec:lines}

After host subtracting (see Sect. \ref{subsec:opt_spec}) and dereddening the spectra, we can identify the emission lines that are present in the spectra. There are strong and broad Balmer lines in the spectra with H$\alpha$ being the most prominent. We also identify \ion{He}{II} $\lambda$4686, typical in TDE spectra and, especially after $\sim$ 20 days post-peak, we start seeing \ion{He}{I} $\lambda$5876. In the two spectra at $+17.7$\,d and $+20.9$\,d, H$\alpha$ develops a sharp peak centred at $\sim$ zero velocity on top of the broad profile, which remains centred around zero velocity until our last spectrum ($+125.7$\,d). In the spectrum at $+20.9$\,d, a sharp peak also emerges at $\sim$ \ion{He}{I} $\lambda$6678. We robustly identify this line as \ion{He}{I} $\lambda$6678 because, apart from being in the \ion{He}{I} wavelength, it emerges simultaneously with \ion{He}{I} $\lambda$5876 and both these lines start becoming weaker in the $+50.7$\,d spectrum and onward. A peculiarity of the \ion{He}{I} $\lambda$5876 line however is that, during the first epochs at least, it seems redshifted and an absorption feature is present on the blue side of the rest wavelength. The latter might be an artefact of the host subtraction, especially since the prominent \ion{Na}{I} D line is in this region of the spectrum as well. Furthermore, the (non host subtracted) NIR-arm spectra (Fig. \ref{fig:nir_spectra}) show the \ion{He}{I} $\lambda$10\,830 line but no evident hydrogen lines from the Paschen series. The \ion{He}{I} line is more prominent in the intermediate phases spectra ($+20.9$\,d, $+31.9$\,d, $+42.8$\,d), coincident with the emergence of the optical \ion{He}{I} lines. Finally, we cannot rule out neither robustly identify the Bowen Fluorescence \ion{N}{III} lines as one ($\lambda4640$) is blended with \ion{He}{II} $\lambda4686$ and a potentially very broad H$\beta$ structure and the other ($\lambda4100$) is blended with the H$\delta$ line in an also heavily blended region, where the continuum is very steep and blue. However, there might be an indication for the latter, as H$\delta$ appears stronger than H$\gamma$ in some of the early spectra but because we do not see that in all the spectra and because this is where the S/N starts deteriorating, we cannot confirm its presence. 

\begin{figure*}
\centering
\includegraphics[width=0.97 \textwidth]{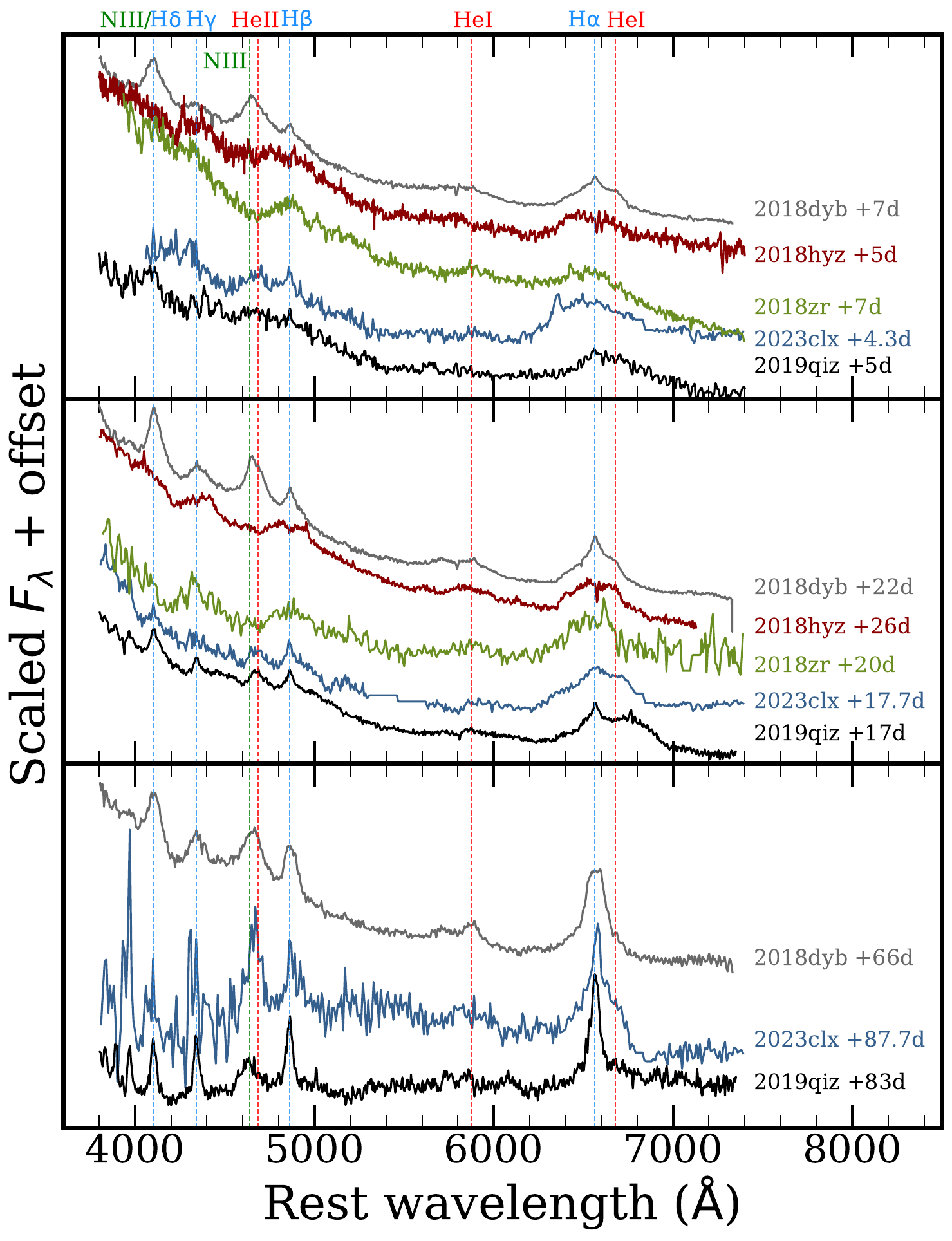}
\caption{Spectral comparison of AT~2023clx with other TDEs at different phases of its evolution. All the spectra presented here are host subtracted. The spectra share many similarities, most importantly a blue continuum that cools slowly with time and broad Balmer and \ion{He}{II} lines. There is a remarkable similarity in line profiles and continuum shapes with AT~2019qiz, a TDE that is of comparable luminosity.}
\label{fig:spec_comp}
\end{figure*}

In Fig. \ref{fig:spec_comp} we present a (host subtracted) spectral comparison of AT~2023clx with other TDEs throughout different phases of its evolution. The spectra share many similarities, most importantly a blue continuum that cools slowly with time and broad Balmer and \ion{He}{II} lines as well as a weak \ion{He}{I} $\lambda$5876 emission, features that are typically seen in TDEs. Furthermore, in the $\sim$ peak epochs (top panel) they all show an extended red wing, attributed by theory and models to early optically thick outflows and electron scattering \citep{Roth2017}. In the intermediate epochs (middle panel) some of the TDEs show a potential second bump on the red side of H$\alpha$, tentatively identified in the literature as \ion{He}{I} $\lambda$6678 (e.g. \citealt{Leloudas2019}). Also, the H$\alpha$ line narrows with time in all the TDEs (evident in the bottom panel; see Sect. \ref{subsubsec:lines} for quantification of the spectral properties and comparison with other TDEs). Finally, we note that there is a remarkable similarity in line profiles and continuum shapes with AT~2019qiz, a TDE that AT~2023clx has comparable luminosities to.

\begin{figure*}
    \centering
  \begin{minipage}[t]{.24\linewidth}
    \includegraphics[width=\linewidth]{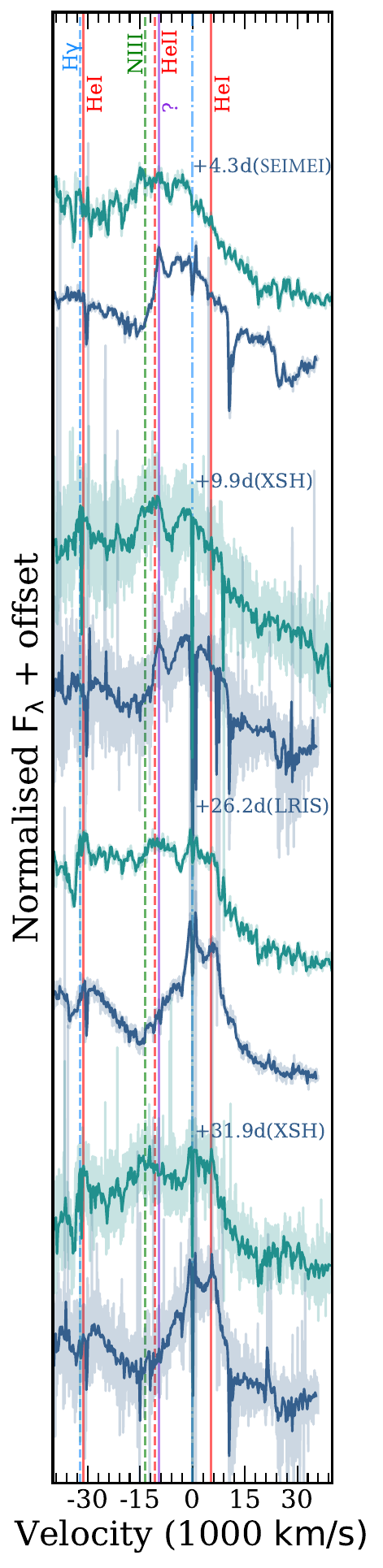}
  \end{minipage}\hfil
  \begin{minipage}[t]{.24\linewidth}
    \includegraphics[width=\linewidth]{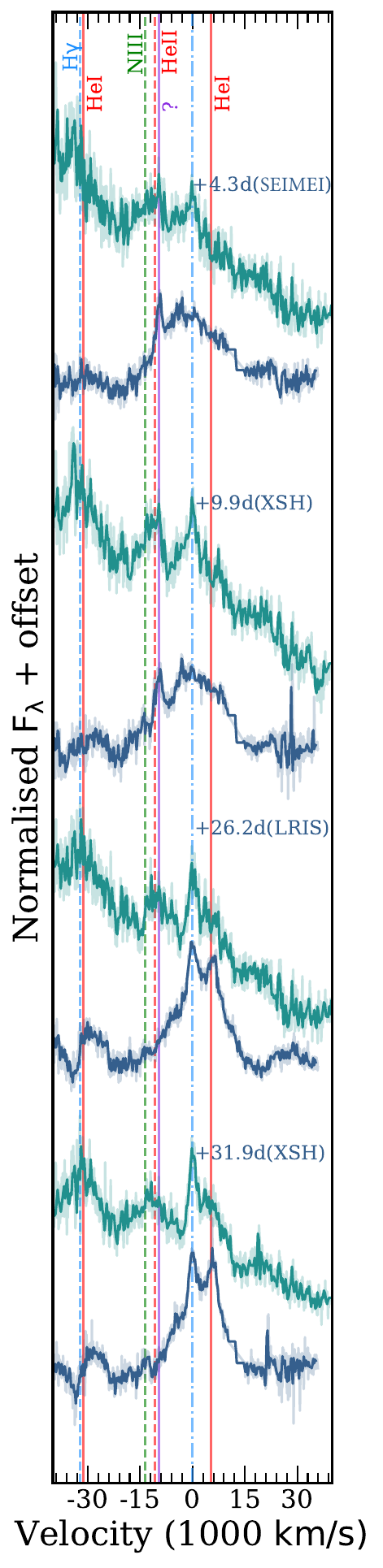}
  \end{minipage}\hfil
  \begin{minipage}[t]{.24\linewidth}
    \includegraphics[width=\linewidth]{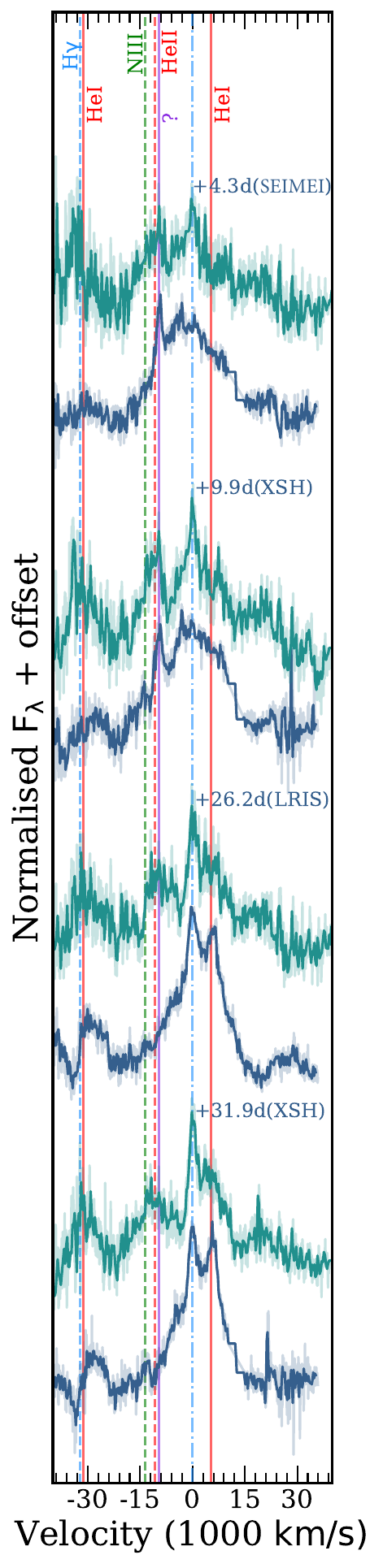}
  \end{minipage}%
\caption{Evolution of H$\alpha$ (blue) and H$\beta$ (green) region of the spectra in velocity space. Panel a: Before host subtraction. Panel b: Host subtracted. Panel c: Host subtracted and continuum subtracted. The dash-dotted vertical line shows the central wavelength. The colours of the vertical lines denote the different elements, blue is for hydrogen, red for helium, green for the \ion{N}{III} and in purple we show the sharp feature at 6353 \AA, present only until $\sim$ $+10$\,d post-peak. If this feature is a fast moving component connected to  H$\alpha$, then its blueshift corresponds to $\sim$ 9600$\rm km\,s^{-1}$. If H$\beta$ has a similar component with the same velocity, it would appear at $\sim$ 4706\,\AA. A line is indeed seen in these wavelengths but could be associated with \ion{He}{II} 4686\,\AA.}
\label{fig:spec_three_cols}
\end{figure*}

 A very interesting and puzzling feature of AT~2023clx is a sharp, narrow emission peak at a rest wavelength of $\sim$ 6353 \AA, on top of the blue side of the broad H$\alpha$ profile. This feature is clearly present in the first four spectra ($+4.3$\,d, $+5.0$\,d, $+9.8$\,d, $+9.9$\,d) taken with three different telescopes and spectrographs, showing that the feature is real and not an artefact. In the two spectra at $+17.7$\,d and $+20.9$\,d, the sharp peak is not there anymore; however it seems there is a bump in the blue wing of H$\alpha$. Then, from $\sim$ a month post-peak until our last spectrum, there seems to be a much fainter and less prominent feature at $\sim$ 6450 \AA. The short-lived emission feature at $\sim$ 6353 \AA\, does not seem to coincide with any known wavelength of any emission line, at least prominent in TDEs. The wavelength coincides with \ion{Si}{II} $\lambda\lambda$ 6347,6371 however this line usually appears in supernovae (SNe) and in absorption. So we discard the silicon interpretation. Another option is that the feature is a fast moving component of H$\alpha$ directed towards the observer with $\sim$ 9\,584 $\rm km\,s^{-1}$. We look for a similar component with the same velocity offset on the blue side of H$\beta$ but this wavelength is almost identical with \ion{He}{II} $\lambda4686$, hence it is hard to confirm or disregard the scenario because the \ion{He}{II} peak is already there. In order to look at the emission features without the underlying continuum slopes, we subtracted the continuum by fitting fourth and fifth degree polynomials in the line-free regions of the spectra. In order to let the reader visually inspect all the above probabilities and make their own judgement, we present in Fig. \ref{fig:spec_three_cols}, the H$\alpha$ and H$\beta$ regions in velocity space of four high S/N spectra, two that include the feature at 6353 \AA\, ($+4.3$\,d and $+9.9$\,d) and two that do not ($+26.2$\,d and $+31.9$\,d). From left to right, we show the original spectra without any subtraction, the host subtracted only spectra, and the host subtracted and continuum subtracted spectra. A longer discussion about the peculiar emission feature at $\sim$ 6353 \AA\, can be found in Sect. \ref{subsec:blip}.

\subsubsection{Line fitting} \label{subsubsec:lines}

\begin{figure*}
\centering
\includegraphics[width=.75 \textwidth]{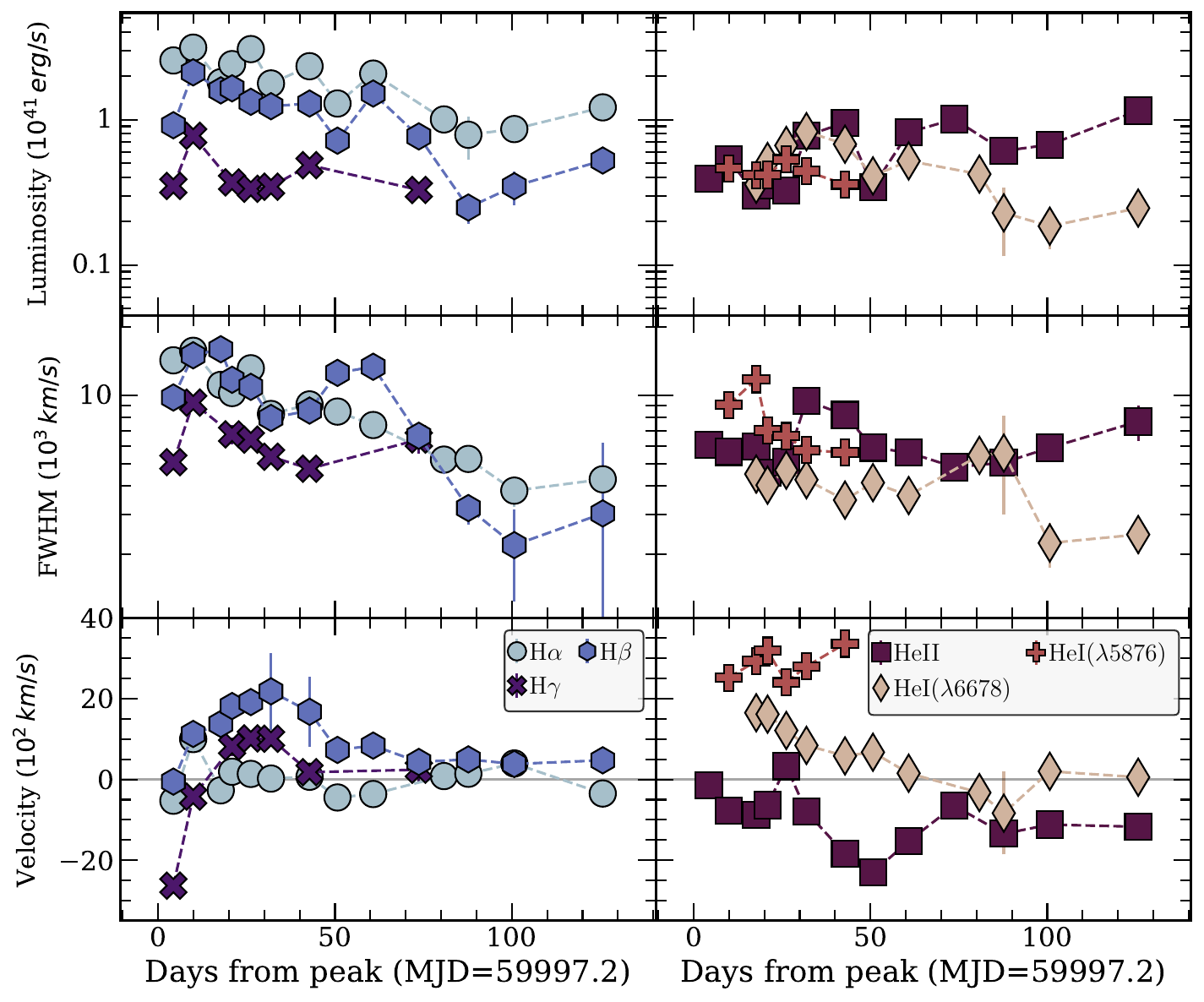}
\caption{Evolution of the emission line properties of AT~2023clx. Top panel: luminosity, middle panel: velocity width, bottom panel: velocity offset. The left panels present the properties of the Balmer lines, while the right panels the properties of the helium lines.}
\label{fig:line_props}
\end{figure*}

We decided to study and fit the following lines: H$\alpha$, H$\beta$, H$\gamma$, \ion{He}{II} $\lambda4686$, \ion{He}{I} $\lambda5876$ and \ion{He}{I} $\lambda6678$. In the first four spectra, we also fit the feature at 6353 \AA. Following the procedures of \citet{Charalampopoulos2022}, we fit the continuum subtracted spectra with both Gaussian and Lorentzian profiles and we find that the Lorentzians provide the better fit, as they both reproduce better the sharp peaks of, for example H$\alpha$ and \ion{He}{I} $\lambda6678$, and they give a better fit to the wings of the lines, especially when they are blended. The Lorentzians provided, in general, better chi-square values as well. In this way, we quantify the line luminosities, the velocity widths and the velocity offsets of the six lines mentioned above. We present the results in Fig. \ref{fig:line_props}. The Balmer lines seem to have identical behaviour, their luminosity and width rise between $\sim$ 5--10 days post peak before they start to decline smoothly onward. The lines slowly become narrower with time as their luminosity drops, in contrast with AGNs where a decrease in luminosity is accompanied by an increase in line widths (e.g. \citealt{Peterson2004, Denney2009}). In Fig. \ref{fig:ha_vs_sample} in the Appendix, we show the comparison of the H$\alpha$ luminosity and width of AT~2023clx, against the TDEs of the sample of \citet{Charalampopoulos2022}. We find that AT~2023clx has one of the brightest H$\alpha$ compared to other TDEs and that the luminosity of the line declines smoothly and not very fast. We note that during the temperature minimum discussed in Sect. \ref{subsubsec:Bol}, H$\alpha$ is very strong in luminosity, potentially leading to an overestimation of the blackbody temperature during those epochs.

\section{Discussion} \label{sec:discussion}

In the first three subsections of this section, we discuss three distinct features of AT~2023clx in order of appearance, and attempt to combine them in a single scenario in Sect. \ref{subsec:scenario}.

\subsection{The fast rise} \label{subsec:fast_rise}

\begin{figure}
\centering
\includegraphics[width=0.5 \textwidth]{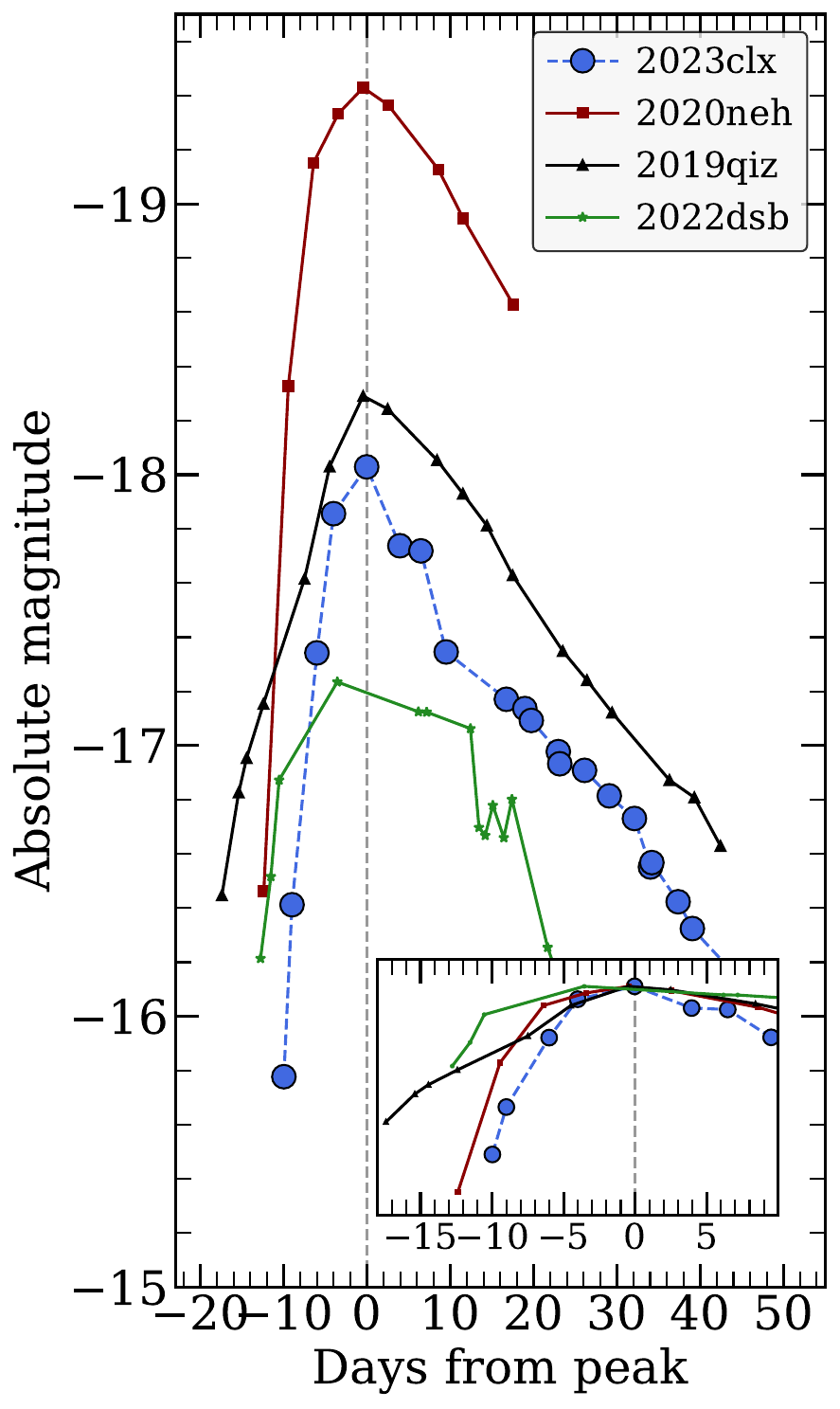}
\caption{Comparison of early $g$-band light curves of AT~2023clx, AT~2019qiz, AT~2022dsb (only ATLAS $o$-band available) and AT~2020neh, previously the fastest rising TDE to date. The peak epochs are taken from \citet{Nicholl2020}, \citet{Malyali2024}, and \citet{Angus2022} respectively. The inset shows the same data but normalised to peak light. AT~2023clx rises faster than all previously reported TDEs.}
\label{fig:g_comp}
\end{figure}

It has been suggested that the rapid rise of some TDEs could be correlated with low black hole mass (e.g. \citealt{Gezari2022}). That means that IMBHs could be probed by fast-rising TDEs, and such candidates have been discovered; the spectacular fast blue optical transient AT~2018cow was suggested to be a TDE from an IMBH \citep{Perley2019} and AT~2020neh was proposed to be the fastest rising TDE recorded, around an IMBH \citep{Angus2022}. In Fig. \ref{fig:g_comp} we compare the early $g$-band light curves of AT~2023clx against those of AT~2019qiz which is a fast TDE of comparable luminosity to AT~2023clx, AT~2020neh which is the fastest TDE to date, and the fast rising AT~2022dsb \citep{Malyali2024} for which only ATLAS $o$-band data nicely cover the rise. The inset figure shows the same data but normalised to peak light. AT~2023clx rises faster than all those fast TDEs making it the fastest rising TDE recorded. Our average SMBH estimate (see Table \ref{tab:mbh_ledd}) -- excluding the high scatter estimate based on the empirical relation between the total galaxy stellar mass and the BH mass -- is $\overline{M}_{BH}\approx10^{6.0} \msun$. We use this estimate for the rest of the manuscript. This estimate is too high for IMBHs that have masses $\lesssim10^{5} \msun$, and lies firmly towards the lower mass range when compared to other TDEs. Thus, the fast rise of AT~2023clx cannot be directly attributed to the BH mass.

\begin{figure}
\centering
\includegraphics[width=0.5 \textwidth]{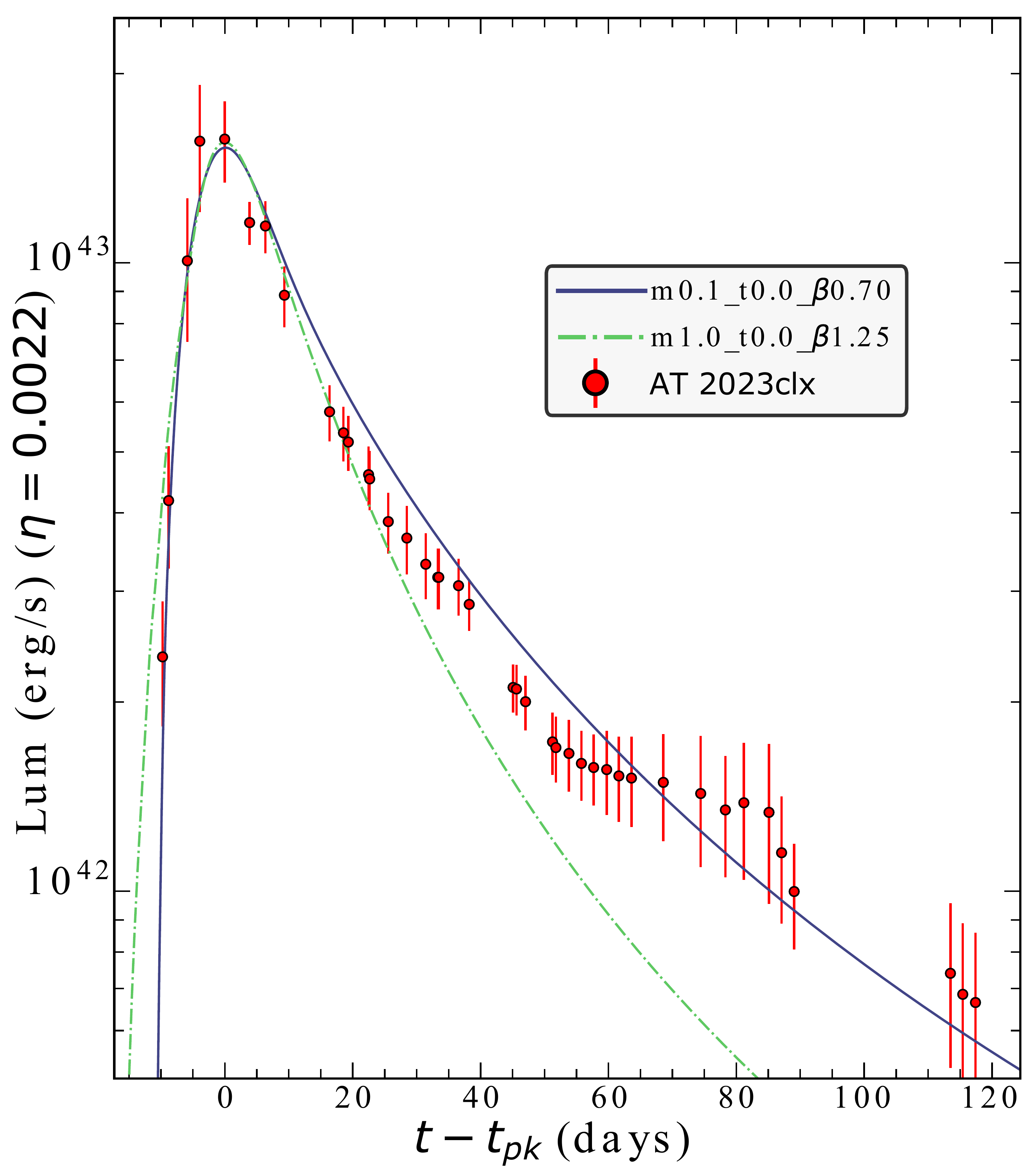}
\caption{Comparison of the bolometric luminosity evolution of AT~2023clx, against mass fallback rate models of \citet{Law-Smith2020}, of a $0.1$ and $1.0 \msun$ ZAMS stars disrupted with an impact parameter of $\beta=0.7$ and $1.25$ ($b\simeq0.8$) by a $10^{6.0} \msun$ BH mass. The data are not normalised; in order to convert the mass fallback rate to luminosity and for the peaks to match, a radiative efficiency parameter of $2.2 \times 10^{-3}$ was needed for the $0.1 \msun$ model. However, the $1.0 \msun$ model required an unrealistically low efficiency parameter of $1.5 \times 10^{-4}$ in order to match the peak luminosity. The $0.1 \msun$ model and the data show an identical rise as well as a shallower decline (after the NUV break).}
\label{fig:JLS_single}
\end{figure}

\citet{Law-Smith2020} presented a library of TDE simulations\footnote{\url{https://github.com/jamielaw-smith/STARS_library}} providing the mass fallback rate to the BH for a main-sequence (MS) star of any stellar mass, stellar age, and impact parameter, with realistic stellar structures and compositions. Conveniently for us, they used a $10^{6.0}$ BH mass in order to calculate their mass fallback rate models. This mass is identical to our average BH mass estimate for AT~2023clx. 
The impact parameter $\beta$ is defined as the ratio of the tidal radius to the pericenter distance; it is a measure of the `strength' of the tidal interaction. The critical impact parameter ($\beta_{crit}$) is defined as the smallest impact parameter for which full disruption can be achieved (where all the star's mass is disrupted); below this critical value a partial disruption will occur. The scaled impact parameter $b$ is the ratio of the impact parameter to the critical one (i.e. $b = \beta/\beta_{crit}$). The mass fallback rate is translated to bolometric luminosity through $L=\eta\dot{m}c^{2}$, where $\eta$ is the radiative efficiency. \citet{Law-Smith2020} find that, in general, less centrally concentrated and less massive stars have steeper mass fallback rate rise slopes and shallower decay slopes. This motivated us to compare the bolometric luminosity evolution of AT~2023clx against the models of \citet{Law-Smith2020} that could simultaneously match the rising luminosity rate and the peak luminosity, after converting the mass fallback rates to luminosity assuming an $\eta$ value. 
Our \texttt{MOSFiT} fits returned a best-fit value of $\eta\simeq6 \times 10^{-4}$ that we used as our initial guess in converting the mass fallback rates. In Fig. \ref{fig:JLS} of the Appendix, we show the comparison of the bolometric luminosity evolution of AT~2023clx, against many of those mass fallback rate models. In the left panels the luminosities are not normalised while in the right panels the luminosities are normalised to peak light. In the top panels, we indeed see that TDEs occurring from higher mass stars have shallower rises while lower mass stars should lead to faster rising light curves, as discussed by \citet{Law-Smith2020}. Concentrating on lower mass and zero-age main sequence (ZAMS) stars that seem to better match with AT~2023clx evolution (bottom panels of the figure), we find models that have similar peak luminosities and rise times with AT~2023clx. Especially a model of a $0.1 \msun$ ZAMS star disrupted with an impact parameter of $\beta=0.7$ ($b\simeq0.8$) shows an almost identical rising luminosity rate. Converting the mass fallback rate of this model with $\eta$ set to $2.2 \times 10^{-3}$, we find also a match in peak luminosity. We also show the comparison with a fast rising $1.0 \msun$ model which still shows a slower rise time than AT~2023clx and that needed an unrealistically low efficiency parameter of $1.5 \times 10^{-4}$ in order to match the peak luminosity. We present the comparison of these models and AT~2023clx in Fig. \ref{fig:JLS_single}. The $0.1 \msun$ model and the data show a nearly identical rising rate. Also, the decline of AT~2023clx after the NUV break, is better described by the shallower decline of the $0.1 \msun$ model. Furthermore, the stellar mass of this model ($0.1 \msun$) agrees with our \texttt{MOSFiT} estimate where, as discussed in Sect. \ref{subsubsec:mosfit}, the stellar mass best-fit value converges to $\sim 0.1 \msun$. Also, we get a best-fit value for the scaled impact parameter $b = 1.03$, in broad agreement with the one of the model ($b=0.80$) as the \texttt{MOSFiT} systematic error was calculated by \citet{Mockler2019} to be $\pm0.35$. Hence, we suggest that AT~2023clx was caused by the disruption of a low-mass star ($\lesssim 0.1 \msun$), with low central concentration ($\rho_{c}/\overline{\rho}\sim5.5$) and a small radius ($\sim 0.12 \rsun$). The TDE could also have been caused by the disruption of a brown dwarf ($0.01\msun\lesssim M_{BD}\lesssim0.08\msun$). Such TDEs should exist and early modelling showed that they should indeed have steep rises and sub-Eddington luminosities \citep{Law-Smith2017a}. In terms of observations, there were two early TDE candidates, suggested among other scenarios to be the disruption of brown dwarfs \citep{Li2002,Nikolajuk2013}. However, those candidates occurred before the era of optical wide-field surveys and were detected only in X-rays. Detailed modelling is needed in order to understand if the unique properties of AT~2023clx are caused by the disruption of a brown dwarf and how different the resulting observables would be in comparison with the disruption of a low-mass MS star.

Another possibility that could potentially explain the fast rise could be the tidal disruption of a star on a bound elliptical orbit (e.g. \citealt{Hu2024}). This naturally reduces the mass return rate and enhances the efficiency of dissipation in stream crossings, hence should lead to much faster rise times than canonical TDEs (occurring from parabolic stellar orbits). Again, detailed modelling is needed in order to test whether more of the observables can be reproduced by such an encounter.

\subsection{The 6353 \AA\, emission feature} \label{subsec:blip}

The 6353 \AA\, emission feature present in our first four spectra ($\sim$ 4 -- 10 days post-peak) is puzzling as it is the first time that it is detected in a TDE. One possibility that we considered is whether this feature is [\ion{O}{I}] caused by far-UV (FUV) pumping. This mechanism requires a source of FUV photons (which exists in TDEs) which is absorbed by oxygen atoms that get excited to higher bound levels, followed by a cascade of radiative or collisional de-excitations leading to the upper levels of the forbidden lines. This interpretation would be consistent with the fact that we see the quick early decline in the NUV bands and the subsequent rise, coinciding with the disappearance of the emission feature. Typically, these [\ion{O}{I}] lines would be at $\lambda\lambda$6300,6364 and $\lambda$5577 \citep{Nemer2020}. However, the central wavelength of the line is at 6353 \AA. It could potentially be a blend of [\ion{O}{I}] $\lambda$6363 and $\lambda$6300; however one would expect that the central wavelength would be closer to $\lambda$6300 since this typically the stronger transition (with a ratio of 3 to 1 compared to the $\lambda$6363, based on the Einstein coefficients). The fact that the line is narrow and there is almost no flux at $\lambda$6300 weakens the possibility that the feature is indeed [\ion{O}{I}] triggered by FUV pumping, although we cannot completely disregard this scenario. 

Another possibility is that this is a high velocity hydrogen component related to H$\alpha$. This would require a velocity of $\sim$ 9\,584 $\rm km\,s^{-1}$ with a direction towards the observer (since the line is blueshifted if indeed related to H$\alpha$). Looking for a similar component with the same velocity offset on the blue side of H$\beta$ does not confirm or reject the scenario because this wavelength is almost identical with \ion{He}{II} $\lambda4686$, and \ion{He}{II} is still there when the 6353 \AA\ feature disappears (see the line identification discussion as well on Sect. \ref{subsubsec:lines}). The feature appears to fade at 6353\,\AA\, but continues to affect the blue wing of H$\alpha$ while it potentially slows down, for epochs out to +42.8\,d. The feature is relatively narrow (compared to typical TDE line widths); the full-width at half-maximum (FWHM) of the line is $\sim 2\,000~\rm km\,s^{-1}$ in the first two spectra ($+4.3$\,d and $+5.0$\,d) and slightly rises to $\sim 2\,700~\rm km\,s^{-1}$ in the next two spectra ($+9.8$\,d and $+9.9$\,d) before it disappears. This makes it $\sim$ six times narrower than the broad H$\alpha$ profile seen in AT~2023clx. Based on these velocity widths, if the emitting material were to predate the TDE, that is, material surrounding the SMBH of the LINER, a Keplerian estimate of the distance from the central object would be $\sim 2.1 \times 10^{15}$~cm. However this cannot explain the $\sim$ 9\,584 $\rm km\,s^{-1}$ blueshift (i.e. high velocity towards the observer). If this material is launched from the TDE, it could be preceding the bulk of an outflow. The bulk of this outflow would set the profile of H$\alpha$, and would match the simulated profiles of \citet{Roth2017} that suggested that electron scattering within an outflow would lead to early blueshifted H$\alpha$ with an extended red wing, similar to what we see in the first epochs of AT~2023clx (see for example the top spectrum in Fig. \ref{fig:spec_three_cols} panel c). Hence, the 6353 \AA\, emission feature could be clumpy material preceding the bulk of the outflow with a velocity of $\sim$ 9\,584 $\rm km\,s^{-1}$, ionised by the light of the rapidly expanding blackbody photosphere, and recombining in order to radiate the Balmer line photons. In the outflow picture of \citet{Metzger2016}, the outflow has a velocity $v_{ej}^{min}$ (see their Equation 12) which, if we solve for the parameters of AT~2023clx ($\mbh\approx10^{6}\msun$, $\mstar\approx0.1\msun$, $\beta\approx0.7$, and $f_{in}=0.05$ \footnote{A main assumption of the authors is that only a small fraction $f_{in}$ of the infalling tidal debris joins the inner Keplerian disc near the circularisation radius and accretes on to the SMBH; they note that reasonable values span $0.01\lesssim f_{in}\lesssim0.1$.}), would give 9\,922 $\rm km\,s^{-1}$. This number ties nicely with the blueshift of the peculiar emission feature ($\sim$ 9\,584 $\rm km\,s^{-1}$) as well as with the spread of velocities (FWHM) that the H$\alpha$ line shows (see Fig. \ref{fig:line_props}; middle panel), dictated by the bulk of the trailing outflow. So, in the picture of \citet{Metzger2016} (see their figure 1), the outflow would be directed towards the observer. Within the viewing angle dependant reprocessing TDE scenario (e.g. \citealt{Metzger2016,Dai2018}), which would explain the lack of detected X-ray emission and a potential very early ($\sim$ $+4.8$\,d post-peak) radio detection \citep{Sfaradi2023}, similar to the radio detections of AT~2019qiz \citep{Nicholl2020}. As the TDE evolves, the bulk of the outflow sweeps up surrounding material. If the emission feature indeed arose in clumpy material, this material would gradually become optically thin while spreading away from the central engine of the TDE. So, the emission feature would be gradually assimilated in the H-dominated outflow, and would eventually disappear.

\citet{Law-Smith2020} also find that for a less centrally concentrated star, the outer layers are
less vulnerable to tidal disruption and this material is then less stretched out post-disruption. This means that some structure in the debris is preserved and that the outflow would be less spherical. The fact that the debris are less spread could create the needed `clumpy' nature of the high velocity component and the lack of sphericity might allow the H$\alpha$ detection of the high velocity clumpy material, not seen in other TDEs. Deviation from outflow sphericity could also be caused by a highly spinning SMBH, as the debris streams might collide with an offset due to Lense-Thirring precession \citep{Jankovic2023}. 
 
Another option would be that this high velocity material is post-disruption unbound material that escapes in unbound orbits and moves quickly towards the observer. In the same way as we described above, the light of the rapidly expanding blackbody photosphere would reach that material before it spreads too much, hence being dense enough to ionise and recombine, leading to emission lines. Based on the traditional picture (e.g. \citealt{Bonnerot2021}), the unbound debris should travel in the plane of the initial parabolic orbit of the star, towards the apocenter of this initial trajectory. This means that the observer should be aligned in this direction, in order to detect the  high velocity blueshift.

\subsection{The temperature break} \label{subsec:UV_dip}

In Sect. \ref{subsubsec:bb_lc} we noted that the broad-band light curves of AT~2023clx show relatively rapid wavelength-dependent evolution, with the bluer bands declining faster (Fig. \ref{fig:colours}). Fitting power-laws to these early epochs resulting in indices monotonically declining from $-2.09$ to $-0.87$, for filters spanning $UVW2$ to $g$ (Fig. \ref{fig:LC_pl}). This is mirrored in the blackbody fits, in the form of a drop in temperature during the first $\sim$ 20 days (Sect. \ref{subsubsec:Bol} and middle panel of Fig. \ref{fig:bol}). This may hint at competing emission mechanisms at play, with one mainly responsible for emission up to a couple of weeks post-peak, before another one takes over, resulting in constant or slowly rising temperature.  Such an early cooling is unusual for TDEs, as they generally show constant colours even during these early epochs, with few exceptions such as ASASSN-14ae \citep{Holoien2014} that shows a similar early temperature evolution to AT~2023clx.

\begin{figure}
  \centering
  \includegraphics[width=0.5 \textwidth]{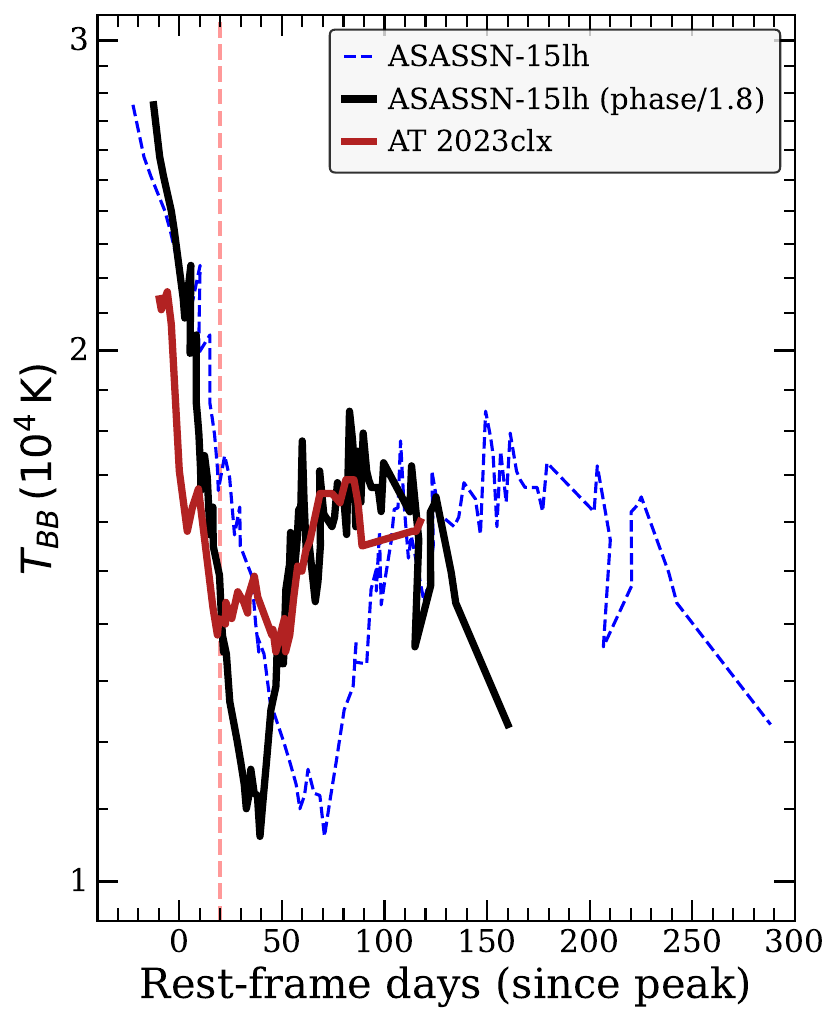}
  \caption{Comparison of the blackbody temperature evolution of AT~2023clx with that of ASASSN-15lh (blue dashed line). We also plot (in black) the temperature evolution of ASASSN-15lh compressed in time by a factor of 1.8, in order temporally align the two temperature breaks with respect to peak light. The vertical red-dashed line is the same as in Fig. \ref{fig:LC_pl}.}
  \label{fig:tbb_comp_15lh}
\end{figure}

Interestingly, the super-luminous transient ASASSN-15lh was argued to be a TDE from a highly spinning, very massive ($ >10^{8} \msun$) BH (\citealt{Leloudas2016}, also see \citealt{Margutti2016}). One of its remarkable features was that it showed an early temperature drop followed by a rebrightening at the UV wavelengths around two months after the peak, which has not been fully understood to date. \citet{Leloudas2016} proposed that this feature was due to the circularisation of the debris, while the rebrightening was due to the prompt formation of an accretion disc. Recently, \citet{Guo2023} came to the same conclusion after studying the inter-band cross-correlation function lags, where the UV consistently peaks earlier than the optical. In Fig. \ref{fig:tbb_comp_15lh}, we show the qualitatively similar temperature evolution of AT~2023clx and ASASSN-15lh. As ASASSN-15lh showed both a more prominent and more prolonged evolution, the drop in AT~2023clx is only seen in three epochs. Therefore, we also show a compressed temperature evolution of ASASSN-15lh in order to roughly match the minimum. In spite of the broad similarities in the early luminosity evolution highlighted above, we stress that there are, of course, huge differences in their respective properties. Nevertheless, we describe below a plausible scenario to account the observed similarity.

\citet{Hong2022} studied the effects of disc formation efficiency ($C$) in TDEs which they express as a function of the BH mass, the disrupted star's mass, and the impact parameter. They find that the efficiency has a monotonically increasing trend towards $\beta\approx0.7$ and lower stellar mass. They depict the above in their figure 2. Assuming $\mbh\approx10^{6}\msun$, $\mstar\approx0.1\msun$, $\beta\approx0.7$, AT~2023clx would be in the $C\approx1$ contour, just within the locus of the highest possible efficiency with the star plunging directly into the BH event horizon without producing a detectable flare. This is the case for a $10^{6}\msun$ BH mass (their middle panels) and is consistent with our interpretation that AT~2023clx must have undergone prompt and efficient disc formation. It is also interesting to note that for a BH mass of $10^{8}\msun$, \citet{Hong2022} find that the only possible disruption would be from a high mass star with low $\beta$, leading to very high efficiency ($C\approx1$). That could be applicable to ASASSN-15lh. The low stellar mass in AT~2023clx will lead to a small amount of available material (e.g. $\mstar/2 \sim 0.05 \msun$) to form a reprocessing layer, making the layer low in density. Hence, the circularisation after the debris stream self-crossing, shocks and/or outflows, dominate the early emission and rapidly cool-down, and then, due to the efficient circularisation, prompt accretion and reprocessing take over (a scenario also previously suggested for ASASSN-15lh).

\begin{figure}
\centering
\includegraphics[width=0.5 \textwidth]{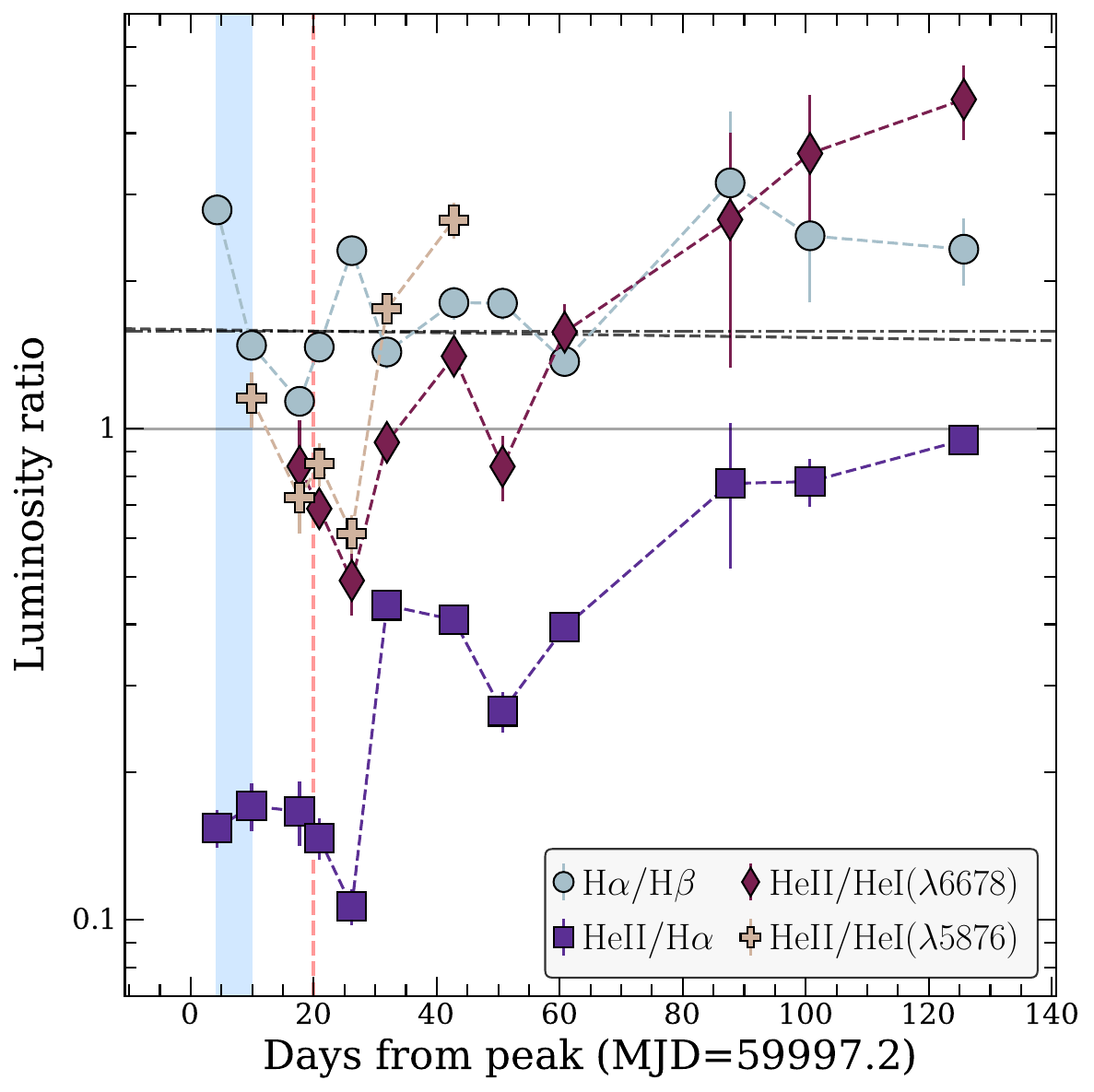}
\caption{Evolution of the emission line luminosity ratios with time. The black dashed line shows a weighted linear fit ($y=ax + b$) to the Balmer decrement ($b=1.60 \pm 0.24)$, while the dash-dotted one ($b=1.58 \pm 0.13)$ is the same but with the assumption that the decrement is flat (i.e. $a\equiv0$). The coloured region and vertical dashed line, are the same as in Fig. \ref{fig:LC_pl}.}
\label{fig:line_ratios}
\end{figure}

In Fig. \ref{fig:line_ratios} we show several line ratios to investigate how they change during and after the temperature break. We see that until $\sim$ +26\,d, all ratios of \ion{He}{II} drop. This is likely a consequence of the dropping temperature with \ion{He}{II} recombining to \ion{He}{I}, with this trend first flattening and then reversing as the temperature rises again, resulting in an increase in the \ion{He}{II} luminosity.

Fig. \ref{fig:line_ratios} also reveals an almost flat Balmer decrement ($L_{H\alpha}$/$L_{H\beta}$). We perform a weighted linear fit ($y=ax + b$) to the line ratio and find $a = -0.0007 \pm 0.0066$ and $b = 1.60 \pm 0.24$. For the same fit, but with the assumption that the ratio is flat (i.e. $a\equiv0$) we find $b = 1.58 \pm 0.13$. Both fits are shown in Fig. \ref{fig:line_ratios}. In a typical AGN broad line region, this ratio is $\sim 3-4$, which is consistent with Case B recombination \citep{Osterbrock1974}. 

Such a comparable and flat Balmer decrement was previously observed only in the TDE AT~2018hyz ($L_{H\alpha}$/$L_{H\beta} \sim 1.5$). For it, \citet{Short2020} suggested that the lines were collisionally excited rather than being produced via photoionisation. Furthermore, the early detection of double-peaked Balmer lines was attributed to prompt accretion disc formation \citep{Hung2020}, and to a reprocessing layer of low optical depth, resulting from the disruption of a low-mass star (invoked by the light curve shape and evolution; \citealt{Gomez1925}). Although we do not detect double-peaked lines in AT~2023clx, this could be due to a different viewing angle of observation. Nevertheless, the early spectra of these two transients share a lot of similarities especially around peak (see top panel of Fig. \ref{fig:spec_comp}) and their blackbody luminosities also have a similar evolution (see Fig. \ref{fig:bol_comp}); both show a break in the luminosity decline, which happens earlier and is more prominent in AT~2023clx. 

Based on the discussion of Sect. \ref{subsec:fast_rise}, we believe it to be probable that the disrupted star had a very low mass ($\sim 0.1 \msun$), and only a portion of this ($\sim$ 50\%) will end up bound around the SMBH. A reprocessing layer encompassing the SMBH predicted by the reprocessing TDE scenario (e.g. \citealt{Guillochon2013,Metzger2016,Dai2018}), would be of low density because of the low available mass, leading to the detection of the disc emission responsible for the early rapid cooling and the temperature break.

\subsection{A plausible scenario} \label{subsec:scenario}

Combining the above three subsections we paint a picture of a plausible scenario for AT~2023clx. 

AT~2023clx likely results from the tidal disruption of a low-mass ($\lesssim0.1 \msun$) small radius ($\sim 0.12 \rsun$) star, with low central concentration ($\rho_{c}/\overline{\rho}\sim5.5$), close to full disruption ($b\approx0.8$). Following \citet{Law-Smith2020}, such a disruption should show sharper rise and shallower decline in the light curves than disruptions occurring from other combinations of these parameters. Our observations are in line with these expectations: AT~2023clx is the fastest rising TDE to date, and also shows a shallower decline after the early `rapid cooling'.

Based on \citet{Hong2022}, the combination of the above parameters (BH mass, stellar mass, and $\beta$) favours highly efficient disc formation after the debris stream self-crossing. This could lead to the manifestation of two different emission mechanisms at play during the evolution, {\it viz.} circularisation after the debris stream self-crossing, with shocks and/or outflows dominating the early emission and rapid cool-down. Then, due to the efficient circularisation (combined with the low density of the reprocessing layer resulting from the low mass of the infalling star), prompt accretion and reprocessing take over.

Two possible mechanisms can be invoked.
The light from the rapidly expanding blackbody ionises either the i) clumpy preceding part of an outflow (as in the picture of \citealt{Metzger2016}), or ii) some post-disruption unbound material. In either case, the ionised material is launched and moves rapidly in our line-of-sight; 
as it cools and recombines, this could account for the observed feature we observe at $\sim$6353\,\AA\. In the former case, the outflow is aligned close to our line-of-sight, while in the latter, unbound debris travels in the plane of the initially parabolic orbit of the infalling star, and towards the apocentre of this orbit.

In both those cases, this would lead to an edge-on view within the reprocessing emission mechanism paradigm, explaining the lack of X-rays. The late ($+97.7$d) XMM-NEWTON detection reported in \citet{Hoogendam2024}, ties in nicely with the viewing angle dependant scenario, where the X-rays are initially obscured by the reprocessing layer that gradually drops in density until it is thin enough for the X-rays to escape. A low late-time X-ray luminosity is what would be expected for an edge-on view \citep{Hayasaki2021}. Therefore, the unique properties of AT~2023clx (and the lack of these in other TDEs) could potentially be described by a combination of disruption parameters and viewing angle effects.

Finally, and for completeness, we considered the possibility that AT~2023clx is due to low-luminosity AGN activity, but deemed this to be unlikely for the following reasons:
i) Four different methods of estimating the central BH mass yield
$\overline{M}_{BH}\approx10^{6.0} \msun$, while typical AGN SMBH masses are $\gtrsim10^{8} \msun$. ii) The fast rise and shallower decline of the light curves resemble neither changing look AGN (CL-AGN; e.g. \citealt{Ricci2023}), nor the stochastic variability of typical AGN flares. iii) Spectroscopically, several indicators point to a TDE rather than AGN activity. For instance, the spectra do not show the characteristic \ion{Mg}{II} and \ion{Fe}{II} \citep{Frederick2020} emission features seen in AGN and CL-AGN. Furthermore, the fact that the lines slowly become narrower with time as their luminosity drops, is a TDE characteristic rather an AGN one; in the latter, a decrease in luminosity is accompanied by an increase in line widths (e.g. \citealt{Peterson2004, Denney2009}). Also, the Balmer lines show a flat Balmer decrement ($L_{H\alpha}$/$L_{H\beta} \sim 1.58$) suggesting that the lines are collisionally excited rather than being produced via photoionisation \citep{Short2020}; this is in contrast to typical AGN that show a ratio of $\sim 3-4$,consistent with Case B recombination \citep{Osterbrock1974}. Finally, the spectra of AT~2023clx are similar to other TDE spectra in the literature (Fig. \ref{fig:line_props}).
iv) The supersoft X-ray spectrum detected by \citet{Hoogendam2024} at late times (see discussion therein) along with the lack of early X-ray detection and stochastic variability, is also more reminiscent of TDEs.

\section{Conclusions} \label{sec:conclusion}

We present UV, optical, and NIR follow-up and detailed analysis of AT~2023clx, the closest optical/UV TDE to date (z=0.01107), in the nucleus of the interacting galaxy NGC~3799.
\begin{enumerate}[label={\arabic*.}]

\item After removing the stellar continuum, we place the host in a BPT diagram and find it to be a LINER galaxy, very close to the AGN and composite region locus, in line with previous classifications.  
\item Strong \ion{Na}{I} D absorption from the host-galaxy spectrum motivated us to quantify the host extinction. Based on our archival SED fits, we find an $\rm E(B-V)_{h}$ = 0.179 mag for the host, which we use when dereddening our data.
\item Employing several standard methods (M--$\rm\sigma$ relation and \texttt{MOSFiT} fits estimate), we estimate the mass of the central supermassive black hole to be $\overline{M}_{BH}\approx10^{6.0} \msun$.
\item We measure a peak absolute magnitude of $-18.03 \pm 0.07$ mag in the $g$-band and a maximum bolometric luminosity of ${L_{\rm pk}=(1.57\pm0.19)\times10^{43} \rm\,erg\,s^{-1}}$.
\item AT~2023clx rose to peak within $10.4 \pm 2.5$ days, making it the fastest rising TDE to date. Our SMBH mass estimate rules out the possibility of an IMBH as the reason for the fast rise. We suggest instead that the fast rise was caused by the disruption of a very low-mass star ($\lesssim0.1\msun$) with an impact parameter of $b\sim0.8$. Theoretical fallback rate estimates based on realistic stellar profiles \citep{Law-Smith2020} can reproduce the rise profile with such disruption parameters. With a radiative efficiency of $\eta\sim0.0022$, the model can also match the peak brightness.
\item Dense spectral follow-up reveals a blue continuum that cools slowly with time and broad Balmer and \ion{He}{II} lines as well as weak \ion{He}{I} $\lambda\lambda$5876,6678 emission, features that are typically seen in TDEs. The lines slowly become narrower with time as their luminosity drops, which is also typically seen in TDEs.
\item The early broad profile of H$\alpha$ matches what is theoretically expected from an optically thick outflow \citep{Roth2017}. The Balmer lines show a flat Balmer decrement ($L_{H\alpha}$/$L_{H\beta} \sim 1.58$) suggesting that the lines are collisionally excited rather than being produced via photoionisation \citep{Short2020}, which is in contrast to typical active galactic nuclei.
\item We find, for the first time in a TDE, a sharp, narrow emission peak at a rest wavelength of $\sim$ 6353 \AA\  on top of the blue side of the broad H$\alpha$ profile. This feature is clearly present in the first four spectra ($\sim 4-10$ days post-peak) and then disappears. We argue that this might be clumpy material preceding the bulk of the outflow ---which is directed towards the observer--- manifesting as a high-velocity component of H$\alpha$ ($v= -9\,584 \rm km\,s^{-1}$). We also discuss other possibilities such as FUV pumping, but we conclude that this seems less likely.
\item Fitting the SED data of AT~2023clx with a blackbody, we find a rapidly expanding blackbody photosphere with a very high constant velocity of $v\approx4\,000\,\rm km\,s^{-1}$ (i.e. $\sim$ 0.014c). Interestingly, there is rapid cooling observed in the early light curves manifesting as a break in the temperature evolution. Such a break was seen in the light curves of the very bright transient ASASSN-15lh (which has been proposed to be an extreme TDE), but the break is much shorter and less prominent in AT~2023clx. 
\item Comparing with theoretical work \citep{Hong2022}, we find that for the disruption parameters (BH mass, stellar mass, and $\beta$) arising from our analysis, highly efficient disc formation after the debris stream self-crossing is favoured.
Combined with the low density of the reprocessing layer (due to the low initial mass of the star), we are able to observe the temperature break caused by the prompt and efficient disc formation.
Thus, even though the properties are markedly different, AT~2023clx appears to share many of the above properties with ASASSN-15lh \citep{Leloudas2016}.
\item In summary, we propose a plausible scenario for AT~2023clx involving the disruption of a very low-mass star ($\lesssim0.1\msun$) with an outflow launched towards our line of sight, and disruption properties that led to circularisation and prompt and efficient accretion-disc formation, observed through a low-density photosphere.
\end{enumerate}

AT~2023clx adds to the diversity of observational properties of TDEs. However, its similarities with ASASSN-15lh and AT~2018hyz allow us to draw parallels by invoking common underlying physical mechanisms likely to be at play in TDEs. Nonetheless, a comprehensive understanding of the diversity in TDE properties necessitates rigorous theoretical interpretation and modelling.
Furthermore, our study serves to demonstrate that ignoring transients based on `{AGN'} contextual classification only in simple selection filters will lead to missing  TDEs, and introduce a bias towards passive galaxies. Indeed, in the local Universe, low-luminosity AGN may well account for a non-negligible fraction of the TDE rates, especially towards the higher BH mass end, and this fraction is only expected to increase with redshift, even for lower BH masses \citep{Polkas2023}. This may be an important consideration given that current empirically determined TDE rates are too low compared to theoretical expectations (e.g. \citealt{Magorrian1999,Wang2004,Stone2016,Stone2020}). Observational tests may be possible during the era of the upcoming time-domain survey of the Rubin Observatory (LSST; \citealt{Ivezic2019}).

\begin{acknowledgements}
We are indebted to an anonymous referee for many insightful comments and suggestions.
We thank K. Uno, K. Maeda and T. Nagao for interesting discussions on the object. P.C. and R.K. acknowledges support via the Research Council of Finland (grant 340613). G.L. is supported by a research grant (19054) from VILLUM FONDEN. PR acknowledges support from STFC grant 2742655. TR and SM acknowledge support from the Research Council of Finland project 350458. AA is supported by the European Research Council (ERC) under the European Union’s Horizon 2020 research and innovation programme (grant agreement No.~948381). This work was funded by ANID, Millennium Science Initiative, ICN12\_009. I.A. acknowledges support from the European Research Council (ERC) under the European Union’s Horizon 2020 research and innovation program (grant agreement number 852097), from the Israel Science Foundation (grant number 2752/19), from the United States - Israel Binational Science Foundation (BSF; grant number 2018166), and from the Pazy foundation (grant number 216312). Y.-Z. Cai is supported by the National Natural Science Foundation of China (NSFC, Grant No. 12303054), the Yunnan Fundamental Research Projects (Grant No. 202401AU070063) and the International Centre of Supernovae, Yunnan Key Laboratory (No. 202302AN360001). TWC acknowledges the Yushan Young Fellow Program by the Ministry of Education, Taiwan for the financial support. L.G and T.E.M.B. acknowledges financial support from the Spanish Ministerio de Ciencia e Innovaci\'on (MCIN), the Agencia Estatal de Investigaci\'on (AEI) 10.13039/501100011033, and the European Union Next Generation EU/PRTR funds under the 2021 Juan de la Cierva program FJC2021-047124-I and the PID2020-115253GA-I00 HOSTFLOWS project, from Centro Superior de Investigaciones Cient\'ificas (CSIC) under the PIE project 20215AT016, and the program Unidad de Excelencia Mar\'ia de Maeztu CEX2020-001058-M , and from the Departament de Recerca i Universitats de la Generalitat de Catalunya through the 2021-SGR-01270 grant. CPG acknowledges financial support from the Secretary of Universities and Research (Government of Catalonia) and by the Horizon 2020 Research and Innovation Programme of the European Union under the Marie Sk\l{}odowska-Curie and the Beatriu de Pin\'os 2021 BP 00168 programme, from the Spanish Ministerio de Ciencia e Innovaci\'on (MCIN) and the Agencia Estatal de Investigaci\'on (AEI) 10.13039/501100011033 under the PID2020-115253GA-I00 HOSTFLOWS project, and the program Unidad de
Excelencia Mar\'ia de Maeztu CEX2020-001058-M. MN is supported by the European Research Council (ERC) under the European Union’s Horizon 2020 research and innovation programme (grant agreement No.~948381) and by UK Space Agency Grant No.~ST/Y000692/1. FO acknowledges support from MIUR, PRIN 2020 (grant 2020KB33TP) ``Multimessenger astronomy in the Einstein Telescope Era (METE)''. M.P. acknowledges support from a UK Research and Innovation Fellowship
(MR/T020784/1). A.N. acknowledges the Warwick Astrophysics PhD prize scholarship made possible thanks to a generous philanthropic donation. SJS  acknowledges funding from STFC Grants ST/X006506/1, ST/T000198/1 and ST/X001253/1. YW acknowledges support from the Strategic Priority Research Program of the Chinese Academy of Sciences (Grant No. XDB0550200).

This work is based (in part) on observations collected at the European Organisation for Astronomical Research in the Southern Hemisphere under ESO DDT program 110.25AX (PI: Wevers) and as part of ePESSTO+ under ESO program ID 108.220C and 111.24PR (PI: Inserra) and on observations made with the Nordic Optical Telescope, owned in collaboration by the University of Turku and Aarhus University, and operated jointly by Aarhus University, the University of Turku and the University of Oslo, representing Denmark, Finland and Norway, the University of Iceland and Stockholm University at the Observatorio del Roque de los Muchachos, La Palma, Spain, of the Instituto de Astrofisica de Canarias under NOT programmes 66-019, 66-506 and 67-021. The NOT data presented here were obtained with ALFOSC, which is provided by the Instituto de Astrofisica de Andalucia (IAA) under a joint agreement with the University of Copenhagen and NOT. Some of the observations from the Nordic Optical Telescope were obtained through the NUTS2 collaboration which are supported in part by the Instrument Centre for Danish Astrophysics (IDA), and the Finnish Centre for Astronomy with ESO (FINCA) via Academy of Finland grant nr 306531. This publication makes use of data products from the Two Micron All Sky Survey, which is a joint project of the University of Massachusetts and the Infrared Processing and Analysis Center/California Institute of Technology, funded by the National Aeronautics and Space Administration and the National Science Foundation.

\end{acknowledgements}

\bibliographystyle{aa}
\bibliography{bib.bib}

\appendix{}
\twocolumn
\centering

\section{Photometry tables} \label{apdx:phot_tab}

\begin{table}[h]
    \def\arraystretch{1.1}%
    \setlength\tabcolsep{3pt}
    \centering
    \fontsize{9}{11}\selectfont
    \caption{Photometry of AT~2023clx.}
   
    \begin{tabular}{c c c c c c}
    \hline
    \hline
        Date & MJD & Phase (d) & Telescope & Band & Mag (Error) \\
    \hline
2023-02-27 & 60002.75 & 5.50 & \textit{Swift} & W2 & 15.85 (0.06) \\ 
2023-03-05 & 60008.37 & 11.12 & \textit{Swift} & W2 & 16.49 (0.09) \\ 
2023-03-09 & 60012.47 & 15.22 & \textit{Swift} & W2 & 16.85 (0.10) \\ 
2023-03-17 & 60020.29 & 23.04 & \textit{Swift} & W2 & 17.06 (0.17) \\ 
2023-03-19 & 60022.53 & 25.28 & \textit{Swift} & W2 & 17.15 (0.14) \\ 
2023-03-22 & 60025.98 & 28.73 & \textit{Swift} & W2 & 17.10 (0.11) \\ 
2023-03-27 & 60030.29 & 33.04 & \textit{Swift} & W2 & 17.25 (0.14) \\ 
2023-04-01 & 60035.05 & 37.80 & \textit{Swift} & W2 & 17.31 (0.16) \\ 
2023-04-06 & 60040.61 & 43.36 & \textit{Swift} & W2 & 17.56 (0.13) \\ 
2023-04-09 & 60043.31 & 46.06 & \textit{Swift} & W2 & 17.64 (0.20) \\ 
2023-04-11 & 60045.70 & 48.45 & \textit{Swift} & W2 & 17.75 (0.15) \\ 
2023-04-24 & 60058.21 & 60.96 & \textit{Swift} & W2 & 17.87 (0.24) \\ 
2023-05-03 & 60067.53 & 70.28 & \textit{Swift} & W2 & 17.97 (0.26) \\ 
2023-05-08 & 60072.04 & 74.79 & \textit{Swift} & W2 & 17.92 (0.25) \\ 
2023-05-13 & 60077.44 & 80.19 & \textit{Swift} & W2 & 17.97 (0.26) \\ 
2023-02-27 & 60002.76 & 5.51 & \textit{Swift} & M2 & 15.50 (0.06) \\ 
2023-03-05 & 60008.37 & 11.12 & \textit{Swift} & M2 & 16.04 (0.08) \\ 
2023-03-09 & 60012.48 & 15.23 & \textit{Swift} & M2 & 16.40 (0.08) \\ 
2023-03-11 & 60014.61 & 17.36 & \textit{Swift} & M2 & 16.46 (0.10) \\ 
2023-03-17 & 60020.29 & 23.04 & \textit{Swift} & M2 & 16.72 (0.19) \\ 
2023-03-19 & 60022.54 & 25.29 & \textit{Swift} & M2 & 16.99 (0.16) \\ 
2023-03-23 & 60026.12 & 28.87 & \textit{Swift} & M2 & 16.99 (0.14) \\ 
2023-03-27 & 60030.30 & 33.05 & \textit{Swift} & M2 & 17.21 (0.17) \\ 
2023-04-01 & 60035.05 & 37.80 & \textit{Swift} & M2 & 17.18 (0.16) \\ 
2023-04-06 & 60040.61 & 43.37 & \textit{Swift} & M2 & 17.73 (0.16) \\ 
2023-04-09 & 60043.32 & 46.07 & \textit{Swift} & M2 & 17.60 (0.22) \\ 
2023-04-11 & 60045.71 & 48.46 & \textit{Swift} & M2 & 17.69 (0.17) \\ 
2023-05-08 & 60072.04 & 74.79 & \textit{Swift} & M2 & 18.06 (0.32) \\ 
2023-02-27 & 60002.75 & 5.50 & \textit{Swift} & W1 & 15.59 (0.06) \\ 
2023-03-05 & 60008.37 & 11.12 & \textit{Swift} & W1 & 16.05 (0.10) \\ 
2023-03-09 & 60012.47 & 15.22 & \textit{Swift} & W1 & 16.25 (0.09) \\ 
2023-03-11 & 60014.60 & 17.35 & \textit{Swift} & W1 & 16.35 (0.10) \\ 
2023-03-17 & 60020.28 & 23.03 & \textit{Swift} & W1 & 16.56 (0.17) \\ 
2023-03-19 & 60022.53 & 25.28 & \textit{Swift} & W1 & 16.78 (0.15) \\ 
2023-03-22 & 60025.97 & 28.73 & \textit{Swift} & W1 & 17.00 (0.13) \\ 
2023-03-27 & 60030.29 & 33.04 & \textit{Swift} & W1 & 17.11 (0.16) \\ 
2023-03-31 & 60034.78 & 37.53 & \textit{Swift} & W1 & 17.18 (0.17) \\ 
2023-04-06 & 60040.61 & 43.36 & \textit{Swift} & W1 & 17.31 (0.13) \\ 
2023-04-09 & 60043.31 & 46.06 & \textit{Swift} & W1 & 17.57 (0.22) \\ 
2023-04-11 & 60045.70 & 48.45 & \textit{Swift} & W1 & 17.62 (0.17) \\ 
2023-04-16 & 60050.47 & 53.22 & \textit{Swift} & W1 & 18.00 (0.36) \\ 
2023-05-03 & 60067.52 & 70.27 & \textit{Swift} & W1 & 18.13 (0.36) \\ 
2023-05-08 & 60072.03 & 74.78 & \textit{Swift} & W1 & 17.77 (0.26) \\ 
2023-02-27 & 60002.75 & 5.50 & \textit{Swift} & U & 15.52 (0.06) \\ 
2023-03-05 & 60008.37 & 11.12 & \textit{Swift} & U & 15.97 (0.11) \\ 
2023-03-09 & 60012.47 & 15.22 & \textit{Swift} & U & 16.01 (0.09) \\ 
2023-03-11 & 60014.60 & 17.35 & \textit{Swift} & U & 16.09 (0.09) \\ 
2023-03-19 & 60022.53 & 25.28 & \textit{Swift} & U & 16.59 (0.16) \\ 
2023-03-22 & 60025.98 & 28.73 & \textit{Swift} & U & 16.67 (0.11) \\ 
2023-03-27 & 60030.29 & 33.04 & \textit{Swift} & U & 16.79 (0.15) \\ 

    \hline
    \hline
    \end{tabular}
\label{tab:phot}
\end{table}

\begin{table}[h]
    \def\arraystretch{1.1}%
    \setlength\tabcolsep{3pt}
    \centering
    \fontsize{9}{11}\selectfont
    \caption*{Table \ref{tab:phot} continued.}
   
    \begin{tabular}{c c c c c c}
    \hline
    \hline
        Date & MJD & Phase (d) & Telescope & Band & Mag (Error) \\
    \hline
2023-03-31 & 60034.78 & 37.53 & \textit{Swift} & U & 16.83 (0.15) \\ 
2023-04-06 & 60040.61 & 43.36 & \textit{Swift} & U & 17.01 (0.13) \\ 
2023-04-09 & 60043.31 & 46.06 & \textit{Swift} & U & 17.22 (0.20) \\ 
2023-04-11 & 60045.70 & 48.45 & \textit{Swift} & U & 17.33 (0.15) \\ 
2023-04-24 & 60058.21 & 60.96 & \textit{Swift} & U & 17.50 (0.24) \\ 
2023-05-03 & 60067.52 & 70.27 & \textit{Swift} & U & 17.57 (0.26) \\ 
2023-05-08 & 60072.03 & 74.78 & \textit{Swift} & U & 17.71 (0.29) \\ 
2023-05-18 & 60082.34 & 85.09 & \textit{Swift} & U & 17.57 (0.26) \\ 
2023-02-27 & 60002.75 & 5.50 & \textit{Swift} & B & 15.72 (0.15) \\ 
2023-03-09 & 60012.47 & 15.22 & \textit{Swift} & B & 16.21 (0.24) \\ 
2023-03-11 & 60014.33 & 17.08 & \textit{Swift} & B & 16.42 (0.29) \\ 
2023-03-14 & 60017.78 & 20.53 & \textit{Swift} & B & 16.54 (0.32) \\ 
2023-03-17 & 60020.29 & 23.04 & \textit{Swift} & B & 16.54 (0.32) \\ 
2023-03-23 & 60026.11 & 28.86 & \textit{Swift} & B & 16.74 (0.38) \\ 
2023-03-13 & 60016.21 & 18.96 & ZTF & g & 16.26 (0.06) \\ 
2023-03-17 & 60020.22 & 22.98 & ZTF & g & 16.42 (0.05) \\ 
2023-03-28 & 60031.29 & 34.04 & ZTF & g & 16.85 (0.06) \\ 
2023-04-02 & 60036.32 & 39.07 & ZTF & g & 17.07 (0.10) \\ 
2023-04-09 & 60043.33 & 46.08 & ZTF & g & 17.31 (0.06) \\ 
2023-04-11 & 60045.32 & 48.07 & ZTF & g & 17.21 (0.06) \\ 
2023-04-16 & 60050.18 & 52.94 & ZTF & g & 17.39 (0.06) \\ 
2023-04-18 & 60052.27 & 55.02 & ZTF & g & 17.42 (0.06) \\ 
2023-04-20 & 60054.26 & 57.01 & ZTF & g & 17.61 (0.07) \\ 
2023-04-22 & 60056.21 & 58.97 & ZTF & g & 17.79 (0.07) \\ 
2023-04-24 & 60058.30 & 61.05 & ZTF & g & 17.65 (0.07) \\ 
2023-04-26 & 60060.25 & 63.01 & ZTF & g & 17.78 (0.06) \\ 
2023-04-28 & 60062.26 & 65.01 & ZTF & g & 17.81 (0.06) \\ 
2023-05-03 & 60067.35 & 70.11 & ZTF & g & 17.74 (0.15) \\ 
2023-05-09 & 60073.34 & 76.09 & ZTF & g & 17.77 (0.08) \\ 
2023-05-13 & 60077.27 & 80.03 & ZTF & g & 17.90 (0.08) \\ 
2023-05-16 & 60080.24 & 82.99 & ZTF & g & 17.90 (0.08) \\ 
2023-05-20 & 60084.25 & 87.01 & ZTF & g & 17.95 (0.06) \\ 
2023-05-22 & 60086.26 & 89.01 & ZTF & g & 18.09 (0.08) \\ 
2023-05-24 & 60088.28 & 91.03 & ZTF & g & 18.21 (0.09) \\ 
2023-06-18 & 60113.27 & 116.02 & ZTF & g & 18.55 (0.10) \\ 
2023-06-20 & 60115.24 & 117.99 & ZTF & g & 18.63 (0.10) \\ 
2023-06-22 & 60117.27 & 120.02 & ZTF & g & 18.66 (0.15) \\ 
2023-02-11 & 59986.28 & -10.96 & ASASSN & g & 19.36 (99.00) \\ 
2023-02-12 & 59987.27 & -9.97 & ASASSN & g & 17.62 (0.49) \\ 
2023-02-13 & 59988.25 & -8.99 & ASASSN & g & 16.99 (0.21) \\ 
2023-02-16 & 59991.24 & -6.01 & ASASSN & g & 16.06 (0.11) \\ 
2023-02-18 & 59993.24 & -4.01 & ASASSN & g & 15.54 (0.08) \\ 
2023-02-22 & 59997.22 & -0.03 & ASASSN & g & 15.37 (0.07) \\ 
2023-02-26 & 60001.20 & 3.96 & ASASSN & g & 15.66 (0.08) \\ 
2023-02-28 & 60003.72 & 6.48 & ASASSN & g & 15.68 (0.08) \\ 
2023-03-03 & 60006.76 & 9.51 & ASASSN & g & 16.05 (0.16) \\ 
2023-03-10 & 60013.99 & 16.74 & ASASSN & g & 16.23 (0.17) \\ 
2023-03-13 & 60016.98 & 19.74 & ASASSN & g & 16.31 (0.11) \\ 
2023-03-17 & 60020.40 & 23.15 & ASASSN & g & 16.47 (0.11) \\ 
2023-03-20 & 60023.38 & 26.13 & ASASSN & g & 16.49 (0.11) \\ 
2023-03-23 & 60026.34 & 29.10 & ASASSN & g & 16.58 (0.12) \\ 
2023-03-26 & 60029.36 & 32.12 & ASASSN & g & 16.67 (0.11) \\ 
2023-03-28 & 60031.46 & 34.21 & ASASSN & g & 16.83 (0.12) \\ 
    \hline
    \hline
    \end{tabular}
\end{table}

\begin{table}[h]
    \def\arraystretch{1.1}%
    \setlength\tabcolsep{3pt}
    \centering
    \fontsize{9}{11}\selectfont
    \caption*{Table \ref{tab:phot} continued.}
   
    \begin{tabular}{c c c c c c}
    \hline
    \hline
        Date & MJD & Phase (d) & Telescope & Band & Mag (Error) \\
    \hline
2023-03-31 & 60034.61 & 37.36 & ASASSN & g & 16.97 (0.25) \\ 
2023-04-09 & 60043.87 & 46.62 & ASASSN & g & 17.16 (0.20) \\ 
2023-04-15 & 60049.64 & 52.39 & ASASSN & g & 17.35 (0.19) \\ 
2023-02-18 & 59993.34 & -3.90 & ATLAS & c & 16.01 (0.05) \\ 
2023-02-22 & 59997.96 & 0.71 & ATLAS & c & 15.60 (0.05) \\ 
2023-02-26 & 60001.70 & 4.45 & ATLAS & c & 15.87 (0.05) \\ 
2023-03-02 & 60005.28 & 8.03 & ATLAS & c & 16.06 (0.05) \\ 
2023-03-17 & 60020.95 & 23.70 & ATLAS & c & 16.85 (0.07) \\ 
2023-03-29 & 60032.95 & 35.70 & ATLAS & c & 17.17 (0.07) \\ 
2023-04-13 & 60047.83 & 50.59 & ATLAS & c & 17.58 (0.12) \\ 
2023-04-15 & 60049.73 & 52.48 & ATLAS & c & 17.94 (0.11) \\ 
2023-04-17 & 60051.39 & 54.15 & ATLAS & c & 17.97 (0.11) \\ 
2023-04-21 & 60055.80 & 58.56 & ATLAS & c & 18.19 (0.15) \\ 
2023-04-23 & 60057.37 & 60.12 & ATLAS & c & 18.28 (0.14) \\ 
2023-05-15 & 60079.33 & 82.08 & ATLAS & c & 18.53 (0.22) \\ 
2023-05-18 & 60082.78 & 85.53 & ATLAS & c & 18.90 (0.25) \\ 
2023-06-10 & 60105.33 & 108.08 & ATLAS & c & 18.85 (0.24) \\ 
2023-06-12 & 60107.55 & 110.30 & ATLAS & c & 18.69 (0.23) \\ 
2023-02-17 & 59992.36 & -4.89 & ZTF & r & 16.29 (0.08) \\ 
2023-03-13 & 60016.35 & 19.11 & ZTF & r & 16.42 (0.06) \\ 
2023-03-28 & 60031.30 & 34.05 & ZTF & r & 16.90 (0.07) \\ 
2023-04-24 & 60058.21 & 60.97 & ZTF & r & 17.93 (0.12) \\ 
2023-05-26 & 60090.18 & 92.94 & ZTF & r & 18.37 (0.21) \\ 
2023-06-04 & 60099.25 & 102.00 & ZTF & r & 18.45 (0.22) \\ 
2023-02-11 & 59986.47 & -10.77 & ATLAS & o & 19.01 (99.00) \\ 
2023-02-13 & 59988.03 & -9.22 & ATLAS & o & 18.60 (0.20) \\ 
2023-02-14 & 59989.24 & -8.00 & ATLAS & o & 18.25 (0.17) \\ 
2023-02-15 & 59990.33 & -6.91 & ATLAS & o & 17.97 (0.12) \\ 
2023-02-27 & 60002.10 & 4.86 & ATLAS & o & 16.24 (0.06) \\ 
2023-02-28 & 60003.94 & 6.69 & ATLAS & o & 16.31 (0.06) \\ 
2023-03-03 & 60006.13 & 8.89 & ATLAS & o & 16.30 (0.07) \\ 
2023-03-04 & 60007.94 & 10.70 & ATLAS & o & 16.27 (0.08) \\ 
2023-03-10 & 60013.75 & 16.50 & ATLAS & o & 16.71 (0.10) \\ 
2023-03-13 & 60016.71 & 19.47 & ATLAS & o & 16.78 (0.07) \\ 
2023-03-14 & 60017.95 & 20.70 & ATLAS & o & 17.07 (0.10) \\ 
2023-03-17 & 60020.93 & 23.69 & ATLAS & o & 16.74 (0.09) \\ 
2023-03-19 & 60022.25 & 25.01 & ATLAS & o & 17.10 (0.07) \\ 
2023-03-21 & 60024.85 & 27.60 & ATLAS & o & 17.39 (0.09) \\ 
2023-03-23 & 60026.45 & 29.21 & ATLAS & o & 17.61 (0.15) \\ 
2023-03-26 & 60029.02 & 31.77 & ATLAS & o & 17.43 (0.09) \\ 
2023-03-27 & 60030.39 & 33.15 & ATLAS & o & 17.38 (0.08) \\ 
2023-03-30 & 60033.76 & 36.52 & ATLAS & o & 17.55 (0.11) \\ 
2023-03-31 & 60034.88 & 37.64 & ATLAS & o & 17.62 (0.12) \\ 
2023-04-07 & 60041.64 & 44.39 & ATLAS & o & 17.62 (0.17) \\ 
2023-04-09 & 60043.91 & 46.66 & ATLAS & o & 17.55 (0.15) \\ 
2023-04-11 & 60045.42 & 48.17 & ATLAS & o & 17.83 (0.16) \\ 
2023-05-08 & 60072.10 & 74.85 & ATLAS & o & 18.51 (0.28) \\ 
2023-05-21 & 60085.29 & 88.05 & ATLAS & o & 18.92 (0.25) \\ 
2023-05-25 & 60089.28 & 92.03 & ATLAS & o & 18.68 (0.20) \\ 
2023-03-15 & 60018.07 & 20.83 & NOT & J & 17.39 (0.11) \\ 
2023-04-25 & 60059.98 & 62.73 & NOT & J & 18.98 (0.09) \\ 
2023-05-30 & 60094.89 & 97.65 & NOT & J & 19.51 (0.12) \\ 
2023-03-15 & 60018.06 & 20.81 & NOT & H & 17.70 (0.11) \\ 
2023-04-25 & 60059.99 & 62.74 & NOT & H & 18.73 (0.15) \\ 
2023-05-30 & 60094.90 & 97.66 & NOT & H & 19.44 (0.21) \\ 
    \hline
    \hline
    \end{tabular}
\end{table}

\begin{table}[h]
    \def\arraystretch{1.1}%
    \setlength\tabcolsep{3pt}
    \centering
    \fontsize{9}{11}\selectfont
    \caption*{Table \ref{tab:phot} continued.}
   
    \begin{tabular}{c c c c c c}
    \hline
    \hline
        Date & MJD & Phase (d) & Telescope & Band & Mag (Error) \\
    \hline
2023-03-15 & 60018.03 & 20.79 & NOT & K & 18.18 (0.12) \\ 
2023-04-25 & 60059.99 & 62.75 & NOT & K & 18.57 (0.13) \\ 
2023-05-30 & 60094.91 & 97.66 & NOT & K & 19.08 (0.12) \\ 
    \hline
    \hline
    \end{tabular}
\tablefoot{All reported magnitudes are in the AB system and are host subtracted and dereddened for both Milky Way ($\rm E(B-V)$ = 0.027 mag) and host $\rm E(B-V)_{h}$ = 0.179 mag) reddening. The table is sorted from bluest to reddest band. A value of 99 in the error refers to an upper-limit.}
\end{table}

\clearpage
\onecolumn
\centering

\section{Supplementary figures} \label{apdx:graphs}

\begin{figure*}[h]
\centering
\includegraphics[width=0.84 \textwidth]{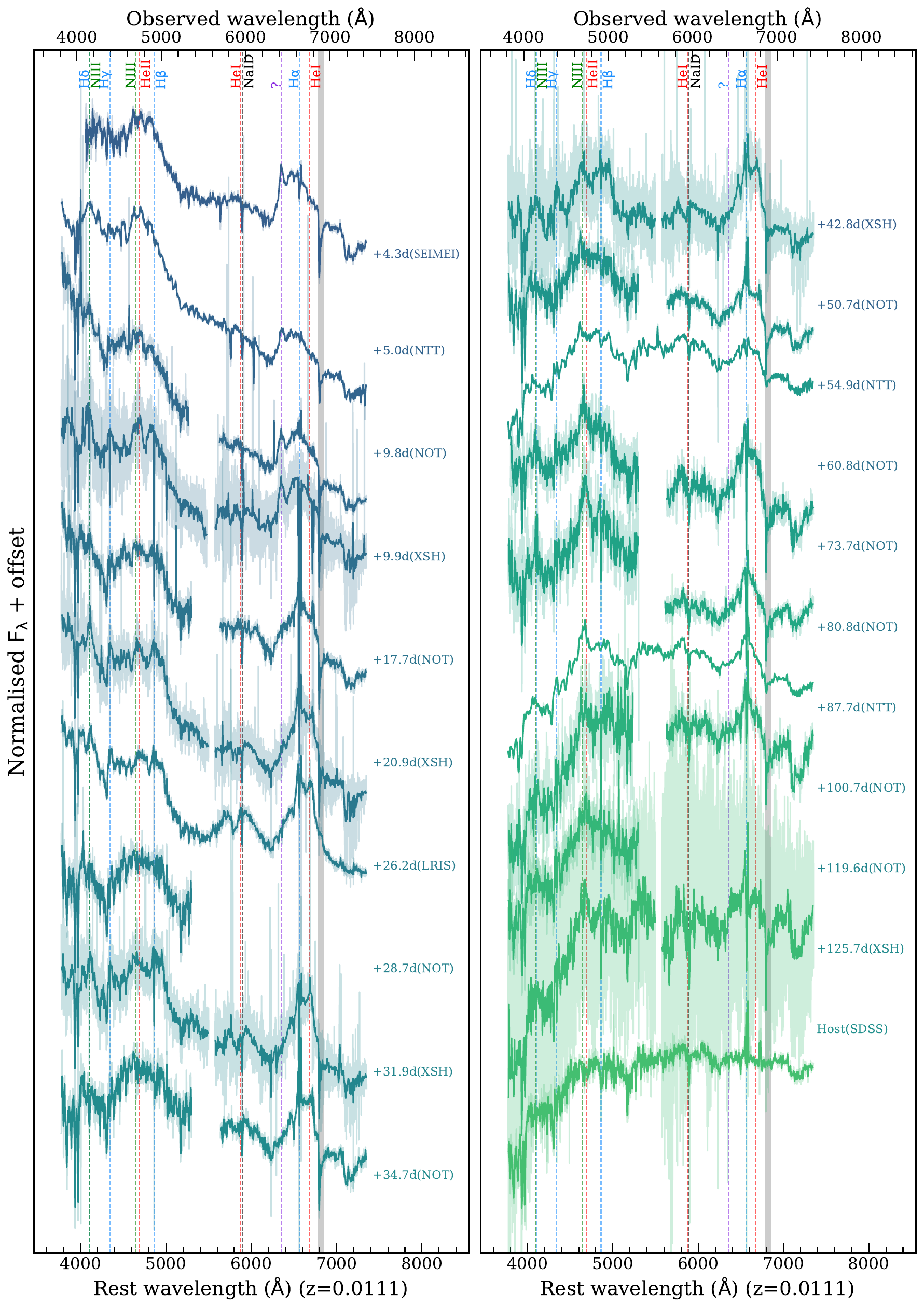}
\caption{Spectra of AT~2023clx before host subtraction (corrected for MW extinction). Emission lines are marked with vertical dashed lines.}
\label{fig:spectra_wo_host_sub}
\end{figure*}

\begin{figure}
  \centering
  \includegraphics[width=0.6 \textwidth]{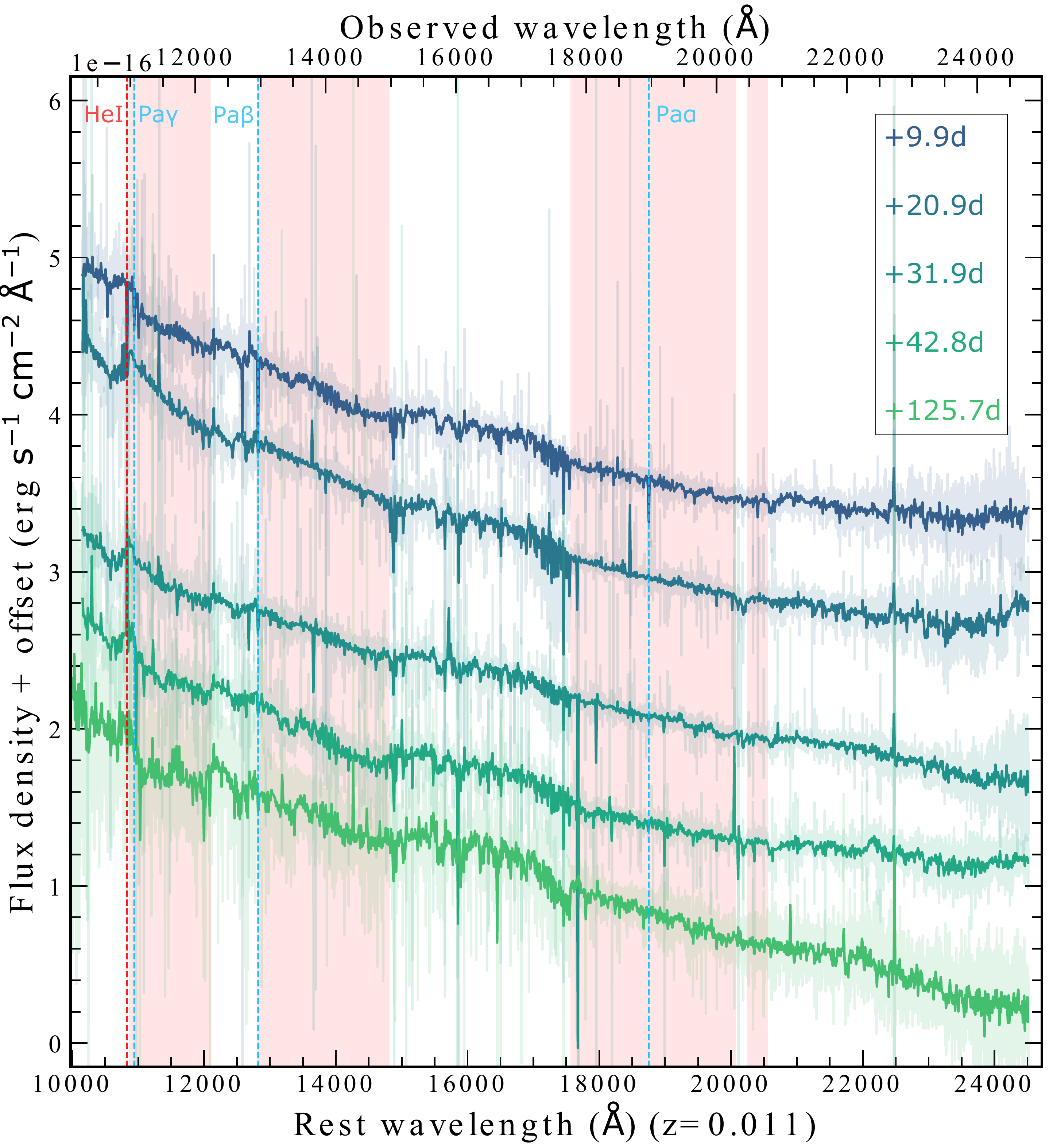}
  \caption{NIR arm X-shooter spectra of AT~2023clx. The telluric features (pink shaded regions) have been removed from the spectra. The spectra show \ion{He}{I} $\lambda$10\,830 but no evident hydrogen lines.}
  \label{fig:nir_spectra}
\end{figure}

\begin{figure}
  \centering
  \includegraphics[width=1 \textwidth]{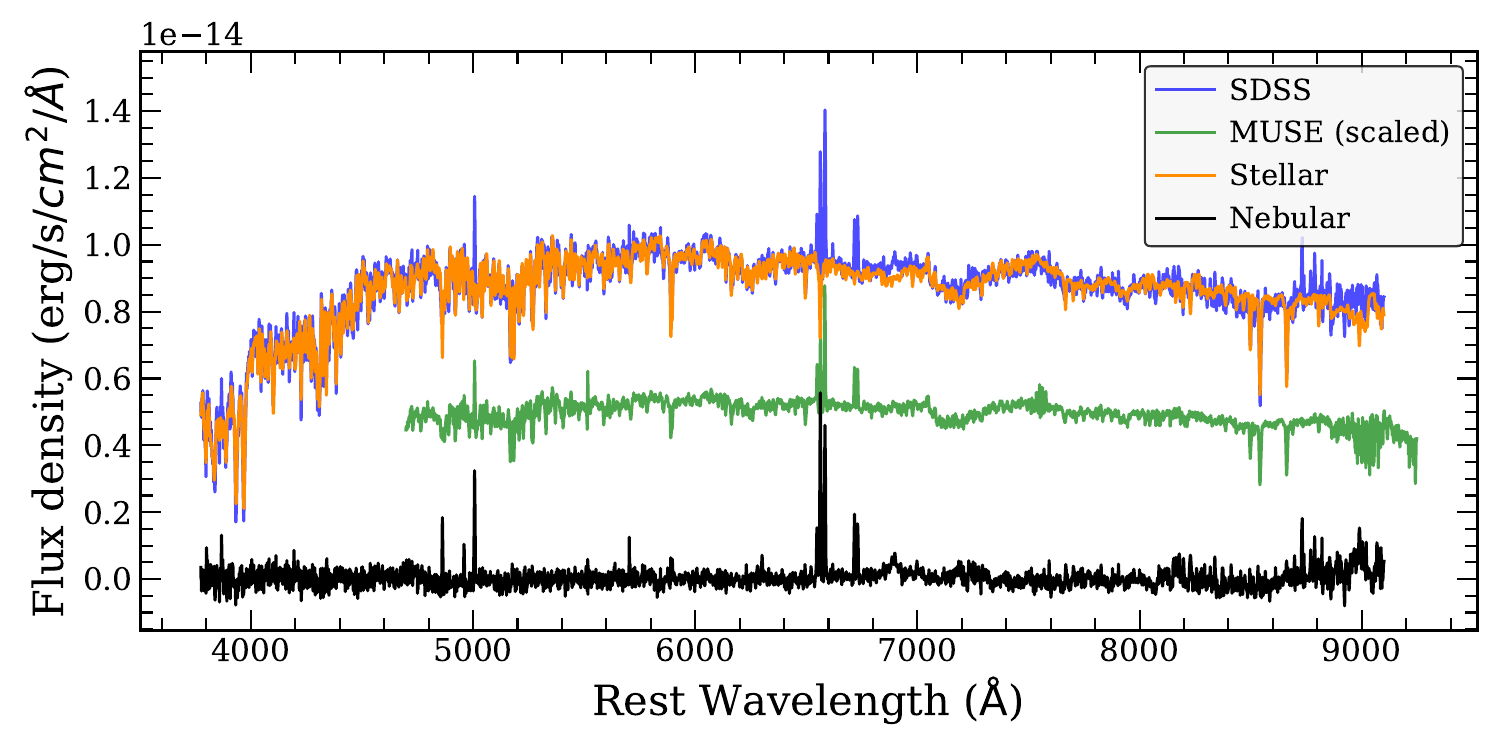}
  \caption{Original SDSS spectrum in blue, stellar continuum derived by our \texttt{STARLIGHT} fit in orange and the residual spectrum in black. We also plot the MUSE host spectrum with green (extracted with a $\sim 0.9''$ diameter aperture).}
  \label{fig:starlight}
\end{figure}

\begin{figure*}[h]
  \centering
  \includegraphics[width=19.5cm]{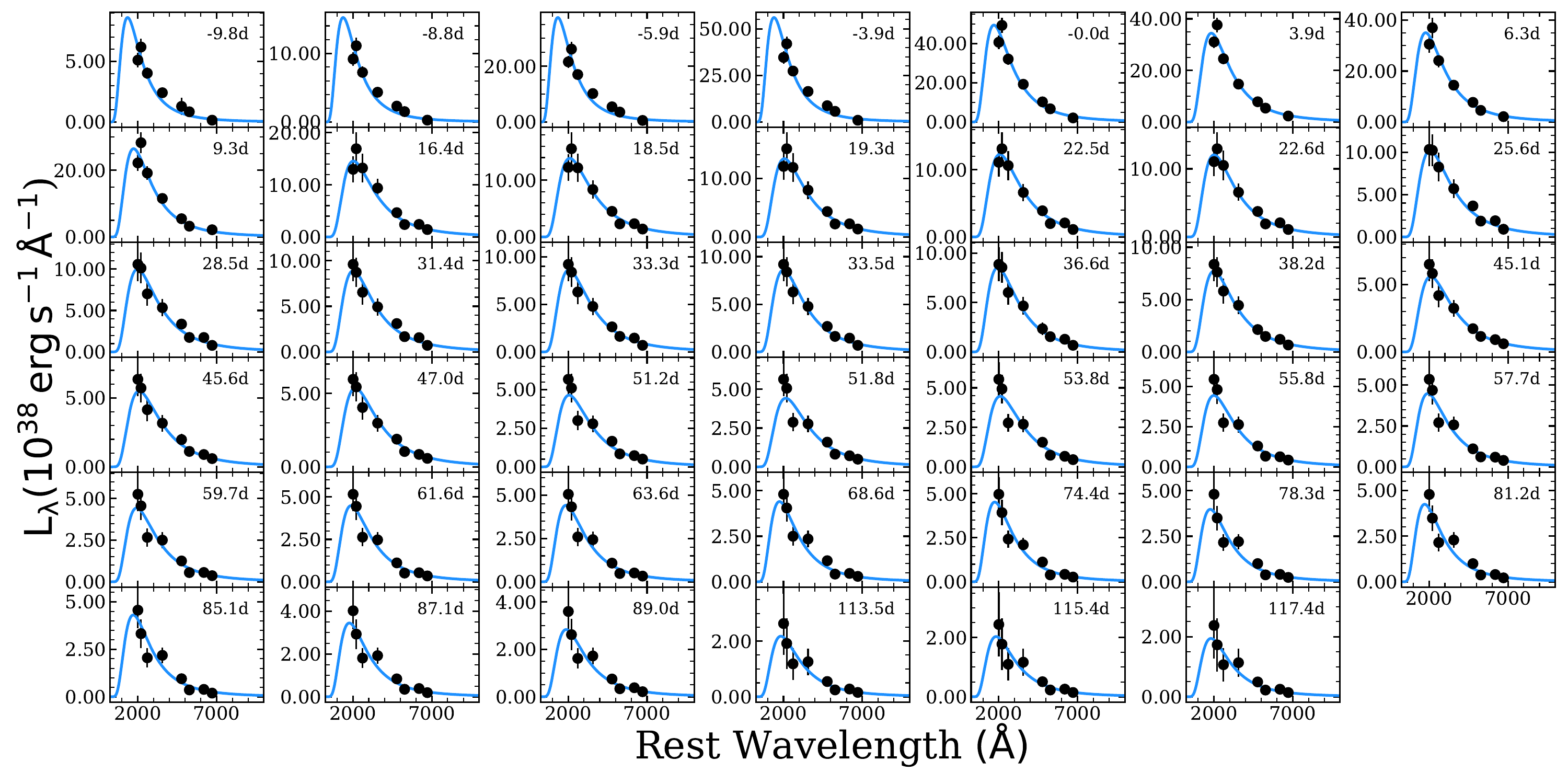}
  \caption{Blackbody fits to the SEDs at each available epoch.}
  \label{fig:BB_fits}
\end{figure*}

\begin{figure}
  \centering
  \includegraphics[width=0.5 \textwidth]{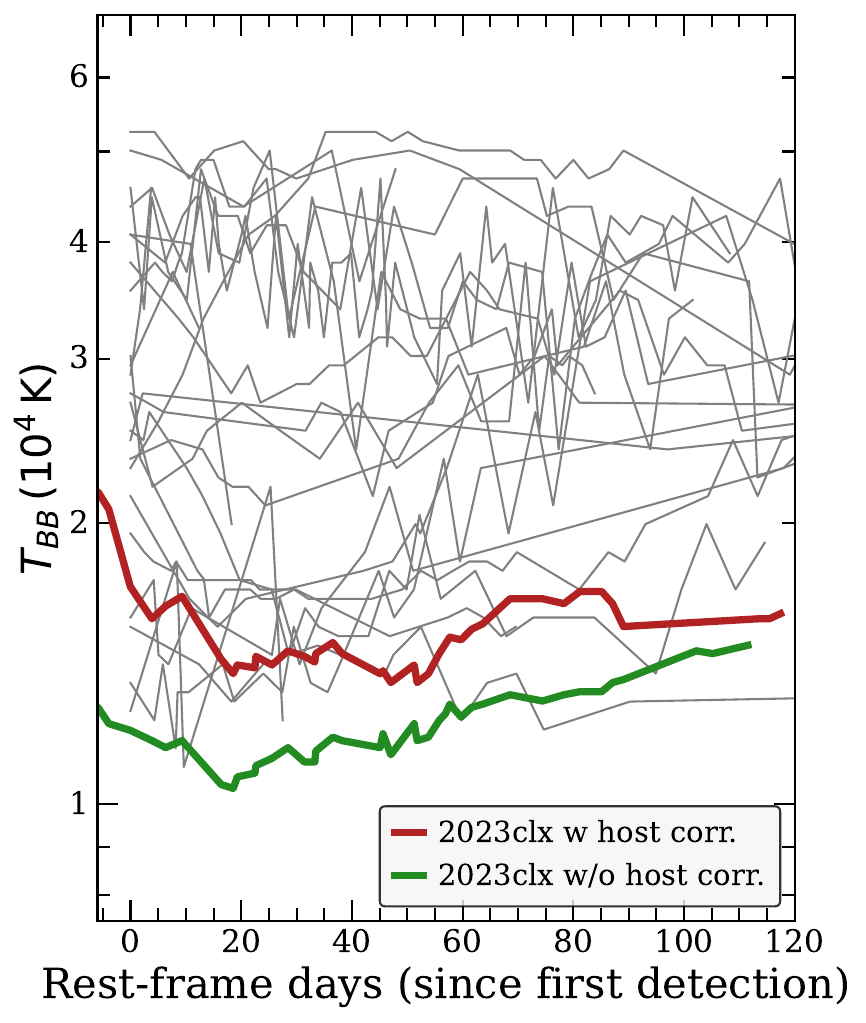}
  \caption{Blackbody temperature evolution of AT~2023clx with (red) and without (green) correcting for host galaxy extinction. The correction brings AT~2023clx in line with the temperature range and evolution of other TDEs -- these are shown in grey, and are from \citet{Hinkle2021a}. We note that although these comparison TDEs have not been corrected for host extinction, their host galaxies (unlike that of AT~2023clx) do not demand it.  }
  \label{fig:tbb_comp}
\end{figure}

\begin{figure*}[h]
  \centering
  \includegraphics[width=19.5cm]{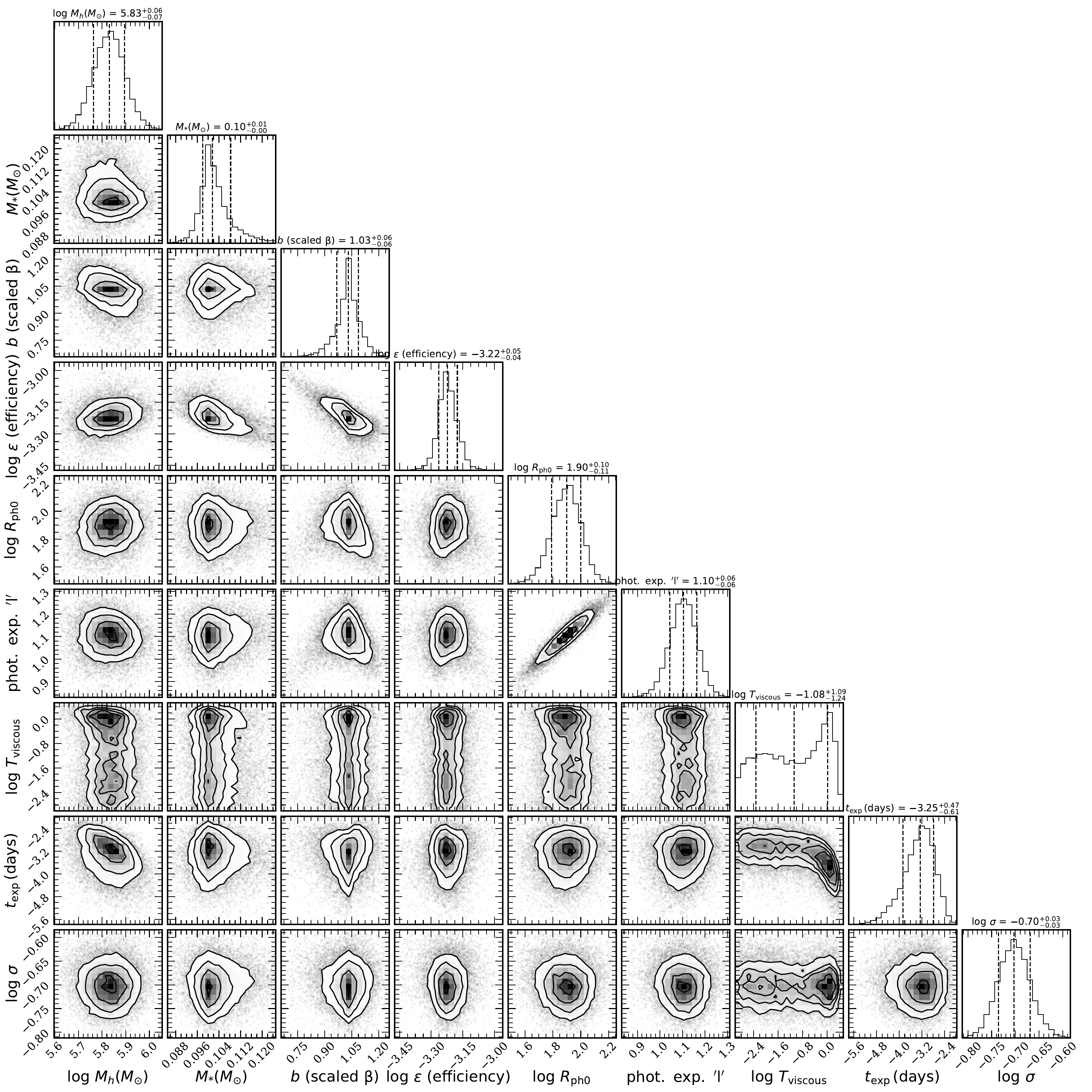}
  \caption{Posterior probability density functions for the free parameters of the model light curves in Fig. \ref{fig:mosfit}.}
  \label{fig:corner}
\end{figure*}

\begin{figure*}
\centering
\includegraphics[width=1 \textwidth]{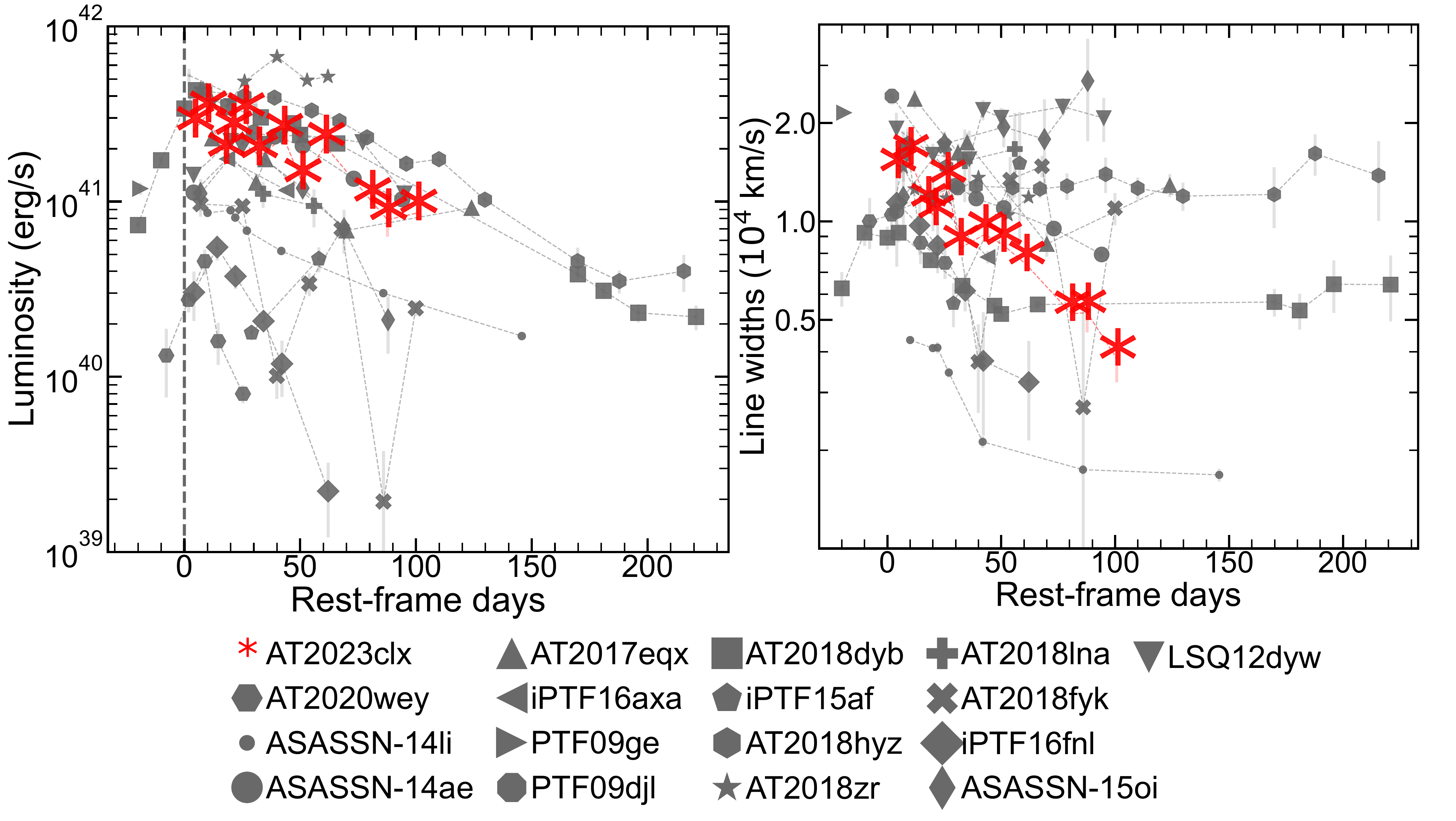}
\caption{H$\alpha$ properties as a function of time. Left panel: H$\alpha$ luminosity evolution of AT~2023clx as a function of time, compared with the sample of \citet{Charalampopoulos2022}. The dashed vertical line denotes the time of peak or discovery of each TDE. Right panel: H$\alpha$ line width evolution of AT~2023clx as a function of time, compared with the sample of \citet{Charalampopoulos2022}.}
\label{fig:ha_vs_sample}
\end{figure*}

\begin{figure*}
\centering
\includegraphics[width=1 \textwidth]{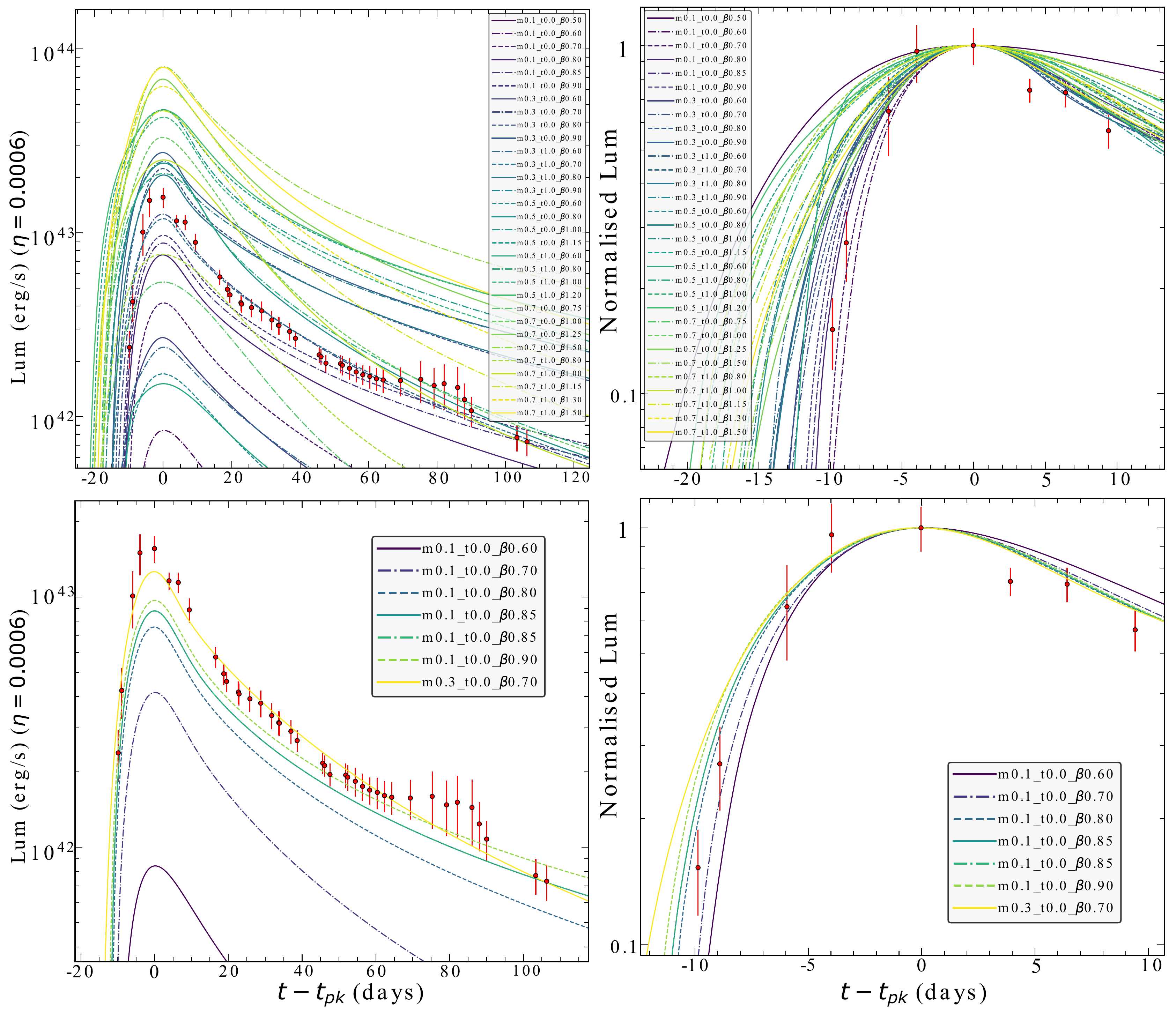}
\caption{Comparison of the bolometric luminosity evolution of AT~2023clx, against mass fallback rate models of \citet{Law-Smith2020} disrupted by a $10^{6.0}$ BH mass (conveniently matching the one of AT~2023clx). The labels of the models have the following format: mXX\_tXX\_$\beta$XX where the first XX is the star's mass, the second is the stellar age and the third is the impact parameter of each displayed simulation. In order to convert the mass fallback rate to luminosity we used a radiative efficiency parameter of $6 \times 10^{-4}$ which was the best-fit result of our \texttt{MOSFiT} fit. In the left panels the luminosities are not normalised while in the right ones the luminosities are normalised to peak light. The top panels are the same as the bottom ones, but in the top panels we show many models, with stellar masses up to $0.7 \msun$. In the top panels it is obvious that TDEs occurring from higher mass stars have shallower rises while lower mass stars should lead to faster rising light curves, as discussed by \citealt{Law-Smith2020}. In the bottom panels we show some models that have similar peak luminosities and rise times with AT~2023clx.}
\label{fig:JLS}
\end{figure*}

\clearpage

\end{document}